\documentclass[12pt]{article}

\newcommand{\cS}{\mathcal{S}}
\newcommand{\pinf}{\mathcal{\ell}_\sS}

\usepackage{amsmath,amssymb,latexsym,amsopn,amsfonts}
\usepackage[boxed,ruled,vlined,linesnumbered,noresetcount]{algorithm2e}
\newcommand\bull{{\operatorname{-\xspace}}}
\usepackage{framed} %microtype
\usepackage[normalem]{ulem}
\usepackage{array,xspace}
\usepackage{graphicx}
\usepackage{color}
\usepackage[dvipsnames]{xcolor}
\definecolor{frenchBlue}{rgb}{0.13, 0.55, 0.13}
\newcommand{\ems}[1]{{#1}} % \textcolor{blue}
\newcommand{\emsB}[1]{{#1}} % \textcolor{blue}
\newcommand{\algSize}{normalsize}%normalsize} %scriptsize %normalsize  
\newcommand{\algSizeSmall}{smaller} %scriptsize %normalsize  
\newcommand{\algSizeVerySmall}{normalsize}

\newcommand{\IM}[1]{[\textcolor{magenta}{IM: #1}]} %{\textcolor{blue}{#1}}
\newcommand{\im}[1]{\textcolor{magenta}{#1}}

\newcommand{\blitza}{\text{\usefont{U}{ulsy}{m}{n}\symbol{'011}}}
\newcommand{\trusted}{\mathsf{progressing}\xspace}

\newcommand{\BBB}{}%{\vspace*{-\bigskipamount}}
\newcommand{\F}{\vspace*{\smallskipamount}}
\newcommand{\FF}{\vspace*{\medskipamount}}

\newcommand{\bigO}{\mathcal{O}\xspace}
\newcommand{\remove}[1]{}

\newcommand{\smallO}{\mathrm{o}\xspace}

\newcommand{\done}{\mathsf{result}\xspace}

\newcommand{\Correct}{\mathit{Correct}\xspace}

\newcommand{\etal}{\emph{et al.}\xspace}
\newcommand{\eg}{\emph{e.g.,}\xspace}

\newcommand{\ie}{\emph{i.e.,}\xspace}
\newcommand{\Ie}{\emph{I.e.,}\xspace}

\newtheorem{remark}{Remark}[section]
\newtheorem{theorem}{Theorem}[section]
\newtheorem{lemma}[theorem]{Lemma}
\newtheorem{definition}{Definition}[section]
\newtheorem{assumption}[theorem]{Assumption}
\newtheorem{corollary}[theorem]{Corollary}
\newtheorem{claim}[theorem]{Claim}
\newcommand{\true}{\mathsf{True}\xspace}

\newcommand{\false}{\mathsf{False}\xspace}

\newcommand{\sS}{\mathcal{S}\xspace}
\newcommand{\sP}{\mathcal{P}\xspace}

\newcommand{\N}{\mathbb{N}\xspace}
\newcommand{\capacity}{\mathsf{channelCapacity}\xspace}

\usepackage{caption}
\SetKwInOut{Input}{input}
\SetKwInOut{Output}{output}

\newenvironment{claimProof}{\par\noindent\textbf{Proof of Claim  \clmcnt\space}}{\hfill $\Box_{Claim ~ \clmcnt}$}

\newenvironment{lemmaProof}{\par\noindent\textbf{Proof of Lemma  \lemcnt\space}}{\hfill $\Box_{Lemma ~ \lemcnt}$}
\newenvironment{lemmaProofSketch}{\par\noindent\textbf{Proof Sketch of Lemma  \lemcnt\space}}{\hfill $\Box_{Lemma ~ \lemcnt}$}
\newenvironment{theoremProof}{\par\noindent\textbf{Proof of Theorem  \thmcnt\space}}{\hfill $\Box_{Theorem ~ \thmcnt}$}
\newcommand{\clmcnt}{0}
\newcommand{\lemcnt}{0}
\newcommand{\thmcnt}{0}

\newcommand{\Section}[1]{\section{#1}}
\newcommand{\Subsection}[1]{\subsection{#1}}
\newcommand{\SubsectionS}[1]{\subsection*{#1}}
\newcommand{\Subsubsection}[1]{\subsubsection{#1}}
\newcommand{\Subsubsubsection}[1]{\paragraph{#1}}
\newcommand{\Subsubsubsubsection}[1]{\subparagraph{#1}}
\usepackage[utf8]{inputenc}
\usepackage{fullpage}

\begin{document}

\title{Loosely-self-stabilizing Byzantine-tolerant Binary Consensus for Signature-free Message-passing Systems}

	\author{Chryssis Georgiou~\footnote{University of Cyprus, Cyprus.~\texttt{\{chryssis,imarco01\}@ucy.ac.cy}} \and Ioannis Marcoullis~$^\ast$  \and Michel Raynal~\footnote{IRISA, Univ. Rennes 1, France,
		and Polytechnic Univ., Hong Kong.~\texttt{michel.raynal@irisa.fr}}  \and Elad Michael Schiller~\footnote{Chalmers University of Technology, Sweden. \texttt{elad@chalmers.se}}}

\date{}

\maketitle

\begin{abstract}
Many distributed applications, such as cloud computing, service replication, load balancing, and distributed ledgers,
\eg Blockchain, require the system to solve \emph{consensus} in which all processes reliably agree on a single value.
\emph{Binary consensus}, where the set of values that can be proposed is either zero or one, is a fundamental
building block for other ``flavors'' of consensus, \eg multivalued, or vector, and of total order broadcast. 
At PODC 2014, Mostéfaoui, Moumen, and Raynal, in short MMR, presented a randomized signature-free asynchronous binary consensus algorithm. 
They demonstrated that their solution could deal with up to $t$ Byzantine processes, where $t < n/3$ and $n$ is the number of processes. 
MMR assumes the availability of a service of random common coins and fair scheduling of message arrivals, which does not depend on the current coin values. It completes within $O(1)$ expected time.

Our study, which focuses on binary consensus, aims at the design of an even more robust consensus protocol. 
We do so by augmenting MMR with self-stabilization, a powerful notion of fault-tolerance. In addition to tolerating process and communication failures, self-stabilizing systems can automatically recover after the occurrence of \emph{arbitrary transient-faults}; these faults represent any violation of the assumptions on which the system was designed to operate (provided that the algorithm code remains intact).

We present the first loosely-self-stabilizing fault-tolerant asynchronous solution to binary consensus in Byzantine message-passing systems.
This is achieved via an instructive transformation of MMR to a self-stabilizing solution that can violate safety requirements with probability $\Pr=\bigO(2^{-M})$, where $M \in \mathbb{Z}^+$ is a predefined constant that can be set to any positive value at the cost of $3 M n + \log M$ bits of local memory; $n$ is the number of processes. 
The obtained self-stabilizing version of the MMR algorithm considers a far broader fault-model since it recovers from transient faults. 
Additionally, the algorithm preserves the MMR's  properties of optimal resilience and termination, \ie $t < n/3$, and $\bigO(1)$ expected decision  time.
{Moreover, any instance of the proposed solution requires a bounded amount of memory.} 

{We also offer a recycling mechanism for these asynchronous objects that allows their reuse once each object completes its task and all non-faulty nodes retrieved the decided values.
This mechanism itself uses synchrony assumptions and is based on a novel composition of existing techniques as well as a new self-stabilizing Byzatine-tolerant multivalued consensus algorithm for synchronous systems.} 
\end{abstract}

%% \linenumbers

	\Section{Introduction}
We propose a loosely-self-stabilizing Byzantine fault-tolerant asynchronous implementation of \emph{binary consensus} objects for signature-free message-passing systems.  

\Subsection{Background and motivation} 
En route to constructing robust distributed systems, rose the need for different (possibly geographically dispersed) computational entities to take common decisions.
Of past and recent contexts in which the need for agreement appeared, one can cherry-pick applications, such as service replication, cloud computing, load balancing, and distributed ledgers (most notably Blockchain).
In distributed computing, the problem of agreeing on a single value after the proposal of values by computational entities, \emph{processes} (sometimes called \emph{nodes} or \emph{processors}), is called \emph{consensus}~\cite{DBLP:journals/toplas/LamportSP82, DBLP:conf/podc/Ben-Or83}.
The most basic form of the consensus problem is for processes to decide between two possible values, \eg zero or one. 
This version of the problem is called \emph{binary consensus}~\cite[Ch. 14]{DBLP:books/sp/Raynal18}. In the absence of faults, solving consensus is straightforward, however, in the presence of faults, even benign ones such as crashes,  and in the face of asynchrony, the problem is not solvable deterministically (cf.~\cite{DBLP:journals/jacm/FischerLP85}). This work aims to fortify consensus protocols with  fault-tolerance guarantees that are more powerful than any existing known solution.
Such solutions are imperative for many distributed systems that run in hostile environments, such as Blockchains.

%	\EMS{For the proceedings version, I think we can skip the next paragraph and just append to the paragraph above with 'Such solutions are imperative for many distributed systems that run in hostile environments, such as Blockchain.' }\cgr{I agree.}

Over the years, research into the consensus problem has tried to exhaust all the different possible variations of the problem by tweaking synchrony assumptions, the range of possible values to be agreed upon, adversarial and failure models, as well as other parameters.  
To circumvent known impossibility results, \eg the celebrated FLP~\cite{DBLP:journals/jacm/FischerLP85}, the system models are also equipped with additional capabilities, such as cryptography, oracles, \eg perfect failure detectors, and randomization~\cite{DBLP:journals/ijccbs/CorreiaVNV11}.
Despite the decades-long research, the consensus problem remains a popular research topic. 
The most recent spike in interest in consensus was triggered by the Blockchain ``rush'' of the past decade. 
Agreement in a common chain of blocks is inherently a consensus problem. 
``Blockchain consensus''~\cite{DBLP:conf/wdag/CachinV17,DBLP:journals/comsur/XiaoZLH20} is a highly-researched topic, and all the ``proof-of-$\ast$'' concepts enclose an underlying consensus-solving mechanism.

\Subsection{Problem definition} 
\label{sec:probDef}
The problem of letting all processes to uniformly select a single value among all the values that they propose is called consensus. When the set, $V$, of values that can be proposed, includes just two values, \ie $V=\{0,1\}$, the problem is called binary consensus.
%(see Definition~\ref{def:consensus} below).  %CG: no need to refer it, it is in the next line!
Otherwise, it is called multivalued consensus. 

%	Definition~\ref{def:consensus} formally states the binary consensus problem.

%	Existing solutions for multivalued consensus often use binary consensus algorithms.

\begin{definition}[Binary Consensus]
	\label{def:consensus}
	Every process $p_i$ has to propose a value $v_i \in V=\{0,1\}$, via an invocation of the $\mathsf{propose}_i(v_i)$ operation. Let $\mathit{Alg}$ be an algorithm that solves binary  consensus. $\mathit{Alg}$ has to satisfy \emph{safety}, \ie BC-validity and BC-agreement, and \emph{liveness}, \ie BC-completion, requirements.
	\begin{itemize}
		\item \textbf{BC-validity.~~} %Let $v \in \{0,1\}$ Suppose that all correct processes propose $v$. All correct processes decide $v$.
%		We say that a process is correct, if it follows its specified protocol for the entire computation
		The value $v \in \{0,1\}$ decided by a non-faulty process is a value proposed by a non-faulty process.
		\item \textbf{BC-agreement.~~} Any two non-faulty processes that decide, do so with identical decided values.
		\item \textbf{BC-completion.~~} All non-faulty processes decide.
	\end{itemize}
\end{definition}

Starting from the algorithm of Mostéfaoui, Moumen, and Raynal~\cite{DBLP:conf/podc/MostefaouiMR14}, from now on MMR, this study proposes  an even more fault-tolerant consensus algorithm, which is a variant on MMR. Note that MMR provides randomized liveness guarantees, \ie with the probability of $1$, MMR satisfies the BC-completion requirement within a finite time that is known only by expectation. The proposed solution satisfies BC-completion within a time that depends on a predefined parameter $M \in \mathbb{Z}^+$. However, it provides randomized safety guarantees, \ie  with the probability of $1-\bigO(2^{-M})$, the proposed solution satisfies the BC-validity and BC-agreement requirements. Since the number of bits that each process needs to store is $3nM+\lceil \log M\rceil$, we note that the probability for violating safety can be made, in practice, to be extremely small, where $n$ is the number of processes, see Remark~\ref{thm:pratSafe} for details.

We note that the literature often refers to BC-completion property as BC-termination. In Section~\ref{sec:SelfStabIntro}, we explain the reason for this deviation.

{Also, Definition~\ref{def:consensus} considers a single instance Binary consensus object. Our implementation considers an extended version of {\em recyclable} Binary consensus objects that can be stored in a $\delta$-size set, where $\delta$ is a predefined constant (Section~\ref{sec:theAlgo}). This set can be repeatedly recycled once all objects complete their task and all non-faulty nodes retrieved their results (Section~\ref{sec:bck}). Thus, the proposed solution can be reused an unbounded number of times (and still, use only a bounded amount of memory).}

\Subsection{Fault model} 
\label{sec:fdIn}
We study solutions for message-passing systems. We model a broad set of %benign 
failures that can occur to computers and networks, \eg due to procrastination, equivocation, selfishness, hostile (human) interference, deviation from the program code, etc. Specifically, our fault model includes up to $t$ process failures, \ie crashed or Byzantine~\cite{DBLP:journals/toplas/LamportSP82}. In detail, a faulty process runs the algorithm correctly, but the adversary completely controls the messages that the algorithm sends, \ie it can modify the content of a message, delay the delivery of a message, or omit it altogether. The adversary's control can challenge the algorithm by creating failure patterns in which a fault occurrence appears differently to different system components. Moreover, the adversary is empowered with the unlimited ability to compute and coordinate the most severe failure patterns. We assume a known maximum number, $t$, of processes that the adversary can capture. We also restrict the adversary from letting a captured process impersonate a non-faulty one. In addition, we limit the adversary's ability to impact the delivery of messages between any two non-faulty processes by assuming fair scheduling of message arrivals \ie Fair Communication (FC) between non-faulty processes is assumed.

%CG: I added this, so to define FC which is given below without being defined. 

\ems{\Subsection{Hybrid synchronous/asynchronous approach } 
	\label{sec:hybridIntro}
	The proposed solution uses a hybridization of two different fault models, which their notations follow Raynal~\cite{DBLP:books/sp/Raynal18}.}

\begin{itemize}
	\item \ems{$\mathsf{BAMP_{n,t}[\mathit{-}FC, t < n/3,RCCs]}$.~~ The studied asynchronous solutions are for message-passing systems where the algorithm cannot explicitly access the local clock or assume the existence of guarantees on the communication delay. These systems are also prone to communication failures, \eg packet omission, duplication, and reordering, as long as fair communication (FC) holds. For the sake of solvability~\cite{DBLP:journals/toplas/LamportSP82,DBLP:journals/jacm/PeaseSL80,DBLP:conf/podc/Toueg84}, we also assume that the number of faulty processes $t<n/3$ is less than one-third of the number of processes in the system. This fault model, $\mathsf{BAMP_{n,t}[\mathit{-}FC, t < n/3,RCCs]}$, is called the Byzantine Asynchronous Message-Passing model with at most $t$ (out of $n$) faulty processes. The array $\mathsf{[\mathit{-}FC, t < n/3,RCCs]}$ denotes the list of all assumptions, \ie FC and $t<n/3$ as well as \emph{random common coins} (RCCs).} 
	
	%	 	\EMS{CHECK IF WE NEED THIS HERE}. 
	
	%	 	\ems{We clarify that our use of RCCs is based on a synchrony assumption, see Section~\ref{sec:boundNeeded} for details. When a model of an asynchronous system makes use of mild synchrony assumptions, Raynal~\cite{DBLP:books/sp/Raynal18} says that the model is time-free since the algorithm does not have access to the clock and there are no bounds on the communication delays.  
		%	 	We use this model for studying the problem of binary consensus. A detailed presentation of $\mathsf{BAMP_{n,t}[\mathit{-}FC, t < n/3,RCCs]}$ appears in Section~\ref{sec:sys}.}  
	
	\item \ems{$\mathsf{BSMP_{n,t}[\kappa\mathit{-}SGC,t < n/3,RCCs]}$.~~ This model is called the Byzantine synchronous message-passing with at most $t$ (out of $n$) faulty processes, and $t < n/3$. The $\mathsf{BSMP_{n,t}[\kappa\mathit{-}SGC,t < n/3,RCCs]}$ model is defined by enriching the  $\mathsf{BAMP_{n,t}[\mathit{-}FC,t < n/3]}$ model with a $\kappa$-state global clock, reliable communication, and a service of RCCs. A detailed presentation of $\mathsf{BSMP_{n,t}[\kappa\mathit{-}SGC,t < n/3,RCCs]}$ appears in Section~\ref{sec:sysS}.}
\end{itemize}

\Subsection{Self-stabilization} 
\label{sec:SelfStabIntro}
In addition to the failures captured by our model, we also aim to recover from \emph{arbitrary transient-faults}, \ie any temporary violation of assumptions according to which the system and network were designed to operate. This includes the corruption of control variables, such as the program counter, packet payload, and indices, \eg sequence numbers, which are responsible for the correct operation of the studied system, as well as operational assumptions, such as that at least a distinguished majority of processes never fail. Since the occurrence of these failures can be arbitrarily combined, we assume that these transient-faults can alter the system state in unpredictable ways. In particular, when modeling the system, Dijkstra~\cite{DBLP:journals/cacm/Dijkstra74} assumes that these violations bring the system to an arbitrary state from which a \emph{self-stabilizing system} should recover, see~\cite{DBLP:series/synthesis/2019Altisen,DBLP:books/mit/Dolev2000} for details. Dijkstra requires recovery after the last occurrence of a transient-fault and once the system has recovered, it must never violate the task specification. 

For the case of the studied problem and fault model, there are currently no known ways to meet Dijkstra's self-stabilizing design criteria. \emph{Loosely-self-stabilizing systems}~\cite{DBLP:journals/tcs/SudoNYOKM12} require that, once the system has recovered, only rarely and briefly can it violate the safety specifications. Although it is a weaker design criterion than the one defined by Dijkstra, the violation occurrence can be made to be so rare, that the risk of breaking the safety requirements of Definition~\ref{def:consensus} becomes negligible.     

\ems{It is well-known that self-stabilizing systems cannot stop sending messages when the system's task has so-called  ``terminated'', see~\cite[Chapter 2.3]{DBLP:books/mit/Dolev2000} for details. This impossibility is, mistakenly, stated as ``self-stabilizing system can never terminate''. However, the system's task can terminate but the system cannot stop sending messages. In order to avoid this confusion, as mentioned, we refer to BC-termination as  BC-completion.}

\Subsection{Related work} 
\ems{In this paper, the design criteria for non-self-stabilizing Byzantine fault-tolerant solutions are called BFT, and the ones for self-stabilizing Byzantine fault-tolerant are called SSBFT. We review the most related BFT and SSBFT solutions for the studied problem.}

\Subsubsection{Impossibilities and lower-bounds} 
\label{sec:impo}
The FLP impossibility result~\cite{DBLP:journals/jacm/FischerLP85} concluded that consensus is impossible to solve deterministically in asynchronous settings in the presence of even a single crash failure. In~\cite{DBLP:journals/ipl/FischerL82} it was shown that a lower bound of $t+1$ communication steps are required to solve consensus deterministically in both synchronous and asynchronous environments. The proposed solution is a randomized one.   
In the presence of asynchrony, transient-faults, and (non-Byzantine) crash failures, there are known problems such as leader election and counting the number of processes in the system, for which there are no (randomized) self-stabilizing solutions~\cite{DBLP:conf/wdag/AnagnostouH93,DBLP:journals/ijsysc/BeauquierK97}. In this work, we consider weaker design criteria than Dijkstra's self-stabilization. 

In the presence of Byzantine faults, the consensus problem is not solvable if a third or more of the processes are faulty~\cite{DBLP:journals/toplas/LamportSP82}. Thus, optimally resilient Byzantine consensus algorithms, such as the one we present, tolerate $t<n/3$ faulty processes.  
The task is also impossible if a process can impersonate some other process in its communication with the other entities~\cite{DBLP:conf/podc/Ben-Or83}. We assume the absence of spoofing attacks and similar means of impersonation.
In the presence of asynchrony, transient-faults, and Byzantine failures, the task of unison is known to be unsolvable (unless the strongest fairness assumptions are made)~\cite{DBLP:journals/jpdc/DuboisPNT12,DBLP:journals/tcs/DuboisPT11}. As indicated by the above impossibility results, the studied problem remains challenging even under randomization and fairness assumptions during the recovery period.

\remove{	
	The FLP impossibility result~\cite{DBLP:journals/jacm/FischerLP85}
	concluded that consensus is impossible to solve deterministically in asynchronous settings in the
	presence of even a single crash failure.  Still, it is possible to
	circumvent the FLP in the following
	ways~\cite{DBLP:journals/ijccbs/CorreiaVNV11}: placing timing
	assumptions, using failure detection or
	wormholes~\cite{DBLP:journals/dc/CorreiaNLV05}, or employing
	randomization.  In the presence of Byzantine faults, consensus is not
	solvable if a third or more of the processes are
	faulty~\cite{DBLP:journals/toplas/LamportSP82}.  Thus, optimally
	resilient Byzantine consensus algorithms, such as the one we present,
	tolerate $t<n/3$ faulty processes.  The task is also impossible if a
	process can impersonate some other process in its communication
	with the other entities~\cite{DBLP:conf/podc/Ben-Or83}.  In
	\cite{DBLP:journals/ipl/FischerL82} it was shown that a lower bound of
	$t+1$ communication steps is required to solve consensus in both
	synchronous and asynchronous environments.
	\im{Solving a problem in asynchronous systems in the presence of (Byzantine) failures and transient-faults can be a very challenging task, even under randomization, and as indicated in \cite{DBLP:conf/wdag/AnagnostouH93,DBLP:journals/ijsysc/BeauquierK97,DBLP:journals/jpdc/DuboisPNT12,DBLP:journals/tcs/DuboisPT11}, it can be impossible.
		
	} 
} % REMOVE

\Subsubsection{Non-self-stabilizing non-BFT solutions} 
Paxos~\cite{DBLP:journals/tocs/Lamport98} is the best-known solution for the consensus problem. 
Despite becoming notorious for being complex~\cite{lamport2001paxos}, Paxos was followed by rich literature~\cite{DBLP:journals/csur/RenesseA15}. 
Raynal~\cite{DBLP:books/sp/Raynal18} offers a family of abstractions for solving a number of well-known problems including consensus. 
This line of research is easier to understand and supports well-organized implementations.
Protocols implementing total order broadcast are usually built on top of consensus since consensus and total order broadcast are equivalent~\cite{DBLP:journals/jacm/ChandraT96,DBLP:journals/tkde/RodriguesR03}. 

%\EMS{Maybe move common coin of Section~\ref{sec:commonCoin} to here.}

%%  LONG Non-self-stabilizing BFT solutions RELATED WORK
%%% CONTAINS NOT DIRECTLY RELEVANT WORK

\Subsubsection{Non-self-stabilizing BFT solutions} 
\label{sec:nonSSBFTsols}
BFT consensus was tackled by many protocols~\cite{DBLP:journals/tdsc/MonizNCV11}. 
Several variants of Paxos consensus tolerate such malicious processes, \eg ~\cite{DBLP:conf/wdag/Lamport11a}.
State machine replication protocols, such as PBFT~\cite{DBLP:journals/tocs/CastroL02} and BFT-SMART~\cite{DBLP:conf/dsn/BessaniSA14} incorporate a BFT consensus mechanism. 

Randomization can circumvent the FLP
impossibility~\cite{DBLP:journals/ipl/FischerL82}, which only entails
deterministic algorithms.  This line of work started with
Ben-Or~\cite{DBLP:conf/podc/Ben-Or83} using a local coin (that
generated a required exponential number of communication steps in the
general case) and resilience $t<n/5$, and by
Rabin~\cite{DBLP:conf/focs/Rabin83} in the same year, which assumes
the availability of RCCs, allowed for a polynomial number of
communication steps and optimal resilience, \ie $t<n/3$. We later discuss more
extensively the notion of RCCs.
Bracha~\cite{Bracha1987} constructed a reliable broadcast protocol
that allowed optimally-resilient binary agreement, but using a local
coin needed an exponential expected number of communication steps.
%	SINTRA~\cite{Cachin2001} Byzantine-tolerant total order reliable broadcast.
%	It employed threshold cryptography for digital signatures, coin-tossing, and public-key encryption. 
%\EMS{The part in the parenthesis is not clear and perhaps can be removed or simplified.}
% (IM: Paranthesis removed)
%	The broadcast protocol is a reduction to multivalued consensus and then to binary consensus. 
Cachin et al.~\cite{DBLP:journals/joc/CachinKS05} solve asynchronous binary consensus using RCCs and cryptographic threshold signatures. 
They achieve optimal resilience ($t<n/3$) and quadratic message-per-round complexity. 

In the sequel, we focus on MMR~\cite{DBLP:conf/podc/MostefaouiMR14} as a signature-free BFT solution for binary consensus.
This algorithm is optimal in resilience,  uses $O(n^2)$ messages per consensus invocation, and completes within $O(1)$ expected time. 
MMR can be combined with a reduction of multivalued consensus to binary consensus~\cite{DBLP:journals/acta/MostefaouiR17} to attain multivalued consensus with the same fault-tolerance properties.

Binary consensus is a fundamental component of total order reliable broadcast, \eg~\cite{Cachin2001,DBLP:journals/cj/CorreiaNV06}  (see Section~\ref{sec:arch}). 
In what appears as a revival of the topic, several Blockchain consensus protocols are also using similar approaches. 
HoneyBadger~\cite{DBLP:conf/ccs/MillerXCSS16} was the first randomized BFT protocol for Blockchain. 
They employ MMR as their binary consensus protocol. The BEAT~\cite{BEAT2018} suite of protocols for blockchain consensus  also uses MMR.

\ems{MMR has PODC 2014~\cite{DBLP:conf/podc/MostefaouiMR14} and JACM 2015~\cite{DBLP:journals/jacm/MostefaouiMR15} variations. The latter variation overcomes an implementation challenge later discussed by Tholoniat and Gramoli~\cite{DBLP:journals/corr/abs-1909-07453}, which raised concerns regarding the liveness of the PODC 2014 variation when the adversary is allowed to control the schedule of message arrivals. Recently, Cachin and Zanolini~\cite{DBLP:conf/wdag/CachinZ21} modified the MMR variation of PODC 2014 with a couple of simple modifications that cope with the above liveness concern. Specifically, they suggest imposing FIFO message delivery and an extra sampling of the arriving values before accessing the RCC. For the sake of a simple presentation, this work considers the PODC 2014 variation and assumes fair scheduling of message arrival (which does not depend on the current coin value).}  Thus, our results do not implement the modifications proposed by Cachin and Zanolini.

{Duvignau, Raynal, and Schiller~\cite[Algorithm 3]{DBLP:journals/corr/abs-2201-12880} explain how to implement the FIFO message ordering in the context of SSBFT.}
{The interested reader is offered to apply the technique for dynamic value reception proposed by Cachin and Zanolini to} {the proposed solution since it preserves MMR's key algorithmic features.}

%%%% SHORT VERSION REMOVAL

\Subsubsubsection{Non-self-stabilizing BFT services for RCCs}
\label{sec:commonCoin}

%[[THE FOLLOWING PARAGRAPH belonged to the section of Byzantine
%Fault-tolerant solutions. %Few works are also very briefly mentioned
%there.]]
Randomized algorithms employ coin flips to circumvent the
FLP impossibility~\cite{DBLP:journals/ipl/FischerL82}, which only entails
deterministic algorithms.  The two known coin flip constructions are
\emph{local coins}, where each process only uses a local random
function, and RCCs, where the $k$-th invocation of the
random function by a non-faulty process, returns the
same bit as to any other non-faulty process.
Ben-Or~\cite{DBLP:conf/podc/Ben-Or83} using a local coin, developed an
asynchronous BFT Binary Consensus algorithm with
$t<n/5+1$ resilience, but (as any local-coin-based algorithm) required
an exponential number of communication steps, unless $t=O(\sqrt{n})$
where a polynomial number can be achieved.
Rabin~\cite{DBLP:conf/focs/Rabin83} was the first to introduce 
RCCs demonstrating the possibility of designing asynchronous
BFT binary consensus algorithms with a polynomial
number of communication steps and with constant expected computational
rounds.  The coin construction is based on Shamir's secret
sharing~\cite{DBLP:journals/cacm/Shamir79} and digital signatures for
authenticating the messages exchanged. Since then, RCCs provision has
become an essential tool, and many subsequent works have devised
randomized coin-flipping algorithms,
\eg~\cite{DBLP:conf/podc/Ben-OrDH08,
	DBLP:books/daglib/0025983, Cachin2001, 		DBLP:journals/joc/CachinKS05,  DBLP:conf/stoc/CanettiR93, DBLP:conf/stoc/FeldmanM88, DBLP:conf/icalp/FeldmanM89, DBLP:conf/eurocrypt/NaorPR99} as building blocks for consensus and
other related problems, such as clock
synchronization. Aspens~\cite{DBLP:journals/jacm/Aspnes98}
demonstrates that agreeing on RCCs is a harder problem than
solving consensus, in the sense that if we can solve it, then we can
solve consensus.

An important feature that RCCs algorithm must provide is \emph{unpredictability}, that is, the outcome of the random bit at a given round should not be predicted by the Byzantine adversary before that round. In this respect, two communication models have been used in devising coin-flipping algorithms. Either \emph{private communication} is assumed, \eg~\cite{DBLP:conf/stoc/FeldmanM88, DBLP:conf/icalp/FeldmanM89, DBLP:conf/stoc/CanettiR93, DBLP:conf/podc/Ben-OrDH08} or digital signatures and other cryptographic tools are employed, \eg~\cite{DBLP:journals/cacm/Shamir79,DBLP:conf/eurocrypt/NaorPR99, Cachin2001, DBLP:journals/joc/CachinKS05}. In the former, the usual assumption is that processes are connected via private channels and the Byzantine adversary can have access to the messages exchanged between faulty and non-faulty processes, but not to the messages exchanged between non-faulty processes, hence providing confidentiality. %~\sout{(cellular networks can be seen as an example of a system providing private channels)}. 
In the latter, cryptographic tools (signatures) conceal the content of a message and only the intended recipient can view its content. Hence, a subtle difference between the two schemes is that with private channels, a third process does not even know whether two other processes have exchanged a message, whereas, with signatures, the third process might be aware of the message exchange, but not the message's content. Feldman and Micali~\cite{DBLP:conf/stoc/FeldmanM88} show how to compile any protocol assuming private channels to a cryptographic protocol not assuming private channels which runs exactly the same.

\Subsubsubsection{Non-self-stabilizing synchronous BFT multivalued consensus}
\sloppypar{As mentioned, self-stabilizing systems are required to use bounded memory, and thus, we are interested in recycling mechanisms for consensus objects {(Section~\ref{sec:probDef}).} 
The proposed recycling mechanism uses an SSBFT multivalued consensus for the $\mathsf{BSMP_{n,t}[\kappa\mathit{-}SGC,t < n/3,RCCs]}$ model, which is based on a non-self-stabilizing BFT multivalued consensus.} Pease, Shostak, and  Lamport~\cite{DBLP:journals/jacm/PeaseSL80} were the first to propose a solution that has optimal resilience to $t<n/3$ and optimal worst-case $t + 1$ synchronous rounds with exponential communication costs. Dolev and Strong~\cite{DBLP:conf/stoc/DolevS82} proposed the first solution that has optimal resilience and polynomial communication costs but not with optimal worst-case rounds.
Garay and Moses~\cite{DBLP:journals/siamcomp/GarayM98} proposed the first solution for binary-valued Byzantine agreement with optimal resilience, polynomial communication costs, optimal $t+1$ rounds, and early termination. Kowalski and Most{\'{e}}faoui~\cite{DBLP:conf/podc/KowalskiM13} proposed the first multivalued optimal resilience, polynomial communication costs, and optimal $t+1$ rounds, but without early stopping. Abraham and Dolev~\cite{DBLP:conf/stoc/AbrahamD15} advance the state of the art by offering also optimal early stopping.
{Unlike the above BFT multivalued consensus solutions, our BFT multivalued solution considers self-stabilization, and its implementation is an} application of the well-known technique of the recomputation of floating outputs~\cite[Chapter 2.8]{DBLP:books/mit/Dolev2000}.

% \CGC{Same comment as above. I believe this sentence along with the paragraph that follows should be move to Section 1.6.5 where we discuss SSBFT solutions.}
%	, and therefore, we do not consider it as a significant independent contribution.}  

\Subsubsection{Self-stabilizing crash-tolerant solutions} 
Lundstr{\"{o}}m, Raynal, and
Schiller~\cite{DBLP:conf/icdcn/LundstromRS21} presented the first
self-stabilizing solution for the problem of binary consensus for
message-passing systems where processes may fail by crashing. They ensure
a line of self-stabilizing
solutions~\cite{DBLP:journals/corr/abs-2104-03129,DBLP:conf/icdcs/LundstromRS20,DBLP:conf/netys/LundstromRS20,DBLP:conf/netys/GeorgiouLS19,DBLP:conf/netys/GeorgiouGLS19}. This
line follows the approach proposed by Dolev, Petig, and
Schiller~\cite{DBLP:conf/podc/DolevPS15,DBLP:journals/corr/abs-1806-03498}
for self-stabilization in the presence of seldom fairness. Namely, in
the absence of transient-faults, these self-stabilizing solutions are
wait-free and no assumptions are made regarding the system's synchrony
or fairness of its scheduler. However, the recovery from transient
faults does require fair execution, \eg to perform a global \ems{restart},
see~\cite{DBLP:journals/corr/abs-1807-07901,DBLP:conf/netys/GeorgiouGLS19},
but only during the recovery period. Our work does not assume
execution fairness either in the presence or absence of arbitrary
transient-faults. As in MMR, our
loosely-self-stabilizing BFT solution assumes
fair scheduling of message arrivals and the accessibility to an independent service for RCCs.

We note the existence of other approaches for recovering from transient faults without assuming execution fairness during the recovery period~\cite{DBLP:conf/netys/SalemS18,DBLP:journals/jcss/DolevGMS18,DBLP:journals/jcss/AlonADDPT15}. However, none of these results consider both Byzantine fault-tolerance and self-stabilization.

Algorithms for loosely-self-stabilizing
systems~\cite{DBLP:journals/ieicet/SudoOKM20,DBLP:journals/tpds/SudoOKMDL19,DBLP:journals/tcs/SudoOKMDL20,DBLP:conf/sirocco/Izumi15,DBLP:conf/sss/DongSIM21}
mainly focus on the task of leader election and population
protocols. {Recently, Feldmann, G{\"{o}}tte, and Scheideler~\cite{DBLP:conf/sss/0001GS19} proposed a loosely-self-stabilizing algorithm for congestion control.} Considering a message-passing system prone to
Byzantine failures, we  implement leaderless binary consensus. Our
loosely-self-stabilizing design criterion is slightly weaker than the
one studied
in~\cite{DBLP:journals/ieicet/SudoOKM20,DBLP:journals/tpds/SudoOKMDL19,DBLP:journals/tcs/SudoOKMDL20,DBLP:conf/sirocco/Izumi15,DBLP:conf/sss/0001GS19}
since it requires the loosely-self-stabilizing condition to hold only
eventually.

%	 also makes no assumption regarding synchrony or fairness as long as no transient fault occurs. Similarly, our recovery from transient-faults is based on synchrony assumptions. Specifically in our work, we use these synchrony assumptions for guaranteeing termination. To the best of our knowledge, this is a novel technique for implementing the above approach by Dolev, Petig, and Schiller~\cite{DBLP:journals/corr/abs-1806-03498,DBLP:conf/podc/DolevPS15}.         

\Subsubsection{Self-stabilizing BFT solutions} 
In the context of this dual design criteria, there are solutions for clock synchronization~\cite{DBLP:journals/corr/abs-2203-14016,perner2013byzantine,DBLP:conf/sss/Malekpour06,DBLP:conf/wdag/DolevH07,DBLP:conf/icpads/YuZY21,DBLP:conf/podc/DaliotDP04,DBLP:conf/sss/DolevH07,DBLP:conf/podc/Ben-OrDH08,DBLP:conf/sss/HochDD06,DBLP:conf/podc/DolevW95,DBLP:journals/jacm/LenzenR19,DBLP:journals/mst/KhanchandaniL19}, storage~\cite{DBLP:journals/tcs/BonomiPP18,DBLP:conf/icdcn/BonomiPP16,DBLP:conf/sss/BonomiPPT18,DBLP:conf/srds/BonomiPPT17,DBLP:conf/podc/BonomiPPT16,DBLP:conf/ipps/BonomiPT15,DBLP:conf/podc/BonomiDPR15}, and gathering of mobile robots~\cite{DBLP:conf/sss/AshkenaziDKKOW21,DBLP:conf/ic-nc/AshkenaziDKOW19,DBLP:journals/corr/DefagoP0MPP16,DBLP:journals/dc/DefagoPP20}.
There are also SSBFT solutions for link-coloring~\cite{DBLP:conf/opodis/MasuzawaT05,DBLP:conf/opodis/SakuraiOM04}, topology discovery~\cite{DBLP:conf/netys/DolevLS13,DBLP:journals/tpds/NesterenkoT09}, overlay networks~\cite{DBLP:conf/opodis/DolevHR07},  exact agreement~\cite{DBLP:conf/podc/DaliotD06} approximate agreement~\cite{DBLP:journals/tcs/BonomiPPT19}, asynchronous unison~\cite{DBLP:journals/jpdc/DuboisPNT12}, communication in dynamic networks~\cite{DBLP:conf/opodis/Maurer20}, and reliable broadcast~\cite{DBLP:journals/corr/abs-2201-12880,DBLP:conf/srds/MaurerT14}. 
The most relevant work is the one by Binun \etal~\cite{DBLP:conf/sss/BinunCDKLPYY16,DBLP:conf/cscml/BinunDH19} and Dolev \etal~\cite{DBLP:conf/cscml/DolevGMS18} for a  deterministic BFT emulation of state-machine replication. Binun \etal present the first self-stabilizing solution for synchronous message-passing systems and Dolev \etal present the first practically-self-stabilizing solution for partially-synchronous settings, utilizing failure detectors. 
We study another problem, which is binary consensus.
%
%\CGC{I have revised the sentence below (the old is commented out for comparison), to make the notion of practically-self-stabilizing, in my opinion, more clear.}
%\ems{Also, practically-self-stabilizing systems have no guarantees to ever recover from the last occurrence of a transient fault, whereas loosely-self-stabilizing systems recover within a bounded (expected) time.}
%
Note that in practically-self-stabilizing systems there can be a bounded number of possible safety violations during any practically infinite period of the system execution, whereas loosely-self-stabilizing systems recover within a bounded (expected) time with no further safety violations. 

To the best of our knowledge, the only SSBFT RCCs construction is the one by Ben{-}Or,
Dolev, and Hoch~\cite{DBLP:conf/podc/Ben-OrDH08}, in short BDH, for synchronous
(pulse-based) systems with private channels. They use a pipeline
technique to transform the non-self-stabilizing synchronous
BFT coin-flipping algorithm of Feldman and
Micali~\cite{DBLP:conf/icalp/FeldmanM89} into a self-stabilizing one;
the work in~\cite{DBLP:conf/icalp/FeldmanM89} assumes private
channels.
In~\cite{DBLP:conf/podc/Ben-OrDH08}, BDH have used their SSBFT RCCs construction as a building block for devising an SSBFT synchronous clock synchronization solution.

%\CGC{The paragraph below used to be at the end of Section 1.6.3. I have moved it here, since this is where we talk about SSBFT, but more importantly, because this is where we introduce BDH -- in the section it used to be, there reader was not yet aware of the BDH work.}

{Our work borrows several mechanisms from BDH, such as SSBFT RCCs and SSBFT clock synchronization. We also borrow proof techniques from their random algorithm for providing SSBFT digital clock synchronization. We note the existence of an earlier SSBFT algorithm for deterministic digital clock synchronization by Dolev and Welch~\cite{DBLP:journals/jacm/DolevW04} rather than BDH's randomized solution. We decided not to base our solution on the one by Dolev and Welch since it has exponential stabilization time.}	
%CG: Since we mention this clock synchronization solution, without mentioning that they did it, and in fact, using the RCCs as a building block.

%	 and consider a stronger design criteria than the one of practically-self-stabilizing system since we guarantee recovery within a bounded time.       

%\Subsubsection{Asymmetric cryptography and transient-faults}
%\label{sec:cryptography}
%%
%Public-key cryptography (PKC) is based on a pair of keys. The public keys is known to all peers whereas the private keys must not be known to ever known by anyone but the key owner. 
%PKC's applications include digital signatures and encryption, \eg for ensuring confidentiality of private channels. In the presence of arbitrary transient-faults, PKC are faced with a non-trivial challenge since the private key (and all of its copies) can be corrupted. For this reason, this work focuses on model that are not enhanced by the above PKC's applications.     

\begin{figure}
%\begin{wrapfigure}{r}{0.235\textwidth}
\begin{center}
	%		\BBB\BBB\BBB
	%		\hspace*{-0.5em}		
	\includegraphics[scale=0.5, clip]{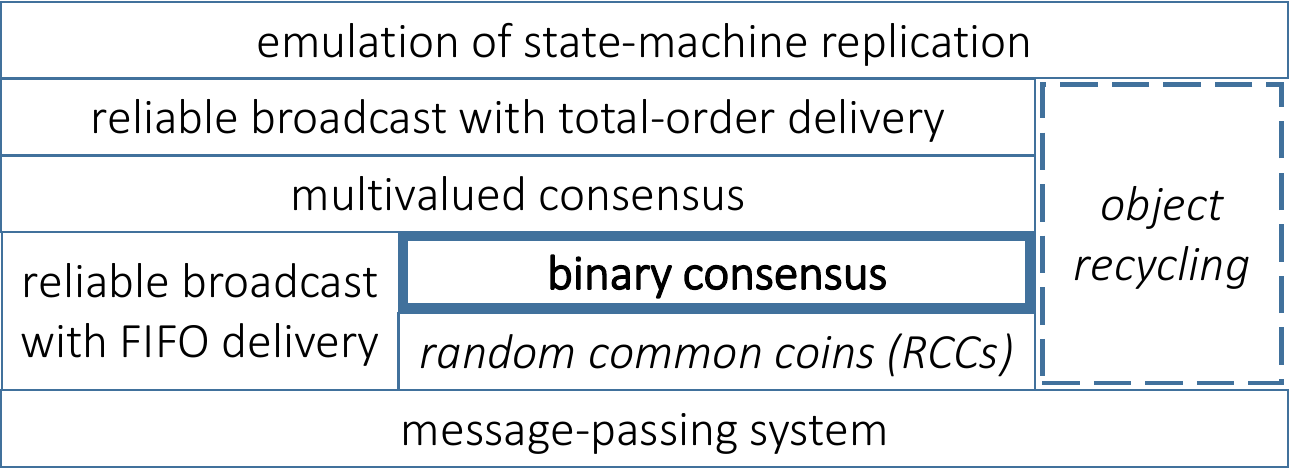}
\end{center}
%	\BBB\B%BB
\caption{\label{fig:suit}\ems{The hybrid architecture of asynchronous and synchronous components. The studied problem (which appears in boldface font and is surrounded by a thick frame) assumes no explicit synchrony, but it requires the availability of a service for RCCs (which appears in italic font) and fair scheduling of message arrival (which does not depend on the current coin value). The object recycling mechanism (which appears in italic font and is surrounded by a dashed frame) assumes synchrony. The other system components mentioned in Section~\ref{sec:arch} are presented in plain font.}}
%	\BBB\BB  
%\end{wrapfigure}
\end{figure}

\Subsection{The studied architecture of asynchronous and synchronous components}
\label{sec:arch}    
A Blockchain can be seen as a replication service for state-machine emulation in extremely hostile environments. The stacking of reliable broadcast protocols can facilitate this emulation, see Figure~\ref{fig:suit} and Raynal~\cite[Ch. 16 and 19]{DBLP:books/sp/Raynal18}. Specifically, the order of all state transitions of the automaton can be agreed by using total order reliable broadcast. The order of the broadcasts is agreed via multivalued consensus~\cite{DBLP:journals/corr/abs-2104-03129}. Whenever multivalued consensus is \ems{invoked, the latter calls} binary consensus for a finite number of times.

\Subsubsection{Using both asynchronous and synchronous components}
Existing solutions for binary consensus use either randomization techniques or synchrony assumptions in order to circumvent the mentioned impossibilities, \eg FLP. The system as a whole can avoid communication-related bottlenecks by making design choices that prefer weaker synchrony assumptions for the components that are more communication demanding. Binary consensus protocols are inherently communication-intensive since a  number of them can be invoked for every transition of the state-machine and each such invocation has to take at least two communication rounds, due to a lower bound by Keidar and Rajsbaum~\cite{DBLP:journals/ipl/KeidarR03}. 
%
% Such rounds incur each at least $\bigO(n^2)$ messages. 
%
Therefore, we select to study the non-self-stabilizing probabilistic MMR algorithm~\cite{DBLP:conf/podc/MostefaouiMR14} for solving binary consensus \ems{(in asynchronous systems) while assuming} access to RCCs. 
%
%As we reviewed in Section~\ref{sec:commonCoin}, the existing implementations of common coin services are often based on the applications of public-key cryptography (PKC).
%%
%%, which is based on a pair of keys. The public key is known to all peers whereas the private keys must not be known to anyone but the key owner. 
%%
%PKC's applications include digital signatures and encryption, \eg for ensuring the confidentiality of private channels. In the presence of arbitrary transient-faults, PKC is faced with a non-trivial challenge since the private key (and all of its copies) can be corrupted. For this reason, this work focuses on models that are not enhanced by the above PKC's applications.
%
% [[Changed upon Chryssis's suggestion]]
\remove{Ben{-}Or, Dolev, and Hoch~\cite{DBLP:conf/podc/Ben-OrDH08}, in short BDH, presented a self-stabilizing synchronous solution for RCCs provision.}%
%
%	BDH, as already mentioned, presented a synchronous SSBFT solution for common coin provision. 
%
%BDH does not use digital signatures nor does it requires confidential communications between any pair of non-faulty processes. To the best of our knowledge, it is the only existing self-stabilizing algorithm for common coin provision. Moreover, we are not aware of any existing (non-self-stabilizing) solution for common coin provision that neither assumes synchrony nor use digital-signatures and confidential communications. \EMS{The last two sentences needs to be checked.}   

\ems{\Subsubsection{Random common coins (RCCs)}
\label{sec:rcc}
As already mentioned, BDH presented a synchronous SSBFT RCCs solution for synchronous message passing systems. Algorithm $\mathcal{A}$, which has the output of $rand_i \in \{0,1\}$, is said to provide an RCC if $\mathcal{A}$ satisfies the following:}

\begin{itemize}
%		\item \textbf{Binary output:} Process $p_i$ has the output of $rand_i \in \{0,1\}$;

\item \ems{\textbf{RCC-completion:} $\mathcal{A}$ provides an output within $\Delta_{\mathcal{A}} \in \mathbb{Z}^+$ synchronous rounds.} 

%	(In this work, we assume that $\Delta_{\mathcal{A}}=4$.)  

\item \ems{\textbf{RCC-unpredictability:} Denote by $E_{x\in \{0,1\}}$ the event that for any non-faulty process, $p_j$, it holds $rand_j=x$ occurs with constant probability $p_x > 0$. Suppose either $E_0$ or $E_1$ occurs at the end of round $\Delta_{\mathcal{A}}$. We require that the adversity can predict the output of $\mathcal{A}$ by the end of round $\Delta_{\mathcal{A}}- 1$ with a probability that is not greater than $1 - \min\{p_0, p_1\}$. 
	Just like MMR's PODC 2014 variation, this work assumes that $p_0=p_1=1/2$.}  
\end{itemize}

{The correctness of our solution depends on the existence of a self-stabilizing RCC service, \eg BDH.}
\ems{BDH considers \emph{(progress) enabling} instances of RCCs if there is $x \in \{0,1\}$ such that for any non-faulty process $p_i$, we have $rand_i=x$. BDH correctness proof depends on the consecutive existence of two enabling RCCs instances.}

%As mentioned, Canetti and Rabin~\cite[Section 8]{DBLP:conf/stoc/CanettiR93} present an asynchronous version (and a matching implementation) of the synchronous requirements above.} 

%		Recall that we assume the fair scheduling of message arrival (which does not depend on the current coin value). Thus, the correctness arguments of the asynchronous MMR algorithm for $\mathsf{BAMP_{n,t}[\mathit{-}FC, t < n/3,RCCs]}$ do not rely on the RCC-unpredictability property to its full extent.}

% Indeed, MMR uses a protection protocol, called $\mathsf{bvBroadcast}()$, that makes sure that if all correct processes proposed the same value, say $v$, the adversary cannot cause the system to decide on a value other than $v$. Otherwise, both $0$ and $1$ are legitimate values to decide on.
% Therefore, for time-free settings, such as $\mathsf{BAMP_{n,t}[\mathit{-}FC, t < n/3,RCCs]}$, we simply require that before the invocation of $\mathsf{propose}()$ by any of the Byzantine processes, the adversity can predict the output of $\mathcal{A}$ for a period of at least $M$ asynchronous rounds with a probability that is not greater than $1 - \min\{p_0, p_1\}$, where  $M \in \mathbb{Z}^+$ is a predefined constant that bounds the number of pseudo-random bits each invocation of binary.

%	\Subsubsection{Periodic re-installation of the common seed {and initialization of consensus objects}}
\Subsubsection{\ems{Recycling and initializing of completed consensus objects}}
\label{sec:boundNeeded}
We clarify the advantage of the studied architecture that considers a hybrid model that is composed of asynchronous, \ie MMR for \ems{the model of $\mathsf{BAMP_{n,t}[\mathit{-}FC, t < n/3,RCCs]}$,} and
synchronous, \ie BDH \ems{for the model of $\mathsf{BSMP_{n,t}[\kappa\mathit{-}SGC,t < n/3,RCCs]}$,} components. 

\ems{The proposed solution uses a synchronous mechanism for object recycling, which we propose in Section~\ref{sec:bck} using a \emph{synchronous} RCCs service, such as BDH. 
That is, whenever an asynchronous consensus object has completed its task, the synchronous recycling mechanism re-initializes the object's state together with the associated instance of an RCCs service{---this synchronous re-initialization facilitates the use of the single instance object in a self-stabilizing manner.}}
As we explain in sections~\ref{sec:initialization} and~\ref{sec:assumptionEasy}, this \ems{simplifies} the correctness proof since it implies that recovery from transient-faults depends only on the completion of all operations after the occurrence of the last transient fault. 
 
{A straightforward extension can further mitigate the effect of the synchronization imposed by the recycling mechanism via the recycling of a predefined number of asynchronous objects at a time (Section~\ref{sec:recyclableVar}).
This way, the communication-intensive components remain asynchronous and synchronization occurs less often.}

\emsB{We point out another (challenging) extension that can be the subject of future work.}
Canetti and Rabin~\cite[Section 8]{DBLP:conf/stoc/CanettiR93} present an asynchronous (non-self-stabilizing) version (and matching implementation) of the synchronous requirements above.  		
The proposed solution could further increase the degree of asynchrony by using a self-stabilizing variation of Canetti and Rabin. This would allow to assume that each asynchronous consensus object has its own instance of an \emph{asynchronous} RCCs service, such as the one by Canetti and Rabin~\cite[Section 8]{DBLP:conf/stoc/CanettiR93}.

\Subsection{Our contribution} 
We present a fundamental module for dependable distributed systems: a loosely-self-stabilizing asynchronous binary consensus algorithm for message-passing systems that are prone to Byzantine process failures. We obtain this new loosely-self-stabilizing algorithm via a transformation of the non-self-stabilizing probabilistic MMR algorithm by Mostéfaoui, Moumen, and Raynal~\cite{DBLP:conf/podc/MostefaouiMR14} for the $\mathsf{BAMP_{n,t}[\mathit{-}FC, t < n/3,RCCs]}$ model. MMR assumes that $t < n/3$ and completes within $O(1)$ expected time, where $t$ is the number of faulty processes and $n$ is the total number of processes. 
%
%MMR assumes the availability of a common coin and uses $O(n^2)$ messages per consensus invocation, which completes within $O(1)$ expected time. 
%
The proposed algorithm preserves these elegant properties of MMR. 

In order to bound the amount of memory required to implement MMR (and our variation of MMR), we use $M \in \mathbb{Z}^+$ as a bound on the number of rounds. This implies that with a probability in $\bigO(2^{-M})$ the safety requirement of Definition~\ref{def:consensus} can be violated. However, as we clarify (Remark~\ref{thm:pratSafe}), by selecting a sufficiently large value of $M$, the risk of violating the safety requirements becomes negligible at affordable costs.

In the absence of transient-faults, our solution achieves consensus within a constant expected time (without assuming execution fairness). 
%
%n asymptotically optimal number of asynchronous (communication) rounds (without assuming fair execution); informally, an asynchronous (communication) round is the shortest execution fragment in which all correct processes have executed at least once the do-forever-loop of a given algorithm, see Section~\ref{sec:asynchronousRounds}). 
%	
After the occurrence of any finite number of arbitrary transient-faults, the system recovers within a {constant time (in terms of asynchronous communication rounds)} while assuming execution fairness. Unlike in MMR, each process uses a bounded amount of memory. Moreover, the communication costs of our algorithm are similar to the non-self-stabilizing MMR algorithm. That is, in every communication round, the proposed solution requires every non-faulty process to complete at least one round-trip with every other non-faulty process.

\ems{For the sake of providing a complete solution, this work also provides an SSBFT mechanism for the model of $\mathsf{BSMP_{n,t}[\kappa\mathit{-}SGC,t < n/3,RCCs]}$ that recycles distributed objects, such as the proposed MMR solution.} 
\ems{The proposed recycling mechanism recovers after the occurrence of the last transient fault within $\bigO(\max\{ \kappa,2(t+1)\})$ synchronous rounds, where $\kappa$ is a predefined constant (Section~\ref{sec:hybridIntro}) and $t$ is an upper bound on the number of Byzantine nodes.}
\ems{We obtain this part of the solution via a novel algorithmic composition of existing solutions, such as  recomputation of floating output, \ems{SSBFT} multivalued consensus, and a modified version of SSBFT clock synchronization. In the context of SSBFT, this composition is of special interest since it can be used not only for object recycling, because it implements SSBFT unison for the $\mathsf{BSMP_{n,t}[\kappa\mathit{-}SGC,t < n/3,RCCs]}$ model.}
\ems{We also present, to the best of our knowledge, the {\em first} SSBFT synchronous multivalued consensus solution, which is needed for the implementation of our SSBFT object recycling mechanism.}

To the best of our knowledge, we propose the {\em first} loosely-self-stabilizing BFT algorithm for solving the problem of binary consensus in \ems{the model of $\mathsf{BAMP_{n,t}[\mathit{-}FC, t < n/3,RCCs]}$ and the SSBFT recycling of these consensus objects in the model of $\mathsf{BSMP_{n,t}[\kappa\mathit{-}SGC,t < n/3,RCCs]}$. As we have explained in Section~\ref{sec:boundNeeded}, the composition of these two parts of the proposed solution has} a long line of distributed applications, such as service replication and Blockchain. Thus, our contribution can facilitate solutions that are more fault-tolerant than the existing implementations which they cannot recover after the occurrence of the last transient fault.  

\Subsection{Document structure} 
The paper proceeds with the system settings (Section~\ref{sec:sys})\remove{ where we specify the failure model, the system's task and complexity metrics}.
Section~\ref{sec:MMR} briefly explains the MMR algorithm.
It then presents a non-self-stabilizing interpretation of MMR that embodies the reliability guarantees for broadcast-based communications that the proposed solution uses.
This non-self-stabilizing algorithm is a steppingstone to our loosely-self-stabilizing algorithm that is featured (along with its correctness poof) in Section~\ref{sec:theAlgo}.
%Section~\ref{sec:correctness} provides the correctness proof, Section~\ref{sec:exte} discusses an extension, 
Section~\ref{sec:bck} presents a SSBFT recycling mechanism for $\mathsf{BSMP_{n,t}[\kappa\mathit{-}SGC,t < n/3,RCCs]}$, and Section~\ref{sec:conclusions} concludes the paper.  

For the reader's convenience, Table~\ref{fig:Glossary} (given before the bibliography) includes the Glossary, where all abbreviations are listed.

\Section{System Settings for $\mathsf{BAMP_{n,t}[\mathit{-}FC, t < n/3,RCCs]}$}
\label{sec:sys}
We consider an asynchronous message-passing system that has no guarantees on the communication delay. Moreover, there is no notion of global (or universal) clocks and the algorithm cannot explicitly access the local clock (or timeout mechanisms). The system consists of a set, $\sP$, of $n$ fail-prone \emph{nodes} (sometimes called \emph{processes} or \emph{processors}) with unique identifiers. Any pair of nodes $p_i,p_j \in \sP$ has access to a bidirectional communication channel, $\mathit{channel}_{j,i}$, that, at any time, has at most $\capacity \in \N$ packets on transit from $p_j$ to $p_i$ (this assumption is due to a well-known impossibility~\cite[Chapter 3.2]{DBLP:books/mit/Dolev2000}). 

%\Subsection{Execution model}
%
%\label{sec:interModel}
%
In the \emph{interleaving model}~\cite{DBLP:books/mit/Dolev2000}, the node's program is a sequence of \emph{(atomic) steps}. Each step starts with an internal computation and finishes with a single communication operation, \ie a message $send$ or $receive$. The \emph{state}, $s_i$, of node $p_i \in \sP$ includes all of $p_i$'s variables and $\mathit{channel}_{j,i}$. The term \emph{system state} (or configuration) refers to the tuple $c = (s_1, s_2, \cdots,  s_n)$. We define an \emph{execution (or run)} $R={c[0],a[0],c[1],a[1],\ldots}$ as an alternating sequence of system states $c[x]$ and steps $a[x]$, such that each $c[x+1]$, except for the starting one, $c[0]$, is obtained from $c[x]$ by $a[x]$'s execution. 

\Subsection{Task specifications}
\label{sec:spec}

\ems{Next, we detail the studied task.}

\Subsubsection{Returning the decided value}
Definition~\ref{def:consensus} considers the $\mathsf{propose}(v)$
operation. We refine the definition of $\mathsf{propose}(v)$ by
specifying how the decided value is retrieved. This value is either
returned by the $\mathsf{propose}()$ operation (as in the studied
algorithm~\cite{DBLP:conf/podc/MostefaouiMR14}) or via the returned
value of the $\done()$ operation (as in the proposed solution). {In
the latter case, the symbol $\bot$ is returned as long as no value
was decided. Also, the symbol $\blitza$ indicates a (transient) error
that occurs only when the proposed algorithm exceeds the bound on the
number of iterations that it may take.}

\Subsubsection{Randomized guarantees}
The studied algorithm has a randomized guarantee with respect to the
liveness requirement, \ie BC-completion. Specifically, MMR states
that each non-faulty node decides with probability $1$. Also,
since MMR is a round-based algorithm, it holds that $\lim_{r
\rightarrow + \infty}(\Pr_{\mathrm{MMR}} [p_i \text{ decides by round }
r])=1$.

In order to bound the amount of memory that the proposed algorithm
uses, the proposed solution allows the algorithm to run for a bounded
number of rounds. Specifically, there is a predefined constant, $M \in
\mathbb{Z}^+$, such that the probability of $\Pr_{\mathit{proposed}}
[p_i \text{ decides by round } M\mathit{+}1]=1$. Due to this,
the proposed algorithm provides a randomized guarantee with
respect to the safety requirements, \ie BC-validity and
BC-agreement. Specifically, $\Pr_{\mathit{proposed}} [p_i
\text{ satisfies the safety
requirements}]=1-\bigO(2^{-M})$. In other words, the
proposed solution has weaker guarantees than the studied
algorithm with respect to the safety requirements.

\Subsubsection{{Invocation by algorithms from higher layers}}
\label{sec:initialization}
We assume that the studied problem is invoked by algorithms that run at higher layers, such as multivalued consensus, see Figure~\ref{fig:suit}. This means that eventually there is an invocation, $I$, of the proposed algorithm that starts from a well-initialized system state. That is, immediately before invocation $I$, all local states of all non-faulty nodes have the (predefined) initial values in all variables and the communication channels do not include messages related to invocation $I$.

For the sake of completeness, we illustrate briefly how the assumption above can be covered~\cite{DBLP:conf/ftcs/Powell92} in the studied hybrid asynchronous/synchronous architecture presented in Figure~\ref{fig:suit}. Suppose that upon the periodic installation of the common seed, the system also initializes the array of binary consensus objects that are going to be used with this new installation. In other words, once all operations of a given common seed installation are done, a new installation occurs, which also initializes the array of binary consensus objects that are going to be used with the new common seed installation. Note that the efficient implementation of a mechanism that covers the above assumption is outside the scope of this work.

\Subsubsection{Legal executions}
The set of \emph{legal executions} ($LE$) refers to all the executions in which the requirements of task $T$ hold. In this work, $T_{\text{binCon}}$ denotes the task of binary consensus, which Definition~\ref{def:consensus} specifies, and $LE_{\text{binCon}}$ denotes the set of executions in which the system fulfills $T_{\text{binCon}}$'s requirements. 

Due to the  BC-completion requirement (Definition~\ref{def:consensus}), $LE_{\text{binCon}}$ includes only finite executions. In Section~\ref{sec:loosely}, we consider executions $R=R_1\circ R_2 \circ,\ldots$ as infinite compositions of finite executions, $R_1, R_2,\ldots \in LE_{\text{binCon}}$, such that $R_x$ includes one invocation of task $T_\text{binCon}$, which always satisfies the liveness requirement, \ie BC-completion, but, with an exponentially small probability, it does not necessarily satisfy the safety requirements, \ie BC-validity and BC-agreement.

\Subsection{The fault model and self-stabilization}
A failure occurrence is a step that the environment takes rather than the algorithm.

\Subsubsection{Benign Failures}  
\label{sec:benignFailures}
When the occurrence of a failure cannot cause the system execution to lose legality, \ie to leave $LE$, we refer to that failure as a benign one. % (Figure~\ref{fig:self-stab-SDN}).

\Subsubsubsection{Communication failures and fairness}
We consider solutions that are oriented towards asynchronous
message-passing systems and thus they are oblivious to the time at
which the packets arrive and depart. We assume that any message can
reside in a communication channel only for a finite period. Also, the
communication channels are prone to packet failures, such as omission,
duplication, and reordering.  However, if $p_i$ sends a message
infinitely often to $p_j$, node $p_j$ receives that message infinitely
often. We refer to the latter as the \emph{fair communication}
assumption. We also follow the assumption of
MMR  regarding the fair scheduling
of message arrivals (also in the absence of transient-faults) that
does not depend on the current coin's value. \Ie the adversary does
not control the network's ability to deliver messages to non-faulty
nodes.

We note that MMR assumes reliable communication channels
whereas the proposed solution does not make any assumption regarding
reliable communications. Section~\ref{sec:intermediateMMR} provides
further details regarding the reasons why the proposed solution cannot
make this assumption.

\Subsubsubsection{Arbitrary node failures}
\label{sec:arbitraryNodeFaults}
Byzantine faults model any fault in a node including crashes, arbitrary behavior, and malicious behavior~\cite{DBLP:journals/toplas/LamportSP82}. Here the adversary lets each node receive the arriving messages and calculate its state according to the algorithm. However, once a node (that is captured by the adversary) sends a message, the adversary can modify the message in any way, delay it for an arbitrarily long period or even remove it from the communication channel. Note that the adversary has the power to coordinate such actions without any limitation on his computational or communication power. 

We also note that the studied algorithm, MMR, assumes the absence of spoofing attacks, and thus authentication is not needed. Also, the adversary cannot change the content of messages sent from a non-faulty node. Since MMR assumes the availability of a RCCs service, and since the only available, to the best of our knowledge, self-stabilizing RCCs algorithm, BDH~\cite{DBLP:conf/podc/Ben-OrDH08}, assumes private channels, we also assume that the communications between any two non-faulty nodes are private. That is, it cannot be read by the adversary.

For the sake of solvability~\cite{DBLP:journals/toplas/LamportSP82,DBLP:journals/jacm/PeaseSL80,DBLP:conf/podc/Toueg84}, the fault model that we consider limits only the number of nodes that can be captured by the adversary. That is, the number, $t$, of Byzantine failure needs to be less than one-third of the number, $n$, of nodes in the system, \ie $3t+1\leq n$. The set of non-faulty nodes is denoted by $\Correct$ and called the set of non-faulty nodes.

\Subsubsection{Arbitrary transient-faults}
\label{sec:arbitraryTransientFaults}
We consider any temporary violation of the assumptions according to which the system was designed to operate. We refer to these violations and deviations as \emph{arbitrary transient-faults} and assume that they can corrupt the system state arbitrarily (while keeping the program code intact). The occurrence of an arbitrary transient fault is rare. Thus, our model assumes that the last arbitrary transient fault occurs before the system execution starts~\cite{DBLP:books/mit/Dolev2000}. Also, it leaves the system to start in an arbitrary state.

\Subsubsection{Dijkstra's self-stabilization}
\label{sec:Dijkstra}
An algorithm is \emph{self-stabilizing} with respect to the task of $LE$, when every (unbounded) execution $R$ of the algorithm reaches within a finite period a suffix $R_{legal} \in LE$ that is legal. Namely, Dijkstra~\cite{DBLP:journals/cacm/Dijkstra74} requires $\forall R:\exists R': R=R' \circ R_{legal} \land R_{legal} \in LE \land |R'| \in \mathbb{Z}^+$, where the operator $\circ$ denotes that $R=R' \circ R''$ is the concatenation of $R'$ with $R''$. The part of the proof that shows the existence of $R'$ is called the \emph{convergence} (or recovery) proof, and the part that shows that $R_{legal} \in LE$ is called the \emph{closure} proof. The main complexity measure of a self-stabilizing system is the length of the recovery period, $R'$, which is counted by the number of its asynchronous communication rounds during fair executions, as we define in Section~\ref{sec:asynchronousRounds}.

\Subsection{Execution fairness and wait-free guarantees}

\label{sec:fairnessEx}

{We say that a system execution is \emph{fair} when every step of a correct node that is applicable infinitely often is executed infinitely often and fair communication is kept. Self-stabilizing algorithms often assume that their executions are fair~\cite{DBLP:books/mit/Dolev2000}. Wait-free algorithms guarantee that any operation (that was invoked by non-failing nodes) is always complete in the presence of asynchrony and any number of node failures. 
%
%Note that fair executions do not consider node failures (that were not detected by the system which then excluded these failing nodes from reentering the system, as in~\cite{DBLP:conf/netys/DolevGMS17}). 
%
This work assumes execution fairness during the period in which the system recovers from the occurrence of the last arbitrary transient fault. In other words, the system is wait-free only during legal executions, which are absent from arbitrary transient-faults. Moreover, the system recovery from arbitrary transient-faults is not wait-free, but this bounded recovery period occurs only once throughout the system execution.}

\Subsection{Asynchronous communication rounds}

\label{sec:asynchronousRounds}

\ems{As explained in Section~\ref{sec:SelfStabIntro},} it is well-known that self-stabilizing algorithms cannot \ems{(stop their execution and)} stop sending messages~\cite[Chapter 2.3]{DBLP:books/mit/Dolev2000}. Moreover, their code includes a do-forever loop. The proposed algorithm uses $M$ communication round numbers. Let $r \in \{1,\ldots, M\}$ be a round number. We define the $r$-th \emph{asynchronous (communication) round} of {an algorithm's} execution $R=R'\circ A_r \circ R''$ as the shortest execution fragment, $A_r$, of $R$ in which {\em every} correct node $p_i \in \sP:i \in \Correct$ starts and ends its $r$-th iteration, $I_{i,r}$, of the do-forever loop\remove{ (lines~\ref{ln:isActive} to~\ref{ln:rM})}. Moreover, let $m_{i,r,j,\mathit{ackReq}=\true}$ \remove{$m_{i,r,j,\true}=\mathrm{EST}(\true,r,\bullet)$} be a message that $p_i$ sends\remove{sent in line~\ref{ln:ESTrepeatB}} to $p_j$ during $I_{i,r}$, where the field $\mathit{ackReq}=\true$ implies that an acknowledgment reply is required. \remove{Note that in line~\ref{ln:estArrival}, $p_j$ acknowledge the message $m_{i,r,j,\true}$ by sending the message $m_{j,r,i,\false}=\mathrm{EST}(\false,r,\bullet)$ to $p_j$.} Let $a_{i,r,j,\true},a_{j,r,i,\false} \in R$ be the steps in which $m_{i,r,j,\true}$ and $m_{j,r,i,\false}$ arrive to $p_j$ and $p_i$, respectively. We require $A_r$ to also include, for every pair of correct nodes $p_i,p_j\in \sP:i,j \in \Correct$, the steps $a_{i,r,j,\true}$ and $a_{j,r,i,\false}$. We say that $A_r$ is \emph{complete} if every correct node $p_i \in \sP:i \in \Correct$ starts its $r$-th iteration, $I_{i,r}$, at \remove{line~\ref{ln:isActive}}{the first line of the do-forever loop}. The latter definition is needed in the context of arbitrary starting system states.

\begin{remark}
\label{ss:first asynchronous cycles}
For the sake of simplifying the  presentation of the correctness proof, when considering fair executions, we assume that any message that arrives in $R$ without being transmitted in $R$ does so within $\bigO(1)$ asynchronous rounds in $R$. 
\end{remark}

\Subsubsection{{Demonstrating recovery of consensus objects invoked by higher layers' algorithms}}
\label{sec:assumptionEasy}
Note that the assumption made in Section~\ref{sec:initialization} simplifies the challenge of meeting the design criteria of self-stabilizing systems. Specifically, demonstrating recovery from transient-faults, \ie convergence proof, can be done by showing completion of all operations in the presence of transient-faults. This is because the assumption made in Section~\ref{sec:initialization} implies that, as long as the completion requirement is always guaranteed, then eventually the system reaches a state in which only initialized consensus objects exist.

% REMOVE

\Subsubsection{Loosely-self-stabilizing systems}
\label{sec:loosely}
Satisfying the design criteria of Dijkstra's self-stabilizing systems
is non-trivial since it is required to eventually satisfy strictly
always the task's specifications. These severe requirements can lead
to some impossibility conditions, as in our case of solving binary
consensus without synchrony
assumptions~\cite{DBLP:conf/wdag/AnagnostouH93,DBLP:journals/ipl/FischerL82,DBLP:journals/jpdc/DuboisPNT12} 

%	\remove{%%% REMOVED COMMENTS
%		\IM{CITE?}. \EMS{IM, maybe you can propose a citation here.} \IM{I
%			am not sure of the most relevant
%			citation. \cite{DBLP:journals/taas/AngluinAFJ08} is the one given
%			by Izumi \etal, but this is leader election and does not perfectly
%			match our case. Is there a specific impossibility result?}\EMS{I
%			think that we need something that appears in
%			Section~\ref{sec:impo}.}  }

To circumvent such challenges, Sudo
\etal~\cite{DBLP:journals/tcs/SudoNYOKM12} proposed the design criteria
for loosely-self-stabilizing systems, which relaxes Dijkstra's
criteria by requiring that, starting from any system state, the system
(i) reaches a legal execution within a relatively short period, and
(ii) remains in the set of legal for a relatively long period. The
definition of loosely-self-stabilizing systems by Sudo \etal
considers the task of leader election, which any system state may, or
may not, satisfy. This paper focuses on an operation-based task that
has both safety and liveness requirements. Only at the end of the task
execution, can one observe whether the safety requirements were
satisfied. Thus, Definition~\ref{def:practSelf} presents a variation
of Sudo \etal's definition that is operation-based and requires
criterion (i) to hold within a finite time rather than within `a short
period'.

To that end, Definition~\ref{def:operation} says what it means for a
system $\cS$ that implements operation $\mathsf{op}()$ to satisfy
task $T_{\mathsf{op}()}$'s safety requirements with a probability
$p_\cS$. Definition~\ref{def:operation} uses the term correct
invocation of operation $\mathsf{op}()$. Recall that
in Section~\ref{sec:spec} we define what a correct invocation of binary
consensus is, \ie it is required that all correct nodes invoke
the $\mathsf{propose}()$ operation exactly once during any execution
that is in $LE_{\text{binCon}}$.

Definition~\ref{def:operation} specifies probabilistic satisfaction of repeated invocations of operation $\mathsf{op}()$.

\begin{definition}%[Probabilistic satisfaction of repeated invocations of operation $\mathsf{op}()$]
\label{def:operation}
For a given system $\cS$ that aims at satisfying task $T_{\mathsf{op}()}$ in a probabilistic manner, denote by $\mathit{IE}_\cS(LE_{\mathsf{op}()})$ the set of all infinite executions that system $\cS$ can run, such that for any $R \in \mathit{IE}_\cS(LE_{\mathsf{op}()})$ it holds that $R=R_1\circ R_2 \circ,\ldots$ is an infinite composition of finite executions, $R_1, R_2,\ldots \in LE_{\mathsf{op}()}$. Moreover, each $R_x:x\in \mathbb{Z}^+$ includes the correct invocation of $\mathsf{op}()$ that always satisfies $T_{\mathsf{op}()}$'s liveness requirements. 

We say that $R$ satisfies task $T_{\mathsf{op}()}$'s safety requirements with probability $\Pr_R$ if (i) for any $x\in \mathbb{Z}^+$ it holds that $R_x \in LE_{\mathsf{op}()}$ with  probability $\Pr_{R_x} \leq \Pr_R$ and (ii) for any $x,y \in \mathbb{Z}^+$ the event of $R_x \in LE_{\mathsf{op}()}$ and $R_y \in LE_{\mathsf{op}()}$ are independent. Furthermore, we say system $\cS$ satisfies task $T_{\mathsf{op}()}$ with  probability $\Pr_\cS$ if $\forall R \in \mathit{IE}_\cS(LE_{\mathsf{op}()}) :\Pr_R \leq \Pr_\cS$.	  
\end{definition}

Definition~\ref{def:practSelf} specifies probabilistic operation-based eventually-loosely-self-stabilizing systems.

\begin{definition}[Eventually-loosely-self-stabilizing systems]
\label{def:practSelf}
Let $\cS$ be a system that implements a probabilistic solution for task $T_{\mathsf{op}()}$. Let $R$ be any unbounded execution of $\cS$, which includes repeated sequential and correct invocations of $\mathsf{op}()$, such that task $T_{\mathsf{op}()}$ completes within a period of $\pinf$ steps in $R$. Suppose that within a finite number of steps in $R$, the system $\cS$ reaches a suffix of $R$ that satisfies $T_{\mathsf{op}()}$'s safety requirements with the probability $\Pr_\cS =1-p:p \in o(\pinf)$. In this case, we say that system $\cS$ is eventually-loosely-self-stabilizing, where $\pinf$ is the complexity measure.
\end{definition}

Definition~\ref{def:practSelf} says that any eventually-loosely-self-stabilizing system recovers within a finite period. After that period, the probability to violate safety-requirement is exponentially small. This work shows that the studied algorithm has an eventually-loosely-self-stabilizing variation for which the probability to violate safety can be made so low that it becomes negligible (Remark~\ref{thm:pratSafe}).

% REMOVE

\Section{Non-self-stabilizing MMR for $\mathsf{BAMP_{n,t}[\mathit{-}FC, t < n/3,RCCs]}$}	
\label{sec:MMR}
We review the MMR algorithm (Section~\ref{sec:MMRdetails}). This algorithm considers a communication abstraction named BV-broadcast, which we bring before we present the details of MMR. Then, we present a non-self-stabilizing \emsB{BFT} algorithm (Section~\ref{sec:intermediateMMR}) that serves as a steppingstone to the proposed \emsB{SSBFT} algorithm (Section~\ref{sec:theAlgo}).

\Subsection{The MMR algorithm}
\label{sec:MMRdetails}
Algorithm~\ref{alg:MMR} presents the MMR algorithm~\cite{DBLP:conf/podc/MostefaouiMR14}, which considers an underlying  communication abstraction named
BV-broadcast.
{Recall that the set $\Correct$ denotes the set of nodes that do not commit failures.}

% REMOVE

\begin{algorithm*}[t!]
\begin{\algSize}
%			\smallskip
%			
%			\textbf{constants:} $[\bot, \ldots ,\bot]:=[\bot, \ldots ,\bot]$\;
%			$M$ a predefined value the bound the number of iterations\;
%			
%			\smallskip
%			
%			\noindent \textbf{local variables:}\\
%			$est[0,..]:=[\bot, \ldots ,\bot]$ \tcc*{array of the received proposals}\label{ln:proposalsBotsV}
%			$\mathit{binValues}[0,..]:=[\bot, \ldots ,\bot]$ \tcc*{array of the received proposals}\label{ln:proposalsBotsV}
%			$aux[0,..]:=[\bot, \ldots ,\bot]$ \tcc*{array of the received proposals}\label{ln:proposalsBotsV}
%			$s[0,..]:=[\bot, \ldots ,\bot]$ \tcc*{array of the received proposals}\label{ln:proposalsBotsV}
%			$r:=0$ \tcc*{\reduce{the asynchronous }round counter}\label{ln:kZero}
%			$BV[0,..]:=[\bot, \ldots ,\bot]$ \tcc*{binary consensus objects (unbounded list)}\label{ln:BCZero}
%			

\smallskip

\textbf{operation} $\mathsf{bvBroadcast}(v)$ \label{ln:bvBradcast}\textbf{do} \textbf{broadcast} $\mathrm{bVAL}(v)$\;

\smallskip			

\textbf{upon} $\mathrm{bVAL}(\mathit{vJ})$ \textbf{arrival from} $p_j$ \Begin{
	\If{$(\mathrm{bVAL}(\mathit{vJ})$ received from $(t+1)$ different nodes and $\mathrm{bVAL}(\mathit{vJ})$ not yet broadcast$)$}{
		\textbf{broadcast} $\mathrm{bVAL}(vJ)$ \tcc{a node echoes a value only once}
	}
	\If{$(\mathrm{bVAL}(\mathit{vJ})$ received from $(2t + 1)$ different nodes$)$}{$\mathit{binValues} \gets \mathit{binValues} \cup \{\mathit{vJ}\}$\label{ln:endOfBVcast} \tcc{local delivery of a value}}
}

\smallskip

\textbf{operation} $\mathsf{propose}(v)$ \label{ln:proposeV}\Begin{
	$(est,r) \gets (v,0)$\;\label{lnMMR:init}
	
	\smallskip
	
	\textbf{do forever} \Begin{\label{lnMMR:doForever}
		$r \gets r + 1$\;\label{lnMMR:incrRound}
		$\mathsf{bvBroadcast }~ \mathrm{EST}[r](est)$\;\label{lnMMR:1stBcast}
		$\mathbf{wait} (\mathit{binValues}[r] \neq \emptyset)$\label{lnMMR:ret1stBcast}
		\tcc*{$\mathit{\mathit{binValues}}[r]$ has not necessarily obtained its final value when wait returns}
		\textbf{broadcast} $\mathrm{AUX}[r](w)$ \textbf{where} $w \in \mathit{binValues}[r]$\; \label{lnMMR:2ndBcast}
		\textbf{wait} $\exists$ a set of binary values, $\mathit{vals}$, and a set of $(n\mathit{-}t)$ messages $\mathrm{AUX}[r](x)$, such that
		$\mathit{vals}$ is the set union of the values, $x$, carried by these $(n\mathit{-}t)$ messages $\land$ $\mathit{vals} \subseteq \mathit{binValues}[r]$\; \label{lnMMR:ret2ndBcast} 
		$s[r] \gets \mathbf{randomBit}()$\label{lnMMR:getCoin}\;
		\If(\% \ie $|\mathit{vals}| = 1$ \%){$(\mathit{vals} = \{v\})$\label{lnMMR:oneVal}}{\If{$(v = s[r])$}{decide$(v)$ if not yet done}
			$est \gets v$\;
		}
		\lElse{$est \gets s[r]$} \label{lnMMR:twoVals}
	}\label{lnMMR:endLoop}	
}

\smallskip

\caption{\label{alg:consensusNon}\label{alg:MMR}Non-self-stabilizing MMR algorithm for Binary BFT consensus with $t < n/3$, $\bigO(n^2)$ messages, and $\bigO(1)$ expected time; code for $p_i$}
\end{\algSize}
\end{algorithm*}

\Subsubsection{Broadcasting of binary-values}

MMR uses an all-to-all broadcast operation of binary values. That is, the operation, $\mathsf{bvBroadcast}(v)$, assumes that all the correct nodes invoke
$\mathsf{bvBroadcast}(w)$, where $v,w \in \{0,1\}$. 

\Subsubsubsection{Task definition}
\label{sec:sefBVbrodcast}
The set of values that are BV-delivered to node $p_i$ are stored in the read-only variable $binValues_i$, which is initialized to $\emptyset$. Next, we specify under which conditions values are added to $binValues_i$.

\begin{itemize}

\item  \textbf{BV-validity.} Suppose that $v \in binValues_i$ and $p_i$ is correct. It holds that $v$ has been BV-broadcast by a correct node.

\item \textbf{BV-uniformity.} $v \in binValues_i$ and $p_i$ is correct. Eventually $\forall j \in \Correct: v \in binValues_j$.

\item \textbf{BV-completion.} Eventually $\forall i \in \Correct: binValues_i \neq \emptyset$ holds.

\end{itemize}

The above requirements imply that eventually $\exists s \subseteq
\{0,1\}: s \neq \emptyset \land \forall i \in \Correct: binValues_i=s$
and the set $s$ does not include values that were BV-broadcast only by
Byzantine nodes.
%\textcolor{blue}{[CG: Here and in line 22 of Algorithm 2: I suggest we use capital $S$ instead of $s$. Usually, capital letters are used for sets/data structures and small letters for numbers etc. If you agree, then I can go through the document very carefully and make the necessary changes wrt to $s$. This affects also Algorithm 3, eg. line 48.]}
%\EMS{I prefer that we do not do this. I want us to focus on more principle issues and the point that you are raising is not about correctness rather than personal style.  }
%\cgr{OK}

\Subsubsubsection{Implementation}

MMR uses the $\mathsf{bvBroadcast}(v)$ operation
(line~\ref{ln:bvBradcast}) to reliably deliver a $\mathrm{bVAL}(v)$
message containing a single binary value, $v$.  Such values are
propagated via a straightforward ``echo'' mechanism that repeats any
arriving value at most once per sender.  In detail, the mechanism
invokes a broadcast of the proposed value $v$.  Upon the arrival of
value $\mathit{vJ}$ from at least $t+1$ distinct nodes,
$\mathit{vJ}$ is replayed via broadcast (but only if this was not done
earlier).  Also, if $\mathit{vJ}$ was received by at least $2t+1$
different nodes, then $\mathit{vJ}$ is added to a set
$\mathit{binValues}$. On round $r$ of MMR's operation
$\mathsf{propose}(v)$, the set $\mathit{binValues}$ appears as
$\mathit{binValues}[r]$.

Note that no correct node can become aware of when its local copy of the set $binValues$ has reached its final value. Suppose this would have been possible, consensus can be solved by instructing each node deterministically select a value from the set $binValues$ and by that contradict FLP~\cite{DBLP:journals/jacm/FischerLP85}.

\Subsubsection{MMR's binary randomized consensus algorithm}

\Subsubsubsection{Variables}
Algorithm~\ref{alg:MMR} uses variable $r$ (initialized by zero) for
counting the number of asynchronous communication rounds. The variable
$est$ holds the current estimate of the value to be decided. As
mentioned in Section~\ref{sec:rcc}, the operation
$\mathrm{randomBit}(r)$ retrieves the value of the RCC on
round~$r$. The set $vals \subseteq \{0,1\}$ holds the value received
during the current round.
%The algorithm makes use of the $\mathsf{bvBroadcast}()$ operation. 
Recall that node $p_i \in \sP$ stores the binary values received
in a round $r$ via a $\mathsf{bvBroadcast}()$ in the read-only set
$binValues_i[r]$.

%During round $r$, every node $p_i \in \sP$ stores in a local variable $est_i[r]$ its estimated decision values, where $est_i[0]=v$ stores its own proposal.
% When a node invokes $\mathsf{bvBroadcast}~ \mathrm{MSG}(v)$, we say that it BV broadcasts the message $\mathrm{MSG}(v)$, where $\mathrm{MSG}$
%is the "type" of the message and $v$ is a binary value ($0$ or $1$).  

%Since nodes exchange these estimations, $est_i[r][j]$ stores the last estimation that $p_i$ received from $p_j$. Note that $est_i[r][j] \subseteq \{0,1\}$ holds a set of values and is initialized by the empty set, $\emptyset$. 
%At the end of round $r$, node $p_i \in \sP$ tests whether it is ready to decide after it selects a single value $w \in est_i[r][i]$ to be exchanged with other nodes. 
%In order to ensure reliable broadcast in the presence of packet loss, there is a need to store $w$ in an auxiliary storage, $aux_i[r][i]$, so that $p_i$ can retransmits $w$.

\Subsubsubsection{Detailed description} %of MMR (Algorithm~\ref{alg:MMR})}

MMR's main algorithm (appearing as the $\mathsf{propose}(v)$ operation) comprises three phases.
After initialization (line~\ref{lnMMR:init}), Algorithm~\ref{alg:consensusNon} enters a do forever loop (lines~\ref{lnMMR:doForever}--\ref{lnMMR:endLoop}) that executes endlessly, reflecting the non-deterministic nature of its completion guarantees. 
Every iteration signifies a new round of the protocol by initiating with a round number increment (line~\ref{lnMMR:incrRound}) and is performed via the following phases.

\begin{itemize}
\item \emph{Query the estimated binary values
(lines~\ref{lnMMR:1stBcast}--\ref{lnMMR:ret1stBcast})}: The
estimate $est$ is broadcast via the $\mathsf{bvBroadcast}()$
protocol. Due to the BV-completion property, eventually,
the set $\mathit{binValues}[r]$ is populated with at least
one binary value, $w$. Even though the system might not
reach the final value of the set during round $r$, by
BV-validity we know that any value in the set is an
estimated value during round $r$ of at least one correct
node.

\item \emph{Inform about the query results (lines~\ref{lnMMR:2ndBcast}--\ref{lnMMR:ret2ndBcast})}: The auxiliary message, $AUX(w)$, carrying the value of  $\mathit{binValues}[r]$ is broadcast. Note that all the correct nodes, $p_j$, broadcast $w\in values_j[r]$, \ie a value that is estimated by at least one correct node. However, arbitrary binary values can be broadcast by the Byzantine nodes. 

Processor $p_i$ then waits for the arrival of $AUX(w)$ messages from $n-t$ distinct nodes, and gathers their attached values, $w$, in the set $\mathit{vals}$. By waiting for {$n-t$ arrivals of these} $AUX()$ messages, Algorithm~\ref{alg:MMR} can:
\begin{itemize}
\item Sift out values that were sent only by Byzantine nodes, cf. $vals_i \subseteq \mathit{binValues}_i[r]$ at line~\ref{lnMMR:ret2ndBcast}.
\item Guarantee that, for a given round $r$, it holds that $\exists i \in \Correct: vals_i = \{v\} \implies \forall j \in \Correct:v \in  vals_j$. Also, $vals_i \subseteq  
\{0, 1\}$ and any $v \in vals_i$ is an estimated value that was BV-broadcast by at least one correct node.
\end{itemize}

\item \emph{Try-to-decide (lines~\ref{lnMMR:oneVal}--\ref{lnMMR:twoVals})}: If there is a single value in $\mathit{vals}$, then this value serves as the estimated value for the next round. This is also the decided value if it coincides with the output of the RCC and the node has not yet decided. If $\mathit{vals}$ contains both of the binary values, the RCC output serves as the estimated value for the next round. Note that deciding on a value does not mean that any node can stop executing Algorithm~\ref{alg:MMR}. (The non-self-stabilizing version of MMR can be found in~\cite{DBLP:conf/podc/MostefaouiMR14}.)
\end{itemize}

We end the description of Algorithm~\ref{alg:consensusNon} by bringing
a couple of examples that illustrate how the try-to-decide phase
works. Note that if all correct nodes estimate the same value
during round $r$, then $\exists x \in \{0,1\}:\forall i \in \Correct: x
\in \mathit{binValues}[r]_i$ holds, which means that $\exists x \in
\{0,1\}:\forall i \in \Correct: vals_i=\{x\}$ holds during round
$r$. Moreover, the proof of MMR~\cite{DBLP:conf/podc/MostefaouiMR14}
shows that $\exists x \in \{0,1\}:\forall i \in \Correct: vals_i=\{x\}$
holds for any round $r'\geq r$. Thus, the decision of $x$ depends only
on the value of the RCC. In other words, the RCC has
the ``correct value'' with probability $1/2$ and the algorithm
decides.

Now suppose that, for any reason, $\exists x \in \{0,1\}:\forall i \in
Correct: vals_i=\{x\}$ does not hold during round~$r$. Then, any
node that decides on round $r$ decides the value of the common
coin. Also, the ones that do not decide on round $r$, since $vals
=\{0,1\}$, estimate for round $r+1$ the value of the common
coin. Therefore, BC-agreement holds in this case. Moreover, all the
nodes for which $vals =\{0,1\}$ holds during round $r$ select
``the correct'' estimated value from the set $vals$ with probability
$1/2$, and thus, the system reaches a state in which all nodes
have the same estimated value. As discussed above, this state leads to
agreement with probability $1/2$. More details can be found
in~\cite{DBLP:conf/podc/MostefaouiMR14}.

\Subsection{The non-self-stabilizing yet bounded version of the studied algorithm}
\label{sec:intermediateMMR}
After reviewing MMR, we transform the code of Algorithm~\ref{alg:MMR}
into Algorithm~\ref{alg:intermediateMMR}, which has a bound, $M$, on
the number of iterations of the do-forever loop in
lines~\ref{lnMMR:doForever} to~\ref{lnMMR:endLoop}. In this paper,
Algorithm~\ref{alg:intermediateMMR} serves as a steppingstone towards
the proposed solution, which appears in
Algorithm~\ref{alg:consensus}. We start the presentation of
Algorithm~\ref{alg:intermediateMMR} by weakening the assumptions that
the studied solution has about the communication channels. This will
help us later when presenting the proposed solution.

\begin{algorithm*}[t!]
\begin{\algSizeSmall}
%		\smallskip
%		
%		\textbf{constants:} 
%		$M$ a predefined value the bound the number of iterations\;
%				$B:4 (\theta+1)(n+1)$\;
%		
%		% $[\bot, \ldots ,\bot]:=[\bot, \ldots ,\bot]$; $\emptys:=[\emptyset, \ldots ,\emptyset]$\;
%		
%		\smallskip
%		
%		
%		\noindent \textbf{local variables:}\\
%		$r:=0$ \tcc*{the asynchronous round counter}\label{ln:NkZero}
%		$est[0,..,M\mathrm{+}1][0,..,n\mathrm{-}1]:=[[\emptyset, \ldots ,\emptyset],\ldots]$ \tcc*{array of estimated decision}\label{ln:NproposalsBotsV}
%		$aux[0,..,M][0,..,n\mathrm{-}1]:=[[\bot, \ldots ,\bot],\ldots]$ \tcc*{array of exchanged estimated decisions\label{ln:NproposalsBotsV}}
%		$ct[0,..,1][0,..,n\mathrm{-}1]$\;
%		$\mathit{rt}[0,..,n\mathrm{-}1][0,..,n\mathrm{-}1]$\;
%		$\mathit{phs}[0,..,n\mathrm{-}1]:=[\mathit{-}1,\ldots,\mathit{-}1]$\;

\smallskip

\textbf{operations:} $\mathsf{propose}(v)$ \label{ln:MproposeV}\textbf{do} $\{(est[0][i],aux[0][i] \gets(\{v\},\bot)\}$;

%		$\mathsf{init}()$ \label{ln:NproposeV}\textbf{do} $\{(r,est,aux) \gets$ $(0,[[\emptyset, \ldots ,\emptyset],\ldots],[[\bot, \ldots ,\bot],\ldots]$;\}
\smallskip

$\mathsf{result}()$ \label{ln:NproposeV}\textbf{do} \{\textbf{if} $(est[M\mathit{+}1][i] = \{v\})$ \textbf{then return} {$v$} {\textbf{else if} $(r \geq M\land \mathsf{infoResult}() \neq \emptyset)$\textbf{ then return}{$\blitza$}}\textbf{else return} {$\bot$};\}

\smallskip

%		\textbf{macros:} $\mathit{binValues}(r,x)$ \textbf{return} {$\{y \in \{0,1\}:\exists s\subseteq \sP:|s|=x \land y\in  \cup_{p_j \in s:\mathit{est}[r][j]\neq \emptyset} \mathit{est}[r][j] \}$}\label{ln:NbinVal}\; 
\textbf{macros:} $\mathit{binValues}(r,x)$ \textbf{return} {$\{y \in \{0,1\}:\exists s\subseteq \sP:|\{p_j \in s: y \in \mathit{est}[r][j]\}| \geq x \}$}\label{ln:NbinVal}\; 
%\textcolor{blue}{[CG: I propose $s\rightarrow S$. Also, from my understanding it could be the case that there are nodes that have $0$ and $1$. For $t+1$, line 33, then it is the case to return both $0$ and $1$? Also isn't it possible that the empty set is returned, e.g., when not enough witnesses exist for 0 or 1? I think we need to clarify these issues.]}\EMS{Yes, it can be any subset of $\{0,1\}$.}\;

{$\mathsf{infoResult}()$ \textbf{do} \{\textbf{if} $(\exists s \subseteq \sP:n\mathrm{-}t \leq |s| \land ( \forall p_j \in s :\mathit{aux}[r][j] \in  \mathit{binValues}(r,2t\mathrm{+}1)))$ \textbf{then} \Return{$\{\mathit{aux}[r][j] \}_{p_j \in s}$} \textbf{else} \Return{$\emptyset$}\}}\;

\smallskip

\textbf{functions:} 		$\mathsf{decide}(x)$ \Begin{
\lIf{$(est[M\mathit{+}1][i] = \emptyset \lor aux[M\mathit{+}1][i] = \bot)$}{$(est[M\mathit{+}1][i],aux[M\mathit{+}1][i]) \gets (\{x\},x)$\label{ln:NNdecideX}}
}

$\mathsf{tryToDecide}(values)$\label{ln:NtryToDecide}  \Begin{\lIf{\label{ln:NcupEq}$(values \neq \{v\})$}{$est[r][i] \gets \{\mathbf{randomBit}(r)\}$\label{ln:NestRiGetsRandom}}
\lElse{\{$est[r][i] \gets \{v\}$\label{ln:NestRiGetsV}; \textbf{if} {$(v = \mathbf{randomBit}(r))$} \textbf{then} {\label{ln:NcupEqMore}$\mathsf{decide}(v)$}\}}}

%\textcolor{blue}{[CG: What does returning $\bot$ represent? Do we need it? I guess it depends when result() is called. If it is triggered when $r$ reaches $M$, then I think we do not need $\bot$]}\EMS{Changed. I hope that now it is clearer and simpler. The symbol $\bot$ represents 'not ready', see the revised text in Section~\ref{sec:spec}. }
%\cgr{Yes, not is clear.}

\smallskip

\textbf{do forever} \Begin{ \label{lnMidAlg:doStartM}
\If{{$(est[0][i] \neq \emptyset)$}\label{ln:isActiveM}}{
	
	$r \gets \min \{r\mathrm{+}1,M\}$\label{ln:NplusOneM}\;
	
	%	\textcolor{blue}{[CG: This ensures that $r$ never get larger than $M$. So, in line 28, the check if $(r \ge M)$ could be made just equality. Of course, a transient fault could make $r$ larger than $M$, so the check for larger than $M$ is possibly put in line 28 to prepare for Algorithm 3. If I am right, then it is good to mention this, otherwise we might have the reader confused (especially one not familiar with self-stab).]}\EMS{Yes, it is becuase of Algorithm 3.}

	\Repeat{$aux[r][i] \neq \bot$\label{ln:NESTrepeat}}{\label{ln:NESTrepeatStart}
		
		\textbf{foreach} $p_j \in \sP$ \textbf{do send} $\mathrm{EST}(\true,r,est[r\mathrm{-}1][i]\cup\mathit{binValues}(r,t\mathrm{+}1))$ \textbf{to} $p_j$\label{ln:NESTrepeatB}
		
		\lIf{$(\exists w \in \mathit{binValues}(r,2t\mathrm{+}1))$\label{ln:NauxIf}}{$aux[r][i]\gets w$\label{ln:NauxThen}}			
	}
	
	%				\textcolor{blue}{[CG: I propose $s\rightarrow S$] }
	
	\Repeat{{$\mathsf{infoResult}()\neq \emptyset$}\label{ln:NESTrepeatC}}{\label{ln:NESTrepeatCstart}
		
		\textbf{foreach} $p_j \in \sP$ \textbf{do send} $\mathrm{AUX}(\true,r,aux[r][i])$ \textbf{to} $p_j$\label{ln:NESTrepeatD}}

	{$\mathsf{tryToDecide}(\mathsf{infoResult}())$}\label{ln:NrM}\;
	
	%				\lIf{$\exists w \in \mathit{binValues}(M\mathit{+}1,t\mathrm{+}1)$\label{ln:NfPlusOneDecideIf}}{$\mathsf{decide}(w)$\label{ln:NfPlusOneDecideThen}}
}
\label{lnMidAlg:doEnd}
}
\smallskip
\textbf{upon} $\mathrm{EST}(\mathit{aJ},\mathit{rJ},\mathit{vJ})$\label{ln:NuponEst} \textbf{arrival from} $p_j$ \textbf{do} \Begin{
{$est[\mathit{rJ}][j]\gets est[\mathit{rJ}][j] \cup\mathit{vJ}$\label{ln:NestUpdate}}\;
\lIf{$(\mathit{aJ})$}{\textbf{send} $\mathrm{EST}(\false,\mathit{rJ},est[\mathit{rJ}\mathit{-}1][i])$ \textbf{to} $p_j$\label{ln:NestArrival}}
}
\smallskip
\textbf{upon} $\mathrm{AUX}(\mathit{aJ},\mathit{rJ},\mathit{vJ})$\label{ln:NuponAUX} \textbf{arrival from} $p_j$ \textbf{do} \Begin{
\lIf{$(\mathit{vJ}\neq \bot)$}{$aux[\mathit{rJ}][j] \gets \mathit{vJ}$\label{ln:NauxUpdate}}
\lIf{$(\mathit{aJ})$}{\textbf{send} $\mathrm{AUX}(\false,\mathit{rJ},aux[\mathit{rJ}][i])$ \textbf{to} $p_j$\label{ln:NauxArrival}}
}

\smallskip

% \label{alg:consensusNon}

\caption{\label{alg:intermediateMMR}Non-self-stabilizing BFT binary consensus that uses $M$ iterations and violates safety with a probability that is in $\bigO(1/2^M)$; code for $p_i$.}
\end{\algSizeSmall}
\end{algorithm*}

\Subsubsection{Variables} 
%
%$r:=0$ \tcc*{the asynchronous round counter}\label{ln:kZero}
%$est[0,..,M\mathrm{+}1][0,..,n\mathrm{-}1]:=[[\emptyset, \ldots ,\emptyset],\ldots]$ \tcc*{array of estimated decision}\label{ln:proposalsBotsV}
%$aux[0,..,M][0,..,n\mathrm{-}1]:=[[\bot, \ldots ,\bot],\ldots]$ \tcc*{array of exchanged estimated decisions\label{ln:proposalsBotsV}}
%
% (All entires in $est[][]$ are initialized to $\emptyset$.) 
%
Algorithm~\ref{alg:intermediateMMR} uses variable $r$ (initialized to zero) for counting the number of asynchronous communication rounds. During round $r$, every node $p_i \in \sP$ stores in the set $est_i[r][i]$ its estimated decision values, where $est_i[0][i]=\{v\}$ stores its own proposal and $est_i[M\mathit{+}1][i]$ aims to hold the decided value. Since nodes exchange these estimates, $est_i[r][j]$ stores the last estimate that $p_i$ received from $p_j$. Note that $est_i[r][j] \subseteq \{0,1\}$ holds a set of values and it is initialized by the empty set, $\emptyset$. At the end of round $r$, node $p_i \in \sP$ tests whether it is ready to decide after it selects a single value $w \in est_i[r][i]$ to be exchanged with other nodes. In order to ensure reliable broadcast in the presence of packet loss, there is a need to store $w$ in auxiliary storage, $aux_i[r][i]$, so that $p_i$ can retransmit $w$. Note that all entries in $aux[][]$ are initialized to $\bot$.

\Subsubsection{Transforming the assumptions about the communication channels} 
\label{sec:transformChannels}
MMR assumes
reliable communication channels when broadcasting in a
quorum-based manner, \ie sending the same message to all nodes in
the system and then waiting for a reply from the maximum number of
nodes that guarantee never to block forever. After explaining
why the proposed algorithm cannot make this assumption, we present how
Algorithm~\ref{alg:intermediateMMR} provides the needed communication
guarantees.

\Subsubsubsection{The challenge} 
Without a known bound on the capacity of the communication channels, self-stabilizing end-to-end communications are not possible~\cite[Chapter 3]{DBLP:books/mit/Dolev2000}. In the context of self-stabilization and quorum systems, Dolev, Petig, and Schiller~\cite{DBLP:journals/corr/abs-1806-03498} explained that one has to avoid situations in which communicating in a quorum-based manner can lead to a contradiction with the system assumptions. Specifically, the asynchronous nature of the system can imply that there is a subset of nodes that are able to complete many round-trips with a given sender, while the other nodes in $\sP$ accumulate messages in their communication channels, which must have bounded capacity. If such a scenario continues, the channel capacity might drive the system either to block or remove messages from the communication channel before their delivery. Therefore, the proposed solution weakens the required properties for FIFO reliable communications when broadcasting in a quorum-based manner.

% (instead of simply using existing solutions~\cite{DBLP:conf/sss/DolevHSS12} for self-stabilizing reliable  end-to-end communications). 

%\EMS{The details of these embedding also motivates the explicit allocation (and in an a priori manner) of all stored information. This explicit allocation is necessary due to the need of nodes that are at communication round $r$ to retransmit information from earlier rounds $r'<r$.}    

\Subsubsubsection{Self-stabilizing communications} 
One can consider advanced automatic repeat request (ARQ) algorithms
for reliable end-to-end communications, such as the ones by Dolev
\etal~\cite{DBLP:conf/sss/DolevHSS12,DBLP:journals/ipl/DolevDPT11}. However,
our variation of MMR requires only communication fairness. Thus, we
can address the above challenge by looking at simple mechanisms for
assuring that, for every round $r$, all correct nodes eventually
receive messages from at least $n-t$ nodes (from which at least
$n-2t$ must be correct).
For the sake of a simple presentation, we start by reviewing these considerations for the $\mathrm{AUX}()$ messages before the ones for the $\mathrm{EST}()$ messages.  

%For given round number, $r$, sender $p_i$, and receiver $p_j$, the information sent in line~\ref{lnMMR:2ndBcast} does not change. So, an acknowledge from $p_j$ to $p_i$ of the arrival of an $\mathrm{AUX}[r](w)$ message from $p_i$ to $p_j$ assures for $p_i$ that $\mathrm{AUX}[r](w)$ has arrived to $p_j$. The repeat-until loop in lines~\ref{ln:NESTrepeatD} to~\ref{ln:NESTrepeatC} makes sure that this occurs even in the presence of packet loss. Note that duplication is not a challenge since, for a given round number $r$, $p_i$ always sends the same $\mathrm{AUX}[r](w)$ message. Algorithm~\ref{alg:intermediateMMR} deals with packet reordering by storing all information arriving via $\mathrm{AUX}[]()$ messages in the array $aux[][]$. FIFO nodeing is practiced by since on the $r$-th iteration of the do-forever loop in lines~\ref{lnMMR:doForever} to~\ref{lnMMR:endLoop}, node $p_i$ considers only the messages stored in $aux[r][]$.       

\Subsubsubsubsection{$\mathrm{AUX}()$ messages} 
For a given round number, $r$, sender $p_j$, and receiver $p_i$, the
repeat-until loop in lines~\ref{ln:NESTrepeatD}
to~\ref{ln:NESTrepeatC} makes sure, even in the presence of packet
loss, that $p_i$ receives at least $(n-t)$ messages of
$\mathrm{AUX}(\bullet,rnd=r,aux=w):aux_j[r][j]=w$ from distinguishable
senders. This is because line~\ref{ln:NESTrepeatD} broadcasts the
message $\mathrm{AUX}(ack=\true,rnd=r,\bullet)$ and upon its arrival
to $p_j$, line~\ref{ln:NauxArrival} replies with
$\mathrm{AUX}(ack=\false,rnd=r,\bullet)$. Note that duplication is not
a challenge since, for a given round number $r$, $p_j$ always sends
the same $\mathrm{AUX}(\bullet,rnd=r,aux=w):aux_j[r][j]=w$
message. Algorithm~\ref{alg:intermediateMMR} deals with packet
reordering by storing all information arriving via $\mathrm{AUX}[]()$
messages in the array $aux[][]$. We observe from the code of
Algorithm~\ref{alg:intermediateMMR} that FIFO processing is practiced
since during the $r$-th iteration of the do-forever loop in
lines~\ref{lnMidAlg:doStartM} to~\ref{lnMidAlg:doEnd}, node $p_i$
nodes only the values stored in $aux_i[r][]$.

\Subsubsubsubsection{$\mathrm{EST}()$ messages.} 
% 
%When it comes to $\mathrm{EST}[r](est[r])$ messages, r
%
Recall that Algorithm~\ref{alg:consensusNon} uses the $\mathsf{bvBroadcast}()$ operation for broadcasting $\mathrm{EST}[r](est[r])$ messages (line~\ref{lnMMR:1stBcast}). The operation $\mathsf{bvBroadcast}()$ sends $\mathrm{bVAL}(v)$ messages, where $v=est[r\mathit{-}1]$ and possibly also the complementary value $v' \in \{0,1\}\setminus \{est[r\mathit{-}1]\}$. 

For the sake of a concise presentation, Algorithm~\ref{alg:intermediateMMR} embeds the code of operation $\mathsf{bvBroadcast}()$ into its own code. Thus, in Algorithm~\ref{alg:intermediateMMR},  node $p_i$ sends $\mathrm{EST}(\bullet,rnd=r,est=e)$ messages, where the value $e$ of the field $est$ is a set that includes $p_i$'s estimated value, $v:est_i[r\mathit{-}1][i]=\{v\}$, from round number $r-1$ and perhaps also the complementary value, $v' \in \mathit{binValues}(r,t\mathrm{+}1) \setminus \{v\}$, see line~\ref{ln:NESTrepeatB} for details {($\mathit{binValues}()$ may return any subset of $\{0,1\}$)}. Note that once $p_i$ adds the complementary value, $v'$, to the field $est$, the value $v'$ remains in the field $est$ in all future broadcasts of $\mathrm{EST}(\bullet,rnd=r,est=e)$.

Thus, the repeat-until loop in lines~\ref{ln:NESTrepeatStart} to~\ref{ln:NESTrepeat} has at least one value, $v$, that appears in the field $est$ of every $\mathrm{EST}(\bullet,rnd=r,est=e)$ message, and a complementary value, $v'$, that once it is added, it always appears in $e$. Thus, eventually, $p_i$ broadcasts the same $\mathrm{EST}(\bullet,rnd=r,est=e)$ message. Therefore, packet loss is tolerated due to the broadcast repetition in lines~\ref{ln:NESTrepeatStart} to~\ref{ln:NESTrepeat}. Duplication is tolerated due to the union operator that $p_i$ uses for storing arriving information from $p_j$ (line~\ref{ln:NestUpdate}). Concerning reordering tolerance, 
%
%we note that property of BV-uniformity (Definition~\ref{def:bvb}) is specified as requirement that needs to hold within a finite time. Therefore, 
%
the value $est_i[r\mathit{-}1][i]$ always appears in $e$. Thus, once the value $v$ is added to $est_j[r\mathit{-}1][i]$ due to the arrival of a $\mathrm{EST}(\bullet,rnd=r,est=\{v,\bullet\})$ message from $p_i$ to $p_j$ (line~\ref{ln:NestUpdate}), $v$ is always present in $est_j[r\mathit{-}1][i]$. The same holds for any complementary value, $v'$, that $p_i$ adds to later on to $e$ due to the union operation (line~\ref{ln:NestUpdate}). This means, that reordering of $\mathrm{EST}(\bullet,rnd=r,est=\{v,\bullet\})$ messages that do, and do not, include the complementary value, $v'$, does not play a role. 

% \EMS{Need to be clearer why reordering does not play a role.}

\Subsubsection{Detailed description}
\remove{\EMS{IM, do you feel conferrable to help with this?} \IM{Yes.}\\
\IM{Please advise on this description}\textcolor{blue}{[CG: I made some adjustments, including some alignment between the code and the line references.]}\\
}

{As in MMR, Algorithm~\ref{alg:intermediateMMR} includes the following three stages.}

\begin{enumerate}
\item \emph{Invocation.} An invocation of operation $\mathsf{propose}(v)$  (line~\ref{ln:MproposeV}) initializes $est_i[0][i]$ with the estimated value $v$.
No communication or decision occurs before such an invocation occurs.
These actions are only possible through the lines enclosed in the do forever loop (lines~\ref{lnMidAlg:doStartM} to~\ref{ln:NrM}). 
These lines are not accessible before such an invocation, because of the condition of line~\ref{ln:isActiveM}. 
Each iteration of the do forever loop is initiated with a round increment (line~\ref{ln:NplusOneM}); {this line ensures that $r$ is bounded by $M$.}

\item \emph{Communication.} The communication mechanism is detailed in Section~\ref{sec:transformChannels}. 
The first communication phase, which queries the estimated binary values, is implemented in the repeat-until loop of lines~\ref{ln:NESTrepeatStart}--\ref{ln:NESTrepeat}.
The receiver's side of this communication is given in the code of lines~\ref{ln:NuponEst}--\ref{ln:NestArrival}.
Similarly, the second communication phase, which informs about the query results through the use of auxiliary messages, is given in the repeat-until loop of  lines~\ref{ln:NESTrepeatCstart}--\ref{ln:NESTrepeatC}.
Lines~\ref{ln:NuponAUX}--\ref{ln:NauxArrival} are the receiver side's actions for this phase.

\item \emph{Decision.} The decision phase (line~\ref{ln:NrM}) is a call to function $\mathsf{tryToDecide}()$. 
Lines~\ref{ln:NtryToDecide} to~\ref{ln:NestRiGetsV} are the implementation of $\mathsf{tryToDecide}()$. 
This exactly maps the \emph{Try-to-decide} phase of MMR: (i) If the $values$ set that was composed of the auxiliary messages that were received is a single value, then this is the estimate of the next round. 
(ii) If this is also the output of $\mathrm{randomBit}()$ then this is the value to be decided. 
(iii) If $values$ is not a single value then the estimate for the next round is the $\mathrm{randomBit}()$  output.
The actual decision action (line~\ref{ln:NNdecideX}) is for both $est[M+1][i]$ and $aux[M+1][i]$ to be assigned the decided value.
\end{enumerate}

As specified in Section~\ref{sec:spec}, the function $\mathsf{result}()$ (line~\ref{ln:NproposeV}) aims to return the decided value. However, the $\bot$ symbol is returned when no value was decided. Also, it indicates whether $r$ has exceeded the limit $M,$ in which case it returns the error symbol $\blitza${, laying the ground for the proposed self-stabilizing algorithm presented in Section~\ref{sec:theAlgo} (Algorithm~\ref{alg:consensus}).} 
%\sout{If the estimated value for round $M+1$ is $v$, \ie the desired outcome, then $\mathsf{result}()$ returns $v$.} [CG: Here we need to explain what returning $\bot$ represents, if it is needed -- see comment in the code.]}

\Subsubsection{Bounding the number of iterations}
Algorithm~\ref{alg:intermediateMMR} preallocates $\bigO(M)$ of memory space for every node in the system, where $M \in \mathbb{Z}^+$ is a predefined constant that bounds the maximum number of iterations that Algorithm~\ref{alg:intermediateMMR} may take. Lemma~\ref{thm:NprobOneL} shows that Algorithm~\ref{alg:intermediateMMR} may exceed the limit $M$ with a probability that is in $\bigO(2^{-M})$. Once that happens, the safety requirements of Definition~\ref{def:consensus} can be violated. As an indication of this occurrence, the $\mathsf{result}()$ operation returns the transient error symbol, $\blitza$, which some nodes might return. Remark~\ref{thm:pratSafe} explains that it is possible to select a value for $M$, such that the probability for a safety violation is negligible.

\begin{lemma}
\label{thm:NprobOneL}
%	Let $R$ be an execution of Algorithm~\ref{alg:intermediateMMR}.
By the end of round $r$, with probability  $\Pr(r) = 1- (1/2)^r$, we have {$\mathsf{result}_i()\in \{0,1\}:p_i \in \sP:i \in \Correct$.} 
%\textcolor{blue}{[CG: Is it $p_i \in \sP$ or $p_i \in \Correct$?]} \EMS{The former is correct.}\cgr{A Byzantine node can return whatever it wants, so, shouldn't the lemma provide guarantees only for correct nodes, as Claim 3.2 does?}\EMS{To make things simple, Byzantine nodes run the algorithm and the adversary controls only the messages that they send --- in any case, I have updated the text to make it clearer also readers that skipped reading the system settings. }\cgr{Since we only need to reason about correct nodes, it is better to make it explicit, i think}
\end{lemma}
\renewcommand{\lemcnt}{\ref{thm:NprobOneL}}

\begin{lemmaProofSketch}
The proof uses Claim~\ref{thm:NprobOne}.
\begin{claim}
\label{thm:NprobOne}
$\exists v \in \{0,1\}:\forall i \in \Correct: est_i[r][i] =\{v\}$ holds with the probability $\Pr(r) = 1- (1/2)^r$.
\end{claim}
\renewcommand{\clmcnt}{\ref{thm:NprobOne}}
\begin{claimProof}
%
%		The proof considers the execution of lines~\ref{ln:NcupEq} to~\ref{ln:NcupEqMore} during round $r$.
Let $values^r_i$ be the parameter that $p_i$ passes to $\mathsf{tryToDecide}()$ (line~\ref{ln:NtryToDecide}) on round $r$. If $\forall k \in \Correct : values^r_i = \{0,1\}$ or $\forall k \in \Correct :values^r_i = \{v_k(r)\}$ hold, $p_k$ assigns the same value to $est_k[r][k]$, which is $\{\mathrm{randomBit}_k(r)\}$, and resp., $v_k(r)$. The remaining case is when some correct nodes assign $\{v_k(r)\}$ to $est_k[r][k]$ (line~\ref{ln:NestRiGetsV}), whereas others assigns $\{\mathrm{randomBit}_k(r)\}$ (line~\ref{ln:NestRiGetsRandom}).

Recall the assumption that the Byzantine nodes have no control over the network or its scheduler. Due to the RCC properties, $\mathrm{randomBit}_k(r)$ and $\mathrm{randomBit}_k(r')$ are independent, where $r\neq r'$. The assignments of $\{v_k(r)\}$ and $\{\mathrm{randomBit}_k(r)\}$ are equal with the probability of $\frac{1}{2}$. Thus, $\Pr(r)$ is the probability that $[\exists  r' \leq r : \mathrm{randomBit}(r)=v(r)] = \frac{1}{2} + (1-\frac{1}{2})\frac{1}{2} + \cdots + (1-\frac{1}{2})^{r-1}\frac{1}{2} = 1- (\frac{1}{2})^r$.
\end{claimProof}

\medskip

The complete proof shows that the repeat-until loop in lines~\ref{ln:NESTrepeatStart} to \ref{ln:NESTrepeat} cannot block forever and that all the correct nodes $p_i$ keep their estimated value $est_i = \{v\}$ and consequently the predicate $(values^{r'}_i = \{v\})$ at line~\ref{ln:NcupEq} holds for round $r'$, where $values^{r'}_i =\cup_{j \in s} \{\mathit{aux}_i[r][j]\}$. With probability $\Pr(r) = 1- (1/2)^r$, by round $r$, $\mathrm{randomBit}(r)=v$ holds. Then, the if-statement condition of line~\ref{ln:NcupEq} does not hold and the one in line~\ref{ln:NcupEqMore} does hold. Thus, all the correct nodes decide $v$.
\end{lemmaProofSketch}

\begin{remark}[safety in practical settings]
\label{thm:pratSafe}
By Lemma~\ref{thm:NprobOneL}, it is known that, asymptotically speaking, $\Pr(M)$ becomes exponentially small as $M$ grows linearly. Therefore, for a given system, $\cS$, we can select $M \in \mathbb{Z}^+$ to be, say, $150$, so it would take at least $\pinf=10^{100}$ invocations of binary consensus to lead to at most one expected instance in which the requirements of Definition~\ref{def:consensus} are violated. Note that for $M=150$, the arrays $est[]$ and $aux[][]$ require the allocation of $57$ bytes per node, since each node needs only $3nM+\lceil \log M\rceil$ bits of memory. So, $\cS$ can be implemented as a practical system. We believe that one expected violation in every $\pinf$ invocations implies a negligible risk.
\end{remark}

\Section{Self-stabilizing \ems{BFT MMR for $\mathsf{BAMP_{n,t}[\mathit{-}FC, t < n/3,RCCs]}$}}	
\label{sec:theAlgo}

% REMOVE	

Algorithm~\ref{alg:consensus} presents a solution that can recover from transient-faults. %(Section~\ref{sec:recover}). 
We demonstrate the correctness of that solution in Section~\ref{sec:correctness}. The \framebox{boxed lines} in Algorithm~\ref{alg:consensus} are relevant only to an extension (Section~\ref{sec:exte}) that accelerates the notification of the decided value.

\SubsectionS{{Algorithm~\ref{alg:consensus}:} Recovering from transient-faults}
\label{sec:recover}
%
%\ems{The studied architecture (Figure~\ref{fig:suit}) assumes that every invocation of binary consensus is part of an execution of multivalued consensus. Each new invocation of multivalued consensus is responsible for initializing all the binary consensus objects that it is going to use.} Thus, the main remaining concern that we have when designing a loosely-self-stabilizing version of MMR is to make sure that no transient fault can cause the algorithm to not terminate

Recall that by Section~\ref{sec:assumptionEasy}, the main concern that we have when designing a loosely-self-stabilizing version of MMR is to make sure that no transient fault can cause the algorithm to not complete, \eg block forever in one of the repeat-until loops in lines~\ref{ln:NESTrepeatStart} to~\ref{ln:NESTrepeat} and~\ref{ln:NESTrepeatCstart} to~\ref{ln:NESTrepeatC} of Algorithm~\ref{alg:intermediateMMR}.

Recall that Algorithm~\ref{alg:intermediateMMR} is a code transformation of MMR~\cite{DBLP:conf/podc/MostefaouiMR14} that runs for $M$ iterations and violates Definition~\ref{def:consensus}'s safety requirement with a probability that is in $\bigO(2^{-M})$. The proposed solution appears in Algorithm~\ref{alg:consensus}. We obtain this solution via code transformation from Algorithm~\ref{alg:intermediateMMR}. The latter transformation aims to offer recovery from transient-faults. 

Note that a transient fault can corrupt the state of node $p_i \in \sP$ by, for example, setting $est_i[i]$ with $\{0,1\}$. Line~\ref{ln:fixAw} addresses this concern. Another case of state corruption is when the round counter, $r_i$, equals to $r$, but there is $r'<r$ and entries $est_i[r']$ or $aux_i[r']$ that point to their initial values \ie $\exists r' \in \{1,\ldots,r\mathit{-}1\}: est_i[r'][i]=\emptyset \lor aux_i[r'][i]=\bot$. Line~\ref{ln:estFix} addresses this concern. Since we wish not that the for-each condition in line~\ref{ln:estFixIF} to hold when a correct node decides, line~\ref{ln:NdecideX} makes sure that all entries of $est[r']$ and $aux[r']$ store the decided value, where $r'$ is any round number that is between the current round number, $r$, and $M\mathit{+}1$, which is the entry that stores the decided value.

The last concern that Algorithm~\ref{alg:consensus} needs to address is the fact that the repeat-until loop in lines~\ref{ln:NESTrepeatCstart} to~\ref{ln:NESTrepeatC} of Algorithm~\ref{alg:intermediateMMR} depends on the assumption that $aux_i[r][i] \neq \bot$, which is supposed to be fulfilled by the repeat-until loop in lines~\ref{ln:NESTrepeatStart} to~\ref{ln:NESTrepeat} of Algorithm~\ref{alg:intermediateMMR}. However, a transient fault can place the program counter to point at line~\ref{ln:NESTrepeatD} without ever satisfying the requirement of $aux_i[r][i] \neq \bot$. Therefore, Algorithm~\ref{alg:consensus} combines in lines~\ref{ln:ESTrepeatStart} to~\ref{ln:ESTrepeat} the repeat-until loops of lines~\ref{ln:NESTrepeatStart} to~\ref{ln:NESTrepeat} and~\ref{ln:NESTrepeatCstart} to~\ref{ln:NESTrepeatC} of Algorithm~\ref{alg:intermediateMMR}. Similarly, it combines in lines~\ref{ln:uponEst} to~\ref{ln:estArrival} of the upon events in lines~\ref{ln:NuponEst} to
\ref{ln:NestArrival} and lines~\ref{ln:NuponAUX} to \ref{ln:NauxArrival} of Algorithm~\ref{alg:intermediateMMR}.

\Subsection{Extension: eventually silent self-stabilization Byzantine fault-tolerance}
\label{sec:exte}
Self-stabilizing systems can never stop the exchange of messages until the consensus object is deactivated, see~\cite[Chapter 2.3]{DBLP:books/mit/Dolev2000} for details. We say that a self-stabilizing system is \emph{eventually silent} if every legal execution has a suffix in which the same messages are repeatedly sent using the same communication pattern. We describe an extension to Algorithm~\ref{alg:consensus} that, once at least $t\mathit{+}1$ nodes have decided, lets all correct nodes decide and reach the $M$-th round quickly. Once the latter occurs, the system execution becomes silent. This property makes Algorithm~\ref{alg:consensus} a candidate for optimization, as described in~\cite{DBLP:journals/tpds/DolevS03}.  

The extension idea is to let node $p_i$ wait until at least $t\mathit{+}1$ nodes have decided. Once that happens, $p_i$ can notify all nodes about this decision because at least one of these $t\mathit{+}1$ nodes is correct. Algorithm~\ref{alg:consensus} (including the boxed code-lines) does this by setting the round number, $r$, to have the value of $M\mathit{+}1$ when deciding (line~\ref{ln:plusOneMM}) and allowing $r$ to have the value of up to $M\mathit{+}1$ (line~\ref{ln:plusOneM}). Also, line~\ref{ln:fPlusOneDecideThen} decides value $w$ whenever it sees that it was decided by $t\mathit{+}1$ other nodes, since at least one of the must be correct.

Since a transient fault can cause the nodes to exceed their storage limit, there is a need to indicate that to the invoking algorithm. Therefore, the self-stabilizing version of Algorithm~\ref{alg:consensus} uses the operation $\mathsf{result}()$ to return the transient error symbol, $\blitza$. In order to ensure that $\forall p_i \in \sP: i \in \Correct \land \mathsf{result}_i()=\blitza$, the self-stabilizing version of Algorithm~\ref{alg:consensus} runs a completion procedure, that is based on additional synchronization assumptions, which we define next. We clarify that these additional assumptions impact Algorithm~\ref{alg:consensus} only in the presence of transient-faults. Otherwise, the system is assumed to be asynchronous. 

We assume that in the presence of transient-faults, any pair $p_i,p_j \in \sP$ of correct nodes is able to complete at least one round-trip of messages exchange whenever $p_i$ is able to exchange at most $\theta$ round trips with all other correct nodes $p_k \in \sP \setminus \{p_i\}: k \in \Correct$. Note that a faulty node $p_{byz}\in \sP$ can attempt to complete rounds trips with $p_i$ much faster than any correct node $p_k$. For example, $p_{byz}$ can respond to any messages from $p_i$ with all the $\theta$ acknowledgments $p_i$ would need to receive for the perspective messages that it is going to send to $p_{byz}$. By flooding the network with responses, $p_{byz}$ creates scenarios in which $p_i$ believes that it has completed $\theta$ round trips without this ever occurring. For this reason, the proposed completion procedure counts the number of round trips each node, $p_k$, was able to complete with $p_i$ ever since $p_j$ has completed a round trip with $p_i$. Moreover, when summing up the number of these round-trips, $p_i$ ignores the $t$ `fastest' node since they might be faulty. 

By identifying the faulty nodes that are `too slow', \ie the ones that do not complete round trips with $p_i$ according to the above synchronization assumption, $p_i$ can safely avoid blocking when waiting for all trusted nodes to respond. Specifically, during the execution of the completion procedure, only dedicated messages, $\mathrm{tEST}(phs,ct,val)$ are to be used, where $phs$ is a phase number, $ct$ is a round-trip counter, and $val$ is the sender's latest estimated value. The procedure uses three phases. Each phase completes (and the next one starts) when $p_i$ receives an acknowledgment from al trusted nodes that they have entered this phase (or a higher one). This way, when $p_i$ phase number changes from zero to one, we know that all correct nodes were able to share the latest estimation value that they had before starting the completion procedure. Moreover, when $p_i$ phase number changes from one to two, we know that all correct nodes were able to share the estimated values that they received during the first phase. Furthermore, when $p_i$ phase number changes from two to three, we know that all correct nodes are aware that the correct nodes were able to exchange all of their estimated values and the procedure can terminate.

\Subsubsection{Constants and variables:} 
Node $p_i$ store the round-trip counters in the $ct[\false,\true][0,..,n\mathrm{-}1]$ array, where $ct_i[\false][j]$ and $ct_i[\true][j]$ store the sender-side, and resp., receiver-side counters of messages that $p_i$ and $p_j$ exchange. The $\mathit{rt}[0,..,n\mathrm{-}1][0,..,n\mathrm{-}1]$ array stores in $\mathit{rt}_i[j][k]$ the number of replies $p_i$ received from $p_k$ ever since $p_j$ has completed its last round-trip with $p_i$ (or since the procedure invocation). Both $ct[][]$ and $\mathit{rt}[][]$ holds integers of at most $B=4 (\theta+1)(n+1)$ states that are initialized with the zero value. The array $\mathit{phs}[0,..,n\mathrm{-}1]$ holds the phase numbers, where $\mathit{phs}_i[i]$ stores $p_i$'s phase number and $\mathit{phs}_i[j]$ stores the highest value received from $p_j$ ever since the invocation of the procedure.

%:=[\mathit{-}1,\ldots,\mathit{-}1]$\; 

%
%\Subsubsection{The self-stabilizing termination procedure:} 
%%
%\label{ss:termProc}

%		\EMS{OUR NEW STYLE} Self-stabilizing algorithm for 
%  with $t < n/3$, $\bigO(n^2)$ messages, and $\bigO(1)$ expected time
%\noindent \textbf{local variables:}\\
%$B:4 (\theta+1)(n+1)$\;w$.
%$r:=0$ \tcc*{the asynchronous round counter}\label{ln:kZero}
%$est[0,..,M\mathrm{+}1][0,..,n\mathrm{-}1]:=[[\emptyset, \ldots ,\emptyset],\ldots]$ \tcc*{array of estimated decision}\label{ln:proposalsBotsV}
%$aux[0,..,M][0,..,n\mathrm{-}1]:=[[\bot, \ldots ,\bot],\ldots]$ \tcc*{array of exchanged estimated decisions\label{ln:proposalsBotsV}}
%$ct[0,..,1][0,..,n\mathrm{-}1]$\;
%$\mathit{rt}[0,..,n\mathrm{-}1][0,..,n\mathrm{-}1]$\;
%$\mathit{phs}[0,..,n\mathrm{-}1]:=[\mathit{-}1,\ldots,\mathit{-}1]$\;

\renewcommand{\algorithmcfname}{Part A of Algorithm}

\begin{algorithm*}[t!]
\begin{\algSize}
%		\smallskip
%		
%		\textbf{constants:} 
%		$M$ a predefined value the bound the number of iterations\;
%		
%		\noindent \textbf{local variables:}\\
%		$r:=0$ \tcc*{the asynchronous round counter}\label{ln:kZero}
%		$est[0,..,M\mathrm{+}1][0,..,n\mathrm{-}1]:=[[\emptyset, \ldots ,\emptyset],\ldots]$ \tcc*{array of estimated decision}\label{ln:proposalsBotsV}
%		$aux[0,..,M][0,..,n\mathrm{-}1]:=[[\bot, \ldots ,\bot],\ldots]$ \tcc*{array of exchanged estimated decisions\label{ln:proposalsBotsV}}
%		$ct[0,..,1][0,..,n\mathrm{-}1]$\;
%		$\mathit{rt}[0,..,n\mathrm{-}1][0,..,n\mathrm{-}1]$\;
%		$\mathit{phs}[0,..,n\mathrm{-}1]:=[\mathit{-}1,\ldots,\mathit{-}1]$\;

\medskip

\textbf{constants:} 
$\mathit{initState} := (0,[[\emptyset, \ldots ,\emptyset],\ldots,[\emptyset, \ldots ,\emptyset]],[[\bot,\ldots, \bot],\dots,[\bot,\ldots, \bot]])$\;

\medskip

\textbf{operations:} $\mathsf{propose}(v)$ \label{ln:SproposeV} 
\Begin{
$(r, est, aux) \gets \mathit{initState};est[0][i] \gets\{v\}$\;
}

\smallskip

$\mathsf{result}()$ \label{ln:resultV}
\Begin{
 \lIf{$(est[M\mathit{+}1][i] = \{v\})$}{\Return{$v$}} 
 \lElseIf{$(r \geq M\land \mathsf{infoResult}() \neq \emptyset)$}{\Return{$\blitza$}}
 \lElse{\Return{$\bot$}}
}

\medskip

%		$\init()$ \textbf{do} $(r, est, aux) \gets (0,[[\emptyset, \ldots ,\emptyset],\ldots,[\emptyset, \ldots ,\emptyset]],[[\bot,\ldots, \bot],\dots,[\bot,\ldots, \bot]])$\;
%		
%					\smallskip

%		\textbf{macros:} $\mathit{binValues}(r,x)$ \textbf{return} {$\{y \in \{0,1\}:\exists s\subseteq \sP:|s|=x \land y\in  \cup_{p_j \in s:\mathit{est}[r][j]\neq \emptyset} \mathit{est}[r][j] \}$}\label{ln:binVal}\;

\textbf{macros:} $\mathit{binValues}(r,x)$ \Begin{\Return{$\{y \in \{0,1\}:\exists s\subseteq \sP:|\{p_j \in s: y \in \mathit{est}[r][j]\}| \geq x \}$}\label{ln:binVal}} 

\smallskip

$\mathsf{infoResult}()$\label{ln:inforesult} \Begin{ 
		\lIf{$(\exists s \subseteq \sP:n\mathrm{-}t \leq |s| \land( \forall p_j \in s :\mathit{aux}[r][j] \in$ $  \mathit{binValues}(r,2t\mathrm{+}1)))$} {\Return{$\{\mathit{aux}[r][j] \}_{p_j \in s}$}} 
		\lElse{\Return{$\emptyset$};}}

\medskip

\textbf{functions:}
$\mathsf{decide}(x)$ \Begin{
\ForEach{$r'\in \{r,\ldots,M\mathit{+}1\}$}{
\If{$(est[r'][i] = \emptyset \lor aux[r'][i] = \bot)$}{$(est[r'][i],aux[r'][i]) \gets (\{x\},x)$\label{ln:NdecideX}}
}
\fbox{$r \gets M\mathrm{+}1$\label{ln:plusOneMM};}
}

\smallskip

$\mathsf{tryToDecide}(values)$\label{ln:tryToDecide}  \Begin{\lIf{\label{ln:cupEq}$(values \neq \{v\})$}{$est[r][i] \gets \{\mathbf{randomBit}(r)\}$\label{ln:estRiGetsRandom}}
\lElse{\{$est[r][i] \gets \{v\}$\label{ln:estRiGetsV}; \textbf{if} {$(v = \mathbf{randomBit}(r))$} \textbf{then} {\label{ln:cupEqMore}$\mathsf{decide}(v)$}\}}}

\medskip

\caption{\label{alg:Rconsensus}\emsB{SSBFT MMR,} code for $p_i$.}
\end{\algSize}
\end{algorithm*}

\setcounter{algocf}{2}

\renewcommand{\algorithmcfname}{Part B of Algorithm}

\begin{algorithm*}[t!]
\begin{\algSize}

\smallskip

\textbf{do forever} \Begin{
\If{{$((r, est, aux) \neq {\mathit{initState}})$}\label{ln:isActive}}{

$r \gets \min \{r\mathrm{+}1,M$\fbox{$\mathrm{+}1$}$\}$\label{ln:plusOneM};

\Repeat{{$\mathsf{infoResult}()\neq \emptyset$}\label{ln:ESTrepeat}}{
	\label{ln:ESTrepeatStart}
	
	\lIf{$(est[0][i] \neq \{v\})$}{$est[0][i] \gets \{w\}:\exists w \in est[0][i]$\label{ln:fixAw}}
	\ForEach{{$r' \in \{1,\ldots,r\mathit{-}1\}: est[r'][i]=\emptyset \lor aux[r'][i]=\bot$\label{ln:estFixIF}}}{{$(est[r'][i],aux[r'][i])\gets (est[0][i],x):x \in est[0][i]$;\label{ln:estFix}}}

	\If{$((\exists w \in \mathit{binValues}(r,2t\mathrm{+}1)\land (aux[r][i] = \bot\lor aux[r][i] \notin \mathit{binValues}(r,2t\mathrm{+}1)))$\label{ln:auxIf}}{$aux[r][i]\gets w$;\label{ln:auxThen}}
	
	\lForEach{$p_j \in \sP$}{\textbf{send} $\mathrm{EST}(\true,r,$ $est[r\mathrm{-}1][i]\cup\mathit{binValues}(r,t\mathrm{+}1),aux[r][i])$  \textbf{to} $p_j$\label{ln:ESTrepeatB}}}

{$\mathsf{tryToDecide}(\mathsf{infoResult}())$}\label{ln:rM}\;

\fbox{\lIf{$(\exists w \in \mathit{binValues}(M\mathit{+}1,t\mathrm{+}1))$\label{ln:fPlusOneDecideIf}}{$\mathsf{decide}(w)$\label{ln:fPlusOneDecideThen}}}

}

%				{{\lElse{$(r,est,aux) \gets (0,[[\emptyset, \ldots ,\emptyset],\ldots],[[\bot, \ldots ,\bot],\ldots])$}}\label{ln:ElseNot Active}}
}

\smallskip

\textbf{upon} $\mathrm{EST}(\mathit{aJ},\mathit{rJ},\mathit{vJ},\mathit{uJ})$\label{ln:uponEst} \textbf{arrival from} $p_j$ \Begin{

$est[\mathit{rJ}][j]\gets est[\mathit{rJ}][j] \cup\mathit{vJ}; aux[\mathit{rJ}][j] \gets \mathit{uJ}$\label{ln:auxUpdate}\; 

%				$est[M\mathrm{+}1][j] \gets \mathit{wJ}$\;

\lIf{$\mathit{aJ}$}{\textbf{send} $\mathrm{EST}(\false,\mathit{rJ},est[\mathit{rJ}\mathit{-}1][i],aux[r][i])$ \textbf{to} $p_j$\label{ln:estArrival}}

}

\smallskip

%\fbox{at most}

\caption{\label{alg:consensus}\emsB{SSBFT MMR,} code for $p_i$.}
\end{\algSize}
\end{algorithm*}

\renewcommand{\algorithmcfname}{Part A of Algorithm}
\begin{algorithm*}[t!]
\begin{\algSize}
%		\smallskip
%		
%		\textbf{constants:} 
%		$M$ a predefined value the bound the number of iterations\;
%		
%		\noindent \textbf{local variables:}\\
%		$r:=0$ \tcc*{the asynchronous round counter}\label{ln:kZero}
%		$est[0,..,M\mathrm{+}1][0,..,n\mathrm{-}1]:=[[\emptyset, \ldots ,\emptyset],\ldots]$ \tcc*{array of estimated decision}\label{ln:proposalsBotsV}
%		$aux[0,..,M][0,..,n\mathrm{-}1]:=[[\bot, \ldots ,\bot],\ldots]$ \tcc*{array of exchanged estimated decisions\label{ln:proposalsBotsV}}
%		$ct[0,..,1][0,..,n\mathrm{-}1]$\;
%		$\mathit{rt}[0,..,n\mathrm{-}1][0,..,n\mathrm{-}1]$\;
%		$\mathit{phs}[0,..,n\mathrm{-}1]:=[\mathit{-}1,\ldots,\mathit{-}1]$\;

%		\fbox{$delivered[\sP]:=[\false,\ldots,\false]$ delivery indications, where $delivered[i]$ stores the local} 
%		
%		\fbox{indication and $delivered[j]$ stores the indication that was last recived from $p_j \in \sP$\;} 

\smallskip

\textbf{constants:} 

$\mathit{initState} := (0,[[\emptyset, \ldots ,\emptyset],\ldots,[\emptyset, \ldots ,\emptyset]],[[\bot,\ldots, \bot],\dots,[\bot,\ldots, \bot]],$ \fbox{$[\false,\ldots,\false])$\;}

\smallskip

\textbf{provided interfaces:}  	
\fbox{$\mathsf{wasDelivered}()$ \textbf{do} \{\textbf{if} {$\exists S \subseteq\sP:$ $n\mathit{-}t\leq |S|:$ $\forall {p_k \in S} : \exists r'\in \{0,$}} \fbox{$\ldots, r\}:delivered[k]=\true$ \textbf{then} \textbf{return} $1$ \textbf{else return} $0$;\}\label{ln:wasDelivered}}

\fbox{\label{ln:recycle}$\mathsf{recycle}()$ \textbf{do} $(r, est, aux,delivered) \gets \mathit{initState}$\;}

\smallskip

\textbf{operations:} $\mathsf{propose}(v)$ \label{ln:RSproposeV}\textbf{do} \fbox{$\{\mathsf{recycle}();$} $est[0][i] \gets\{v\}\}$\;

\smallskip

$\mathsf{result}()$ \label{ln:RresultV}\textbf{do} \Begin{
\lIf{$(est[M\mathit{+}1][i] = \{v\})$}{\fbox{\{$delivered[i]\gets \true$;} \Return{$v$}\label{ln:nonBot1}\}} 
\lElseIf{$(r \geq M\land \mathsf{infoResult}() \neq \emptyset)$}{\fbox{\{$delivered[i]\gets \true$;} \Return{$\blitza$}\label{ln:nonBot2}\}}
\textbf{else return} {$\bot$};
}

\smallskip

%		$\init()$ \textbf{do} $(r, est, aux) \gets (0,[[\emptyset, \ldots ,\emptyset],\ldots,[\emptyset, \ldots ,\emptyset]],[[\bot,\ldots, \bot],\dots,[\bot,\ldots, \bot]])$\;
%		
%					\smallskip

%		\textbf{macros:} $\mathit{binValues}(r,x)$ \textbf{return} {$\{y \in \{0,1\}:\exists s\subseteq \sP:|s|=x \land y\in  \cup_{p_j \in s:\mathit{est}[r][j]\neq \emptyset} \mathit{est}[r][j] \}$}\label{ln:binVal}\;

\textbf{macros:} $\mathit{binValues}(r,x)$ \textbf{return} {$\{y \in \{0,1\}:\exists s\subseteq \sP:|\{p_j \in s: y \in \mathit{est}[r][j]\}| \geq x \}$}\label{ln:RbinVal}\; 

$\mathsf{infoResult}()$\label{ln:Rinforesult} \Begin{
	\If{$(\exists s \subseteq \sP:n\mathrm{-}t \leq |s| \land ( \forall p_j \in s :\mathit{aux}[r][j] \in $ $ \mathit{binValues}(r,2t\mathrm{+}1)))$}{\Return{$\{\mathit{aux}[r][j] \}_{p_j \in s}$} \textbf{else} \Return{$\emptyset$};\}}}

\smallskip

\textbf{functions:}
$\mathsf{decide}(x)$ \Begin{
\ForEach{$r'\in \{r,\ldots,M\mathit{+}1\}$}{
\lIf{$(est[r'][i] = \emptyset \lor aux[r'][i] = \bot)$}{$(est[r'][i],aux[r'][i]) \gets (\{x\},x)$\label{ln:RNdecideX}}
}
}

$\mathsf{tryToDecide}(values)$\label{ln:RtryToDecide}  \Begin{\lIf{\label{ln:RcupEq}$(values \neq \{v\})$}{$est[r][i] \gets \{\mathbf{randomBit}(r)\}$\label{ln:RestRiGetsRandom}}
\lElse{\{$est[r][i] \gets \{v\}$\label{ln:RestRiGetsV}; \textbf{if} {$(v = \mathbf{randomBit}(r))$} \textbf{then} {\label{ln:RcupEqMore}$\mathsf{decide}(v)$}\}}}

\smallskip

\caption{\label{alg:RconsensusA}\ems{A recyclable variation of Algorithm~\ref{alg:consensus}; code for $p_i$.}}
\end{\algSize}
\end{algorithm*}

\setcounter{algocf}{3}
\renewcommand{\algorithmcfname}{Part B of Algorithm}

\begin{algorithm*}[t!]
\begin{\algSize}

\smallskip

\textbf{do forever} \Begin{
\If{{$((r, est, aux) \neq {\mathit{initState}})$}\label{ln:RisActive}}{

$r \gets \min \{r\mathrm{+}1,M\}$\label{ln:RplusOneM};

\Repeat{{$\mathsf{infoResult}()\neq \emptyset$}\label{ln:RESTrepeat}}{
	\label{ln:RESTrepeatStart}
	
	\lIf{$(est[0][i] \neq \{v\})$}{$est[0][i] \gets \{w\}:\exists w \in est[0][i]$\label{ln:RfixAw}}
	\ForEach{{$r' \in \{1,\ldots,r\mathit{-}1\}: est[r'][i]=\emptyset \lor aux[r'][i]=\bot$\label{ln:RestFixIF}}}{{$(est[r'][i],aux[r'][i])\gets (est[0][i],x):x \in est[0][i]$;\label{ln:RestFix}}}

	\If{$((\exists w \in \mathit{binValues}(r,2t\mathrm{+}1)\land (aux[r][i] = \bot\lor aux[r][i] \notin \mathit{binValues}(r,2t\mathrm{+}1)))$\label{ln:RauxIf}}{$aux[r][i]\gets w$;\label{ln:RauxThen}}
	
	%					\textbf{foreach} $p_j \in \sP$ \textbf{do} \textbf{send} $\mathrm{EST}(\true,r,est[r\mathrm{-}1][i]\cup\mathit{binValues}(r,t\mathrm{+}1),aux[r][i],$\fbox{$delivered[i])$}  \textbf{to} $p_j$\label{ln:RESTrepeatB}}

\ForEach{$p_j \in \sP$}{\textbf{send} $\mathrm{EST}(\true,r,est[r\mathrm{-}1][i]\cup\mathit{binValues}(r,t\mathrm{+}1),aux[r][i],$\fbox{$delivered[i])$}  \textbf{to} $p_j$\label{ln:RESTrepeatB}}}

{$\mathsf{tryToDecide}(\mathsf{infoResult}())$}\label{ln:RrM}\;

}

%				{{\lElse{$(r,est,aux) \gets (0,[[\emptyset, \ldots ,\emptyset],\ldots],[[\bot, \ldots ,\bot],\ldots])$}}\label{ln:ElseNot Active}}
}

%		\smallskip

\textbf{upon} $\mathrm{EST}(\mathit{aJ},\mathit{rJ},\mathit{vJ},\mathit{uJ},$\fbox{$delivered\mathit{J})$} \label{ln:RuponEst} \textbf{arrival from} $p_j$ \Begin{ 

$delivered[i]\gets delivered[i] \lor delivered\mathit{J}$\label{ln:RRESTrepeatB}\;

$est[\mathit{rJ}][j]\gets est[\mathit{rJ}][j] \cup\mathit{vJ}; aux[\mathit{rJ}][j] \gets \mathit{uJ}$\label{ln:RauxUpdate}\; 

%				$est[M\mathrm{+}1][j] \gets \mathit{wJ}$\;

\lIf{$\mathit{aJ}$}{\textbf{send} $\mathrm{EST}(\false,\mathit{rJ},est[\mathit{rJ}\mathit{-}1][i],aux[r][i])$ \textbf{to} $p_j$\label{ln:RestArrival}}

}

\smallskip

%\fbox{at most}

\caption{\label{alg:RconsensusB}\ems{A recyclable variation of Algorithm~\ref{alg:consensus}; code for $p_i$.}}
\end{\algSize}
\end{algorithm*}

\renewcommand{\algorithmcfname}{Algorithm}

\Subsection{A recyclable variation on Algorithm~\ref{alg:consensus}}
\label{sec:recyclableVar}
%
%\CGC{There seems to be an issue with the Algorithm names (regardless of parts A or B). In my understanding, after Algorithm 2, we should give the self-stab version, which should be called Algorithm 3. Then, Algorithm 4 should be the recycable variation of Algorithm 3. I think what is now called Part A of Algorithm 3 is correct, but the caption is incorrect. Then, the one that now is called Part B of Algorithm 4, should be Part B of Algorithm 3. Then, the one called Part A of Algorithm 5 should be Part A of Algorithm 4, which is the recyclable version of Algorithm 3. And Part B of Algorithm 6 should be Part B of Algorithm 4. Algorithm 7 should be called Algorithm 5 and so one. Unless I am missing something?}
%
\ems{Algorithm~\ref{alg:RconsensusA} presents a recyclable variation on Algorithm~\ref{alg:consensus} that is needed for allowing the system to sequentially instantiate and recycle an unbounded number of Algorithm~\ref{alg:consensus}'s objects using an SSBFT recycling mechanism, which we propose in Section~\ref{sec:bck}. The \fbox{boxed} code lines highlight the modified code lines with respect to the code of Algorithm~\ref{alg:consensus}. Also, as before, the line numbers of the latter continue the one of the former. We clarify that the correctness proof (Section~\ref{sec:correctness}) focuses on Algorithm~\ref{alg:consensus} rather than the straightforward added details of Algorithm~\ref{alg:RconsensusA}.}   

\ems{Algorithm~\ref{alg:RconsensusA} uses the array $delivered[\sP]$ (initialized to $[\false,\ldots,\false]$) for delivery indications, where $delivered_i[i]:p_i \in \sP$ stores the local indication and $delivered_i[j]:p_i,p_j \in \sP$ stores the indication that was last received from $p_j$. This indication is set to $\true$ whenever a non-$\bot$ value is returned by $\done()$, see lines~\ref{ln:nonBot1}	and~\ref{ln:nonBot2}. Algorithm~\ref{alg:RconsensusA} updates $delivered[j]$ according to the arriving values from $p_j$ (lines~\ref{ln:RESTrepeatB} and~\ref{ln:RRESTrepeatB}). The interface function $\mathsf{wasDelivered}()$ (line~\ref{ln:wasDelivered}) returns $1$ whenever there is a set of at least $n-t$ entries with the value $\true$ in $delivered[]$. The interface function $\mathsf{recycle}()$ (line~\ref{ln:recycle}) allows the note to restart its local state w.r.t. Algorithm~\ref{alg:RconsensusA}.}

\ems{The approach studied here considers the instantiation of one object at a time. A straightforward extension is to allow the allocation and recycling of a set of objects. Specifically, one can run $\delta$ concurrent MMR instances, where $\delta$ is a parameter for balancing the trade-off between fault recovery time and the number of MMR instances that can be used (before the next $\delta$ concurrent instances can start).} 

\Subsection{Correctness}
\label{sec:correctness}
The correctness proof shows that the solution presented in Section~\ref{sec:theAlgo} recovers from transient-faults without blocking (Section~\ref{sec:revo}) and that any consensus operation always satisfies the liveness requirements of Definition~\ref{def:consensus} (Section~\ref{sec:meeting}). Also, it satisfies the safety requirements of Definition~\ref{def:consensus} in the way that loosely-self-stabilizing systems do (Section~\ref{sec:meeting}), \ie any consensus operation satisfies the  requirements of  Definition~\ref{def:consensus} with  probability $\Pr(r) = 1- (1/2)^{-M}$.

\Subsubsection{Transient fault recovery}
\label{sec:revo}
We say that a system state $c$ is \emph{resolved} if $\forall  i \in \Correct:  \big|est_i[0][i]\big| \in \{0,1\} \land \nexists r' \in \{1,\ldots,r\mathrm{-}1\}: est_i[r'][i]=\emptyset \lor aux_i[r'][i]=\bot$ and no {communication channel that goes out from $p_i \in \sP:i \in \Correct$ to any other correct node includes $\mathrm{EST}(rnd=r,est=W,aux=w): r_i < r \lor  W \nsubseteq est_i[r][i] \lor (w \neq \bot \land w \notin W)$} messages. Suppose that during execution $R$, every correct node $p_i \in \sP$ invokes $\mathsf{propose}_i()$ exactly once. In this case, we say that $R$ includes a \emph{complete invocation} of binary consensus. Theorem~\ref{thm:recovery} shows recovery to resolved system states and termination during executions that include a complete invocation of binary consensus. The statement of Theorem~\ref{thm:recovery} uses the term \emph{active} for node $p_i \in \sP$ when referring to the case of $est_i[0][i]\neq \mathit{initState}$.

\begin{theorem}[Convergence]
\label{thm:recovery}
Let $R$ be an execution of Algorithm~\ref{alg:consensus}. (i) Within one complete asynchronous (communication) round, the system reaches a resolved state. Moreover, suppose that throughout $R$ all correct nodes are active. (ii) Within $\bigO(M)$  asynchronous (communication) rounds, for every correct node $p_i \in P$, it holds that the operation $\mathsf{result}_i()$ returns $v \in \{0,1,\blitza\}$, where $\blitza$ is the transient error symbol.
\end{theorem}
\renewcommand{\thmcnt}{\ref{thm:recovery}}
\begin{theoremProof}
Lemmas~\ref{thm:recoveryI} and~\ref{thm:recoveryII} demonstrate the theorem.   
\begin{lemma}
\label{thm:recoveryI}
Invariant (i) holds.
\end{lemma}
\renewcommand{\lemcnt}{\ref{thm:recoveryI}}
\begin{lemmaProof}
{Let $m$ be a message that in $R$'s starting system state resides in the communication channels between any pair of correct nodes. By Remark~\ref{ss:first asynchronous cycles}, within $\bigO(1)$ asynchronous rounds, the system reaches a state in which $m$ does not appear.} Let us look at $p_i$'s first complete iteration of the do-forever loop (lines~\ref{ln:isActive} to~\ref{ln:rM}) {after $m$ has left the system. Once that happens, for any message $\mathrm{EST}(rnd=r,est=W,aux=w)$ that appears in any communication channel that is going out from $p_i \in \sP: i \in  \Correct$, it holds that $r \leq r_i \land  W \subseteq est_i[r][i] \land (w = \bot \lor w \in W)$ (due to lines~\ref{ln:SproposeV} and~\ref{ln:ESTrepeatB}).}

Let $I_{i,r}$ be $p_i$'s first complete iteration in the first complete asynchronous (communication) round of $R$. Suppose that in the iteration's first system state, it holds that $\forall i \in  \Correct:(r_i, est_i, aux_i) = \mathit{initState}$. In this case, Invariant (i) holds by definition. In case {$\forall i \in  \Correct:(r_i, est_i, aux_i) = \mathit{initState}$ does not hold,} lines~\ref{ln:fixAw} to~\ref{ln:estFix} imply that Invariant (i) holds.
Invariant (i) also holds when the round number $r$ is incremented. Note that regardless of which branch of the if-statement in line~\ref{ln:cupEq} node $p_i$ follows, $est_i[r][i]$ is always assigned a value that is not the empty set at the end of round $r$, cf. lines~\ref{ln:estRiGetsRandom} and~\ref{ln:estRiGetsV}. Moreover, the assignment of $w$ to $aux_i[r][i]$ in line~\ref{ln:auxThen} is always of a value that is not the empty set  due to the if-statement condition in line~\ref{ln:auxIf} and the definition of $\mathit{binValues}()$ (line~\ref{ln:binVal}).
\end{lemmaProof}    

\medskip
Lemma~\ref{thm:repeatDoesNotBlock} is needed for the proof of Lemma~\ref{thm:recoveryII}.

\begin{lemma}
\label{thm:repeatDoesNotBlock}
Suppose that $R$'s states are resolved (Lemma~\ref{thm:recoveryI}). The repeat-until loop in lines~\ref{ln:auxIf} to \ref{ln:ESTrepeat} cannot block forever.	
\end{lemma}	
\renewcommand{\lemcnt}{\ref{thm:repeatDoesNotBlock}}
\begin{lemmaProof}
The proof is by contradiction; to prove the lemma to be true, we begin by assuming it is false and show that this leads to a contradiction, which implies that the lemma holds. Argument 5 shows the needed contradiction and it uses arguments 1 to 4. 

\noindent \textbf{Argument 1:} \emph{Eventually $aux_i[r][i] \in\mathit{binValues}_i(r,2t\mathrm{+}1)$ holds.~~} 
Suppose that in $R$'s starting state, $ (aux[r][i] \neq \bot\land aux[r][i] \in \mathit{binValues}(r,2t\mathrm{+}1))$ does not hold, because otherwise the proof of the argument is done.
There are at least $n\mathrm{-}t\geq2t\mathrm{+}1 = (t\mathrm{+}1) \mathrm{+}t$ correct nodes and each of them sends $\mathrm{EST}(\bullet,rnd=r,est=\{w,\bullet\},\bullet):w\in \{0,1\}$ messages to all nodes (line~\ref{ln:ESTrepeatB}). Therefore, we know that there is $v \in \{0, 1\}$, such that at least $(t\mathrm{+}1)$ correct nodes send $\mathrm{EST}(\bullet,rnd=r,est=\{v,\bullet\},\bullet)$ messages to all other nodes.

Since every correct node receives {$\mathrm{EST}(\bullet,rnd=r,est=\{v,\bullet\},\bullet)$} from at least $(t\mathrm{+}1)$ nodes (line~\ref{ln:estArrival}), we know that eventually every correct node relays the value $v$ via the message $\mathrm{EST}(\bullet,rnd=r,est=\{v,\bullet\},\bullet)$ that line~\ref{ln:ESTrepeatB} sends due to the fact that $v \in \mathit{binValues}_i(r,t\mathrm{+}1)$. 

Since $n\mathrm{-} t\geq 2t\mathrm{+}1$ holds, the clause $(\exists w \in \mathit{binValues}(r,2t\mathrm{+}1))$ in the if-statement condition at line~\ref{ln:auxIf} is eventually satisfied at each correct node $p_i \in \sP$. Thus, if $(aux[r][i] = \bot\lor aux[r][i] \notin \mathit{binValues}(r,2t\mathrm{+}1))$ does not hold, line~\ref{ln:auxThen} makes sure it does.

\noindent \textbf{Argument 2:} \emph{Eventually the system reaches a state in which $\exists i \in \Correct: w \in \mathit{binValues}_i(r_i,2t\mathrm{+}1)\implies \exists s \subseteq \Correct:t\mathit{+}1\leq |s|\land \forall k \in s:w \in est_k[k]$.~~}

We prove the argument by contradiction;  
%\EMS{Note that there are two levels of proof by contradiction --- one at the lemma level and one by the argument level}\cgr{Yes, I realized that. I just feel that people know what a proof by contradiction is and I was trying to remove some text repetition. Perhaps, we can say: We prove the argument by contradiction. Speficifally,... I think this will make it clear.} \EMS{Please put it back to the way it was before.} To prove the argument to be true, 
we begin by assuming the argument is false and show that this leads to a contradiction, which implies that the argument holds. Specifically, suppose that  $\exists i \in \Correct: w \in \mathit{binValues}_i(r_i,2t\mathrm{+}1)$ holds in every system state in $R$ and yet $\forall s \subseteq \Correct:t\mathit{+}1\leq |s|$, it is true that $\exists k \in s:w \notin est_k[k]$.

By lines~\ref{ln:ESTrepeatB} and~\ref{ln:auxUpdate}, the only way in which $w \in \mathit{binValues}_i(r_i,2t\mathrm{+}1)$ hold in every system state $c' \in R$, is if there is a system state $c$ that appears in $R$ before $c'$, such that $\exists s \subseteq \Correct:t\mathit{+}1\leq |s|:\forall k \in s:w \in est_k[k]$. Thus, a contradiction is reached (with respect to the assumption made at the start of this argument's proof), which implies that the argument is true.

\noindent \textbf{Argument 3:} \emph{Eventually the system reaches a state $c'\in R$ in which $\exists s \subseteq \Correct:t\mathit{+}1\leq |s|\land \forall p_k \in s:w \in est_k[k] \implies \forall i \in \Correct: w \in \mathit{binValues}_i(r_i,2t\mathrm{+}1)$.~~}

By line~\ref{ln:ESTrepeatB} and the argument's assumption, there are at least $(t\mathrm{+}1)$ correct nodes that send $\mathrm{EST}(\bullet,rnd=r,est=\{w,\bullet\},\bullet)$ messages to all (correct) nodes. Since every correct node receives $w$ from at least $(t\mathrm{+}1)$ nodes (line~\ref{ln:auxUpdate}), every correct node eventually reply $w$ via the message $\mathrm{EST}(\bullet,rnd=r,est=\{w,\bullet\},\bullet)$ at lines~\ref{ln:ESTrepeatB} and~\ref{ln:estArrival} due to the fact that $w \in \mathit{binValues}_i(r,t\mathrm{+}1)$. Since $n\mathrm{-} t\geq 2t\mathrm{+}1$ holds, we know that $(\exists w \in \mathit{binValues}(r,2t\mathrm{+}1))$ holds and the argument is true.

\noindent \textbf{Argument 4:} \emph{Suppose that the condition \remove{$cond(i):=\exists s \subseteq \sP:n\mathrm{-}t \leq |s| \land  \cup_{p_k \in s}\{\mathit{aux}_i[r][k]\} \subseteq \mathit{binValues}_i(r_i,2t\mathrm{+}1))$}{$cond(i):=\mathsf{infoResult}_i()\neq \emptyset: i \in \Correct$} does not hold in $R$'s starting system state. Eventually, the system reaches a state $c'' \in R$, in which $cond(i) :i \in \Correct$ holds.~~}

We prove the argument by contradiction; we begin by assuming the argument is false and show that this leads to a contradiction, which implies that the argument holds. Specifically, suppose that $cond(i)$ never holds, \ie $c'' \in R$ does not exist. We note that $cond(i)$ must hold if $\mathit{binValues}_i(r_i,2t\mathrm{+}1)=\{0,1\}$. The same can be said for the case of $\mathit{binValues}_i(r_i,2t\mathrm{+}1)=\{v\} \land \exists s \subseteq \sP:n\mathrm{-}t \leq |s| \land (\cup_{p_k \in s}\{\mathit{aux}_i[r][k]\}) = \{w\} \land w= v$. Therefore, we assume that, for any system state, it holds that $\mathit{binValues}_i(r_i,2t\mathrm{+}1)=\{v\}{\subsetneq}\{0,1\}$ and $\forall s \subseteq \sP:n\mathrm{-}t \leq |s| \implies w \in (\cup_{p_k \in s}\{\mathit{aux}_i[r][k]\}):w\neq v$. We demonstrate a contradiction by showing that eventually $w \in \mathit{binValues}_i(r_i,2t\mathrm{+}1)$.

By lines~\ref{ln:ESTrepeatB} and~\ref{ln:auxUpdate}, the only way in which $w \in (\cup_{p_k \in s}\{\mathit{aux}_i[r][k]\})$ holds in every system state $c' \in R$, is if there is a system state $c$ that appears in $R$ before $c'$, such that $\exists p_k \in \sP: \mathit{aux}_k[r][k]=w$. Note that $c'$ and $c$ can be selected such that the following sequence of statements are true. By Argument 1, $aux_k[r][k] \in\mathit{binValues}_k(r,2t\mathrm{+}1)$. By Argument 2, $w \in \mathit{binValues}_k(r_i,2t\mathrm{+}1)\implies \exists s \subseteq \Correct:t\mathit{+}1\leq |s|\land \forall p_k \in s:w \in est_k[k]$ in $c$. By Argument 3, $\exists s \subseteq \Correct:t\mathit{+}1\leq |s|\land \forall k \in s:w \in est_k[k] \implies \forall i \in \Correct: w \in \mathit{binValues}_i(r_i,2t\mathrm{+}1)$ in $c$. Thus, a contradiction is reached (with respect to the assumption made at the start of this argument's proof), which implies that the argument is true.

\noindent \textbf{Argument 5:} \emph{The lemma is true.~~}
Argument 4 implies that a contradiction (with respect to the assumption made in the start of this lemma's proof) was reached since the exist condition in line~\ref{ln:ESTrepeat} eventually holds.
\end{lemmaProof}

\begin{lemma}
\label{thm:recoveryII}
Invariant (ii) holds.
\end{lemma}
\renewcommand{\lemcnt}{\ref{thm:recoveryII}}
\begin{lemmaProof}
Lemma~\ref{thm:recoveryI} shows that $R$'s system states are resolved. Lemma~\ref{thm:repeatDoesNotBlock} says that the repeat-until loop in lines~\ref{ln:auxIf} to \ref{ln:ESTrepeat} does not block. {By line~\ref{ln:plusOneM} and the definition of an asynchronous (communication) round (Section~\ref{sec:asynchronousRounds}), every iteration of the do-forever loop (lines~\ref{ln:isActive} to~\ref{ln:rM}) can be associated with at most one asynchronous (communication) round. Thus, line~\ref{ln:resultV} and Argument (4) of the proof of Lemma~\ref{thm:repeatDoesNotBlock} imply that $(r_i \geq M\land \mathsf{infoResult}_i() \neq \emptyset)$ holds within $\bigO(M)$ asynchronous (communication) rounds. Therefore,} $\mathsf{result}_i()$ returns a non-$\bot$ value {within $\bigO(M)$ asynchronous (communication) rounds.}	
\end{lemmaProof}    
\end{theoremProof} 

\Subsubsection{\ems{Satisfying the task specifications}}
\label{sec:meeting}
We say that the system state $c$ is \emph{well-initialized} if $\forall  i \in \Correct: (r_i,est_i,aux_i) := \mathit{initState}$ holds and no communication channel between two correct nodes includes $\mathrm{EST}()$ messages. Note that a well-initialized system state is also a resolved one (Section~\ref{sec:revo}). Theorem~\ref{thm:terminationSafe} shows that Algorithm~\ref{alg:consensus} satisfies the requirements of Definition~\ref{def:consensus} during legal executions that start from a well-initialized system state and have a complete invocation of binary consensus. The proof of Theorem~\ref{thm:terminationSafe} uses Theorem~\ref{thm:BVtermination}, which demonstrates that Algorithm~\ref{alg:consensus} satisfies the requirements of Definition~\ref{def:bvb}, which adds more details to the one given in Section~\ref{sec:sefBVbrodcast}. Recall that the operation $\mathsf{bvBroadcast}(v)$ of Algorithm~\ref{alg:MMR} is embedded in the code of Algorithm~\ref{alg:consensus}.

\begin{definition}[BV-broadcast]
\label{def:bvb}
Let $p_i \in \sP$, {$r \in \{1,\ldots,M\}$,} and $v \in \{0,1\}$. Suppose that $r_i=r\land est_i[r\mathit{-}1][i]=\{v\}$ holds immediately before $p_i$ executes line~\ref{ln:ESTrepeatB}. In this case, we say that $p_i$ BV-broadcast $v$ during round $r$ in line~\ref{ln:ESTrepeatB}. Let $c \in R$ and suppose that $w \in \mathit{binValues}_i(r,t\mathrm{+}1)$ holds in $c$ (for the first time). In this case, we say that $p_i$ BV-delivers $w$ during round $r$.

\begin{itemize}
\item \textbf{BV-validity.} Suppose that $p_i$ is correct and $v \in \mathit{binValues}_i(r,t\mathrm{+}1)$ {holds} in system state $c \in R$. Then, before $c$ there is a step in $R$ in which a correct node BV-broadcast $v$.
\item \textbf{BV-uniformity.} Suppose that $p_i$ is correct and $v \in \mathit{binValues}_i(r,t\mathrm{+}1)$ {holds} in system state $c \in R$.  Then, eventually, the system reaches a state in which $\forall j \in \Correct: v \in \mathit{binValues}_j(r,t\mathrm{+}1)$ holds.
\item \textbf{BV-completion.} Eventually, the system reaches a state in which $\forall i \in \Correct: \mathit{binValues}_i(r,t\mathrm{+}1)\neq \emptyset$ holds.
\end{itemize}
\end{definition}

\begin{theorem}[BV-broadcast]
\label{thm:BVtermination}
Let $R$ be an execution of Algorithm~\ref{alg:consensus} that starts from a well-initialized system state and includes a complete invocation of binary consensus. Lines~\ref{ln:auxIf} to~\ref{ln:ESTrepeat} and lines~\ref{ln:uponEst} to~\ref{ln:estArrival} of Algorithm~\ref{alg:consensus} implement the BV-broadcast task (Definition~\ref{def:bvb}).
\end{theorem}
\renewcommand{\thmcnt}{\ref{thm:BVtermination}}
\begin{theoremProof} We prove that the {requirements of Definition~\ref{def:bvb}} hold.

\noindent \textbf{BV-validity.~~} Suppose that, during round $r$, merely faulty nodes BV-broadcast $v$. We show that $\nexists c \in R$, such that $(\exists v \in \mathit{binValues}_i(r,2t\mathrm{+}1))$ holds in $c$. Since only faulty nodes BV-broadcast $v$, then no correct node receives $\mathrm{EST}(\bull,rnd=r,est=\{v,\bullet\},\bullet)$ messages from more than $t$ different senders. Consequently, $v \notin \mathit{binValues}_i(r,t\mathrm{+}1)$ in line~\ref{ln:ESTrepeatB} at any correct node $p_i \in \sP$. Similarly, no correct node $p_i \in \sP$ can satisfy the predicate $(\exists w \in \mathit{binValues}(r,2t\mathrm{+}1))$ at line~\ref{ln:ESTrepeat} (via line~\ref{ln:inforesult}). Thus, the {requirement} holds.

\noindent \textbf{BV-uniformity.~~} Suppose that $w \in \mathit{binValues}_i(r,2t\mathrm{+}1)$ holds in $c$. By lines~\ref{ln:binVal} and \ref{ln:auxIf} we know that $p_i$ stores $v$ in at least $(2t\mathrm{+}1)$ entries of $EST[r][]$. Since $R$ starts in a well-initialized system state, this can only happen if $p_i$ received $\mathrm{EST}(\bullet,rnd=r,est=\{v,\bullet\},\bullet)$ messages from at least $(2t\mathrm{+}1)$ different nodes (line~\ref{ln:uponEst}). This means that $p_i$ received this message from at least $(t\mathrm{+}1)$ different correct nodes. Since each of these correct nodes sent the message $\mathrm{EST}(\bullet,rnd=r,est=\{v,\bullet\},\bullet)$ to any node in $\sP$, we know that  $\forall j \in \Correct:\mathit{binValues}_j(r,t\mathrm{+}1)\neq\emptyset$ (line~\ref{ln:ESTrepeatB}) holds eventually. Therefore, every correct node $p_j$ sends $\mathrm{EST}(\bullet,rnd=r,est=\{v,\bullet\},\bullet):v \in \mathit{binValues}_i(r,t\mathrm{+}1)$ to all. Since $n\mathrm{-} t\geq 2t\mathrm{+} 1$, we know that $(\exists w \in \mathit{binValues}_k(r,2t\mathrm{+}1))$ holds eventually at each correct node, $p_k \in \sP$. 

\noindent \textbf{BV-completion.~~} This requirement is implied by Lemma~\ref{thm:repeatDoesNotBlock}. 
\end{theoremProof}

\remove{
Cost of the algorithm 

As far as the cost of the algorithm is concerned, we have the following for each BV-broadcast instance.
\begin{itemize}
\item  If all correct nodes BV-broadcast the same value, the algorithm requires a single communication step. Otherwise, it can require two communication steps.
\item Let $c \geq n\mathrm{-}t$ be the number of correct nodes. The correct nodes send $c n$ messages if they BV-broadcast the same value, and send $2 c n$ messages otherwise. Hence, in a BV-broadcast instance, the correct nodes sends $\bigO(n^2)$ messages.
\item In addition to the control tag B VAL, a message carries a single bit. Hence, message size is constant.
\end{itemize}
}

\begin{theorem}[Closure]
\label{thm:terminationSafe}
Let $R$ be an execution of Algorithm~\ref{alg:consensus} that starts from a well-initialized system state  and includes a complete invocation of binary consensus. Within $\bigO(r):r\leq M$ asynchronous (communication) rounds, with probability $\Pr(r) = 1- (1/2)^r$, and for each correct node $p_i \in \sP$, the operation $\mathsf{result}_i()$ returns $v \in \{0,1\}$.
\end{theorem}

\renewcommand{\thmcnt}{\ref{thm:terminationSafe}}
\begin{theoremProof} Lemmas~\ref{thm:ifitwored} to~\ref{thm:probOneL} show the proof. Lemma~\ref{thm:ifitwored} shows that once all correct nodes estimate the same value in a round $r$, they hold on this estimate in all subsequent rounds. Lemma~\ref{thm:letPiAndPjtwocorrect} shows that correct nodes that pass a singleton  to $\mathsf{tryToDecide}()$, pass the same set. Lemma~\ref{thm:valid} shows that correct nodes can only decide a value that has been previously proposed by a correct node. Lemma~\ref{thm:probOneL} shows that correct nodes have $v\in\{0,1\}$ as a return value from $\mathsf{result}()$. This occurs by round $r\le M$ with the probability of  $1- (1/2)^r$. Putting these together, we obtain the proof of Theorem~\ref{thm:terminationSafe}.

\begin{lemma}
\label{thm:ifitwored}
Suppose that every correct node, $p_i \in \sP$, estimates value $v$ upon entering round $r$, \ie $est_i[r\mathit{-}1][i]=\{v\}\land r_i=r\mathit{-}1$ immediately before executing line~\ref{ln:isActive}. Then, $p_i$ estimates the value $v$ in any round later than $r$, \ie $r'\in \{r,\ldots, M\}:est_i[r'][i]=\{v\}$.
\end{lemma}
\renewcommand{\lemcnt}{\ref{thm:ifitwored}}
\begin{lemmaProof}
There are $n-t > t+1$ correct nodes. By the lemma statement, all of them broadcast $\mathrm{EST}(\bullet,rnd=r,est=\{v\},\bullet)$ (line~\ref{ln:ESTrepeatB}). Thus, $binValues_i(r,2t\mathrm{+}1)=\{v\}$ (BV-completion and BV-validity, Theorem~\ref{thm:BVtermination}) and $values^r_i=  \{v\}$ (lines~\ref{ln:cupEq}), where $values^r_i$ is the parameter that $p_i$ passes to $\mathsf{tryToDecide}()$ (line~\ref{ln:tryToDecide}) during round $r$, cf. $values^r_i=\cup_{j \in s} \{\mathit{aux}[r][j]\}$ (line~\ref{ln:rM} via line~\ref{ln:inforesult}).  

Therefore, $est_i[r][i]=\{v\}$ holds due to the assignment in the start of line~\ref{ln:cupEqMore}. Since there are most $t$ Byzantine nodes, and for an estimate to be forwarded (and hence accepted) it needs a ``support" of $t+1$ nodes (line~\ref{ln:ESTrepeatB}), it follows that the correct {nodes cannot} change their estimate in any round $r'\ge r$.
\end{lemmaProof}

\begin{lemma}
\label{thm:letPiAndPjtwocorrect}
Suppose that there is a system state $c \in R$, such that $(values^r_i = \{v\}) \land (values^r_j	= \{w\})$, where $p_i, p_j \in \sP$ are two correct nodes and $values^r_i$ is the parameter that $p_i$ passes to $\mathsf{tryToDecide}()$ (line~\ref{ln:tryToDecide}) during round $r$. It holds that $v =
w$ in $c$.
\end{lemma}
\renewcommand{\lemcnt}{\ref{thm:letPiAndPjtwocorrect}}
\begin{lemmaProof}
Due to the exit condition of the repeat-until loop in lines~\ref{ln:auxIf} to \ref{ln:ESTrepeat}, $p_i$ had to receive before $c$ identical $\mathrm{EST}(\bullet,rnd=r,\bullet,aux=v,\bullet)$ {messages} from at least $(n\mathit{-}t)$ different nodes. Since at most $t$ nodes are faulty, $(n\mathit{-}t)=(n\mathit{-}2t)$, which means that $p_i$ received $\mathrm{EST}(\bullet,rnd=r,\bullet,aux=v,\bullet)$ messages before $c$ from at least $(t\mathit{+}1)$ different correct nodes, as $n\mathit{-}2t \geq t\mathit{+}1$. Using the symmetrical arguments, we know that $p_j$ had to receive before $c$ identical $\mathrm{EST}(\bullet,rnd=r,\bullet,aux=w,\bullet)$ messages from at least $(n\mathit{-}t)$ different nodes.

Since $(n\mathit{-}t)\mathit{+}(t\mathit{+}1) > n$, the pigeonhole principle implies the existence for at least one correct node, $p_x \in \sP$, from which from $p_i$ and $p_j$ have received the messages $\mathrm{EST}(\bullet,rnd=r,\bullet,aux=v,\bullet)$ and $\mathrm{EST}(\bullet,rnd=r,\bullet,aux=w,\bullet)$, respectively. The fact that $p_x$ is correct implies that it has sent the same $\mathrm{EST}(\bullet,rnd=r,\bullet)$ message to all the nodes in line~\ref{ln:ESTrepeatB}. Thus $v = w$.
\end{lemmaProof}

\begin{lemma}
\label{thm:valid}
Suppose that there is {a system state} $c \in R$, such that $\mathsf{result}_i()=v\in\{0,1\}$ in $c$, where $p_i \in \sP$ is a correct node. There is a correct node $p_j \in \sP$ and a step $a_j \in R$ (between $R$'s starting system state and $c$) in which $p_j$ invokes $\mathsf{propose}_j(v)$.
\end{lemma}
\renewcommand{\lemcnt}{\ref{thm:valid}}
\begin{lemmaProof}
Suppose that $r_i = 1$. Recall (a) the BV-validity property (Theorem~\ref{thm:BVtermination} and line~\ref{ln:ESTrepeatB}), observe (b) the if-statement condition in line~\ref{ln:auxIf}, which selects the value $w_i \in \mathit{binValues}(1,2t\mathrm{+}1)$ (line~\ref{ln:auxThen}) as well as the exit condition in line~\ref{ln:ESTrepeat} of the repeat-until loop in lines~\ref{ln:auxIf} to~\ref{ln:ESTrepeat} in which (c) correct nodes, $p_j \in \sP$, broadcast $\mathrm{EST}(\bullet,rnd=1,\bullet,aux=w_j,\bullet):w_j \in \mathit{binValues}(1,t\mathrm{+}1)$ messages. Thus, the set $values^1_i$ includes only values arriving from correct nodes, where $values^r_i$ is the parameter that $p_i$ passes to $\mathsf{tryToDecide}()$ (line~\ref{ln:tryToDecide}) during round $r$.

Node $p_i$ can decide $v$ (line~\ref{ln:cupEqMore}) when $values^1_i = \{v\} \land v=\mathrm{randomBit}_i(r_i)$ holds. Regardless of the decision, $p_i$ updates its new estimate (line~\ref{ln:estRiGetsV}). Processor $p_i$ updates its estimate $est_i[r_i][i]$ with the value, $\mathrm{randomBit}(r))$, obtained by the RCC (line~\ref{ln:estRiGetsRandom}) whenever $values^1_i = \{0, 1\}$. This means, that $p_i$ updates the estimated value with a value that a correct node has proposed. Note that  the $values^1_i = \{0, 1\}$ case occurs when both $0$ and $1$ were proposed by correct nodes. The same arguments hold also for round numbers $r>1$, and therefore, a decided value must be a value proposed earlier by a correct node {$p_j$, where $i=j$ can possibly hold.}
\end{lemmaProof}		

\begin{lemma}
\label{thm:agreement}
Suppose that there is system state $c \in R$, such that $\mathsf{result}_i()$ and $\mathsf{result}_j()$ are not members of $\{\bot,\blitza\}$ holds in $c$, where $p_i,p_j \in \sP$ are correct nodes. It holds that $\mathsf{result}_i()=\mathsf{result}_j()$.
\end{lemma}
\renewcommand{\lemcnt}{\ref{thm:agreement}}
\begin{lemmaProof}
Suppose, without the loss of generality, that node $p_i$ is the first correct node that decides during $R$ and it does so during round $r$. Suppose that there is another node, $p_j$, that decides also at round $r$. We know that both $p_i$ and $p_j$ decide the same value due to the $v_i = \mathrm{randomBit}_i(r)$ condition of the if-statement in line~\ref{ln:cupEqMore} and the properties of the RCC. We also know that $p_i$ and $p_j$ update their estimates in $est_x[r][x]:x\in \{i,j\}$ to $\mathrm{randomBit}_x(r)$.

Recall that $values^r_i$ denotes the parameter that $p_i$ passes to $\mathsf{tryToDecide}()$ (line~\ref{ln:cupEq}) during round $r$. Lemma~\ref{thm:letPiAndPjtwocorrect} says that $(values^r_i = \{v\}) \land (values^r_j	= \{w\})$ means that $v\neq w$ cannot hold. Moreover, if $p_i$ decides during round $r$ and $p_j$ {is not ready to decide},  it must be the case that $values^r_j = \{v,w\} = \{0, 1\}$, see lines~\ref{ln:estRiGetsRandom} to~\ref{ln:estRiGetsV} and the proof of Lemma~\ref{thm:letPiAndPjtwocorrect}. Therefore, $p_j$ assigns $\mathrm{randomBit}(r)$ to $est_j[r_j][j]$ (line~\ref{ln:estRiGetsRandom}). This means that every correct node starts round $(r+1)$ with $est_j[r_j][j]=\mathrm{randomBit}(r)$ and $\mathrm{randomBit}(r)=v$. Lemma~\ref{thm:ifitwored} says that this estimate never change, and thus, only $v$ can be decided.
\end{lemmaProof}
\begin{lemma}
\label{thm:probOneL}
By the end of round $r \leq M$, for each correct node $p_i \in \sP$, the operation $\mathsf{result}_i()$ returns $v \in \{0,1\}$ with probability  $\Pr(r) = 1- (1/2)^r$.
\end{lemma}

\renewcommand{\lemcnt}{\ref{thm:probOneL}}

\begin{lemmaProof}
The proof uses Claim~\ref{thm:probOne}.
\begin{claim}
\label{thm:probOne}
Let $c_r \in R$ be the state that the system reaches at the end of round $r\leq M$. With probability $\Pr(r) = 1- (1/2)^r$, $\exists v \in \{0,1\}:\forall i \in \Correct: est_i[r][i] =\{v\}$ holds in $c_r$.
\end{claim}
\renewcommand{\clmcnt}{\ref{thm:probOne}}
\begin{claimProof}
%
%			The proof considers the execution of lines~\ref{ln:cupEq} to~\ref{ln:cupEqMore} during round $r$.
Let $values^r_i$ be the parameter that $p_i$ passes to $\mathsf{tryToDecide}()$ (line~\ref{ln:tryToDecide}) on round~$r$.

%\EMS{All three cases had to be written.}

\begin{itemize}

\item \textbf{Case 1:} Suppose that the if-statement condition $values^r_i = \{v_k(r)\}$ (line~\ref{ln:cupEq}) holds for all correct nodes $p_k \in \sP$. Similarly to  the proof of Lemma~\ref{thm:agreement}, any correct node $p_k$ assigns to $est_k[r][k]$ the same value, $v_k(r)$ (line~\ref{ln:NestRiGetsV}).

\item \textbf{Case 2:} Suppose that the if-statement condition $values^r_i = \{v_k(r)\}$ (line~\ref{ln:cupEq}) does not hold for all correct nodes $p_k \in \sP$. By similar arguments as in the previous case, any correct $p_k$ assigns to $est_k[r][k]$ the same value, $\{\mathrm{randomBit}_k(r)\}$ (line~\ref{ln:NestRiGetsRandom}).

\item \textbf{Case 3:} Some correct nodes assign $\{v_k(r)\}$ to $est_k[r][k]$ (line~\ref{ln:estRiGetsV}), whereas others assign $\{\mathrm{randomBit}_k(r)\}$ (line~\ref{ln:estRiGetsRandom}).
\end{itemize}
The rest of the proof focuses on Case 3. Recall the assumption that the Byzantine nodes have no control over the network or its scheduler. Thus, the values $\mathrm{randomBit}_k(r)$ and $\mathrm{randomBit}_k(r')$ are independent (due to the RCC {properties, see Section~\ref{sec:rcc}),} where $r\neq r'$. Therefore, there is probability of $\frac{1}{2}$ that the assignments of the values $\{v_k(r)\}$ and $\{\mathrm{randomBit}_k(r)\}$ are equal. Let $\Pr(r)$ be the probability that $[\exists  r' \leq r : \mathrm{randomBit}(r)=v(r)]$. Then, $\Pr(r)= \frac{1}{2} + (1-\frac{1}{2})\frac{1}{2} + \cdots + (1-\frac{1}{2})^{r-1}\frac{1}{2} = 1- (\frac{1}{2})^r$.
\end{claimProof}

\medskip

Recall that Lemma~\ref{thm:repeatDoesNotBlock} says that the repeat-until loop in lines~\ref{ln:auxIf} to \ref{ln:ESTrepeat} cannot block forever.	It follows from Lemma~\ref{thm:ifitwored} and Claim~\ref{thm:probOne} that all the correct nodes $p_i$ keep their estimated value $est_i = v$ and consequently the predicate $(values^{r'}_i = \{v\})$ at line~\ref{ln:cupEq} holds for round $r'$, where $values^{r'}_i =\cup_{j \in s} \{\mathit{aux}_i[r][j]\}$. With probability $\Pr(r) = 1- (1/2)^r$, by round $r$, it holds that $\mathrm{randomBit}(r)=v$ due to the RCC properties. Then, the if-statement condition of line~\ref{ln:cupEq} does not hold and the one in line~\ref{ln:cupEqMore} does hold. Thus, all the correct nodes decide $v$.
\end{lemmaProof}
\end{theoremProof}	

\F
{We conclude the proof by showing that Algorithm~\ref{alg:consensus} is an eventually loosely-self-stabilizing solution for binary consensus.}

\begin{theorem}
\label{thm:imple}
Let $R$ be an execution of Algorithm~\ref{alg:consensus} that starts in a well-initialized system state and during which every correct node $p_i \in \sP$ invokes $\mathsf{propose}_i()$ exactly once. Execution $R$ implements a loosely-self-stabilizing and randomized solution for binary consensus that can tolerate up to $t$ Byzantine nodes, where $n\geq 3t+1$. Moreover, within four asynchronous (communication) rounds, all correct nodes are expected to decide.
\end{theorem}
\renewcommand{\thmcnt}{\ref{thm:imple}}
\begin{theoremProof} We divide the proof into {four} arguments.

{\noindent \textbf{Argument 1:} \emph{BC-completion is always guaranteed.~~} Lemma~\ref{thm:repeatDoesNotBlock} and~\ref{thm:probOneL} demonstrate BC-completion when starting from an arbitrary, and resp., a well-initialized system state.}

{\noindent \textbf{Argument 2:} \emph{Suppose that, for every correct node $p_i \in \sP$, operation $\mathsf{result}_i()$ returns $v \in \{0,1\}$. A complete and well-initialized invocation of binary consensus satisfies the safety requirements of Definition~\ref{def:consensus}.~~} Lemmas~\ref{thm:valid},~\ref{thm:agreement}, and~\ref{thm:probOneL} imply BC-validity and BC-agreement as long as $\forall i \in \Correct:\mathsf{result}_i()$ returns $v \in \{0,1\}$.}

\noindent \textbf{Argument 3:} \emph{Algorithm~\ref{alg:consensus} satisfies the design criteria of Definition~\ref{def:practSelf}.~~}
By Theorem~\ref{thm:recovery}, we know that any complete invocation of binary consensus terminates within a finite number of steps. Once that happens, the next well-initialized invocation of $\mathsf{propose}()$ can succeed independently of previous invocations. Argument 1 and Lemma~\ref{thm:probOneL} imply that with probability $\Pr(M) = 1- (1/2)^M$, a complete and well-initialized invocation of binary consensus satisfies the requirements of Definition~\ref{def:consensus}. 

\noindent \textbf{Argument 4:} \emph{All correct nodes are expected to decide within four iterations of Algorithm~\ref{alg:consensus}.~~}
{The proof of Claim~\ref{thm:probOne} considers two stages when demonstrating BC-completion (after starting from a well-initialized system state).} That is, all correct nodes need to first use the same value, $v$, as their estimated one, see the assignment to $est_i[r][i]$ in lines~\ref{ln:estRiGetsV} to~\ref{ln:estRiGetsRandom}. Then, each correct node waits until the next round in which the condition, $v_i = \mathrm{randomBit}_i(r_i)$, of the if-statesmen in line~\ref{ln:cupEqMore} holds, where $\mathrm{randomBit}()$ is the interface to the RCC. The rest of the proof is implied via the linearity of expectation and the following arguments regarding the expectation of each stage.

\indent \textbf{Stage I.~~} The proof of Claim~\ref{thm:probOne} reveals the case in which not all correct nodes use the same value (Case 3). This is when the condition, $values  = \{v\}$, of the if-statement in line~\ref{ln:estRiGetsV} is true but not for any correct node $p_i\in\sP$. We show how to bound by two the number of asynchronous rounds in which this situation can happen. Suppose that $values^r_i \neq \{v\}$. Note that, with probability $1/2$, the assignment in line~\ref{ln:estRiGetsRandom} sets the value {$\{v\}$} to $est_i[r][i]$. Once that happens, Stage I is finished and Stage II begins. If this does not happen, with  probability $1/2$, Stage I needs to be repeated and so does the above arguments. Thus, within two rounds, Stage I is expected to end.  

\indent \textbf{Stage II.~~} By the RCC properties {(Section~\ref{sec:rcc}),} we know that $\Pr(v_i = \mathrm{randomBit}_i(r_i))=1/2$ and $E(\Pr(v_i = \mathrm{randomBit}_i(r_i)))=2$.
\end{theoremProof}

% REMOVE

%	We present an SSBFT recycling mechanism that uses a bounded array of objects. At any time, there is at most one active object, \ie an object that has not completed its task. Once an object has completed its task, a new object needs to be allocated on the array. That new object is supposed to be initialized to a predefined state so that it is ready to perform its task. Once the task is completed, the recycling mechanism can move to the next array entry as long as the following space constraint is satisfied.
%	
%	
%	there are no more than a predefined constant number, $logSize$, of objects that are ready to be recycled. In order to satisfy the latter constraint, 

\Section{\ems{SSBFT Recycling Mechanism for $\mathsf{BSMP_{n,t}[\kappa\mathit{-}SGC,t < n/3,RCCs]}$}}
\label{sec:bck}
We present a SSBFT recycling mechanism that uses a bounded array of recyclable objects. 
These objects, for example, can be instances of recyclable objects based on Algorithm~\ref{alg:Rconsensus} (with the boxed code lines), which implements the operations $\mathsf{propose}()$ and $\mathsf{result}()$ as well as $\mathsf{wasDelivered}()$ and $\mathsf{recycle}()$.

The mechanism aims at making sure that, at any time, there is at most a constant number, $logSize$, of active objects, \ie objects that have not completed their tasks. 
Once an object completes its task, the recycling mechanism can allocate a new object by moving to the next array entry as long as some constraints are satisfied. 
Specifically, the proposed solution is based on a synchrony assumption that guarantees that every correct node retrieves at least once the result of a completed object, $x$, within $logSize$ synchronous rounds since the first time in which at least $t+1$ correct nodes have retrieved the result of $x$, and thus, $x$ can be recycled. 

In this section, we refine $\mathsf{BAMP_{n,t}[\mathit{-}FC, t < n/3]}$ model into the model of $\mathsf{BSMP_{n,t}[\kappa\mathit{-}SGC,t < n/3,RCCs]}$ (Section~\ref{sec:sysS}), which is a synchronous model enriched with a random RCC and $\kappa$-state clock. 
We then present the synchrony assumptions (Assumption~\ref{def:logSize}) that we mentioned above and bring an overview of the proposed solution (Section~\ref{sec:over}) before providing the details and correctness proofs (sections~\ref{sec:ssbftMVC} and~\ref{sec:ssbftIndexViaSimRelInc}).  

\Subsection{System Settings for $\mathsf{BSMP_{n,t}[t < n/3, RCCs]}$}
\label{sec:sysS} 
We denote the $\mathsf{BSMP_{n,t}[t < n/3,\kappa\mathit{-}SGC]}$ model, which stands for Byzantine synchronous message-passing with at most $t$ (out of $n$) faulty nodes, and $t < n/3$. 
%
%We use the model of $\mathsf{BSMP_{n,t}[\kappa\mathit{-}SGC,t < n/3,RCCs]}$ for \EMS{studying the RRC problem.} 
%
The $\mathsf{BSMP_{n,t}[\kappa\mathit{-}SGC,t < n/3,RCCs]}$ model is defined by enriching the model of $\mathsf{BAMP_{n,t}[\mathit{-}FC,t < n/3]}$ with a $\kappa$-state global clock (Section~\ref{sec:kappaState}), reliable communications (Section~\ref{sec:relComm}), and RCCs (Section~\ref{sec:rcc}).

\Subsubsection{A $\kappa$-state global clock}
\label{sec:kappaState}
We assume that the algorithm takes steps according to a common global pulse (beat) that triggers a simultaneous step of every node in the system. Specifically, we denote synchronous executions by $R={c[0],c[1],\ldots}$, where $c[x]$ is the system state that immediately precedes the $x$-th global pulse. Also, $a_i[x]$ is the step that node $p_i$ takes between $c[x]$ and $c[x+1]$ simultaneously with all other nodes. We also assume that each node has access to a $\kappa$-state global clock via the local function $clock(\kappa)$, which returns an integer between $0$ and $\kappa-1$. Algorithm 3 of BDH~\cite{DBLP:conf/podc/Ben-OrDH08} offers an SSBFT $\kappa$-state global clock.    

\Subsubsection{Reliable communications}
\label{sec:relComm}
We assume the availability of reliable communications.
We assume that any correct node $p_i \in \sP$ starts any step $a_i[x]$ with receiving all pending messages from all nodes. Also, $p_i$ sends any message during $a_i[x]$, it does so only at the end of $a_i[x]$. We require (i) any message that a correct node $p_i$ sends during step $a_i[x]$ to another correct node $p_j$ is received at $p_j$ at the start of step $a_j[x+1]$, and
(ii) any message that $p_j$ received during step $a_j[x+1]$, was sent before the end of $a_i[x]$. 

\begin{figure}[t!]
	%\begin{wrapfigure}{r}{0.25\textwidth}
	\begin{center}
		%					\BB%\BBB
		%\hspace*{-0.5em}
		%		\includegraphics[scale=0.25, clip]{arcCrop.pdf}
		\includegraphics[scale=0.4, clip]{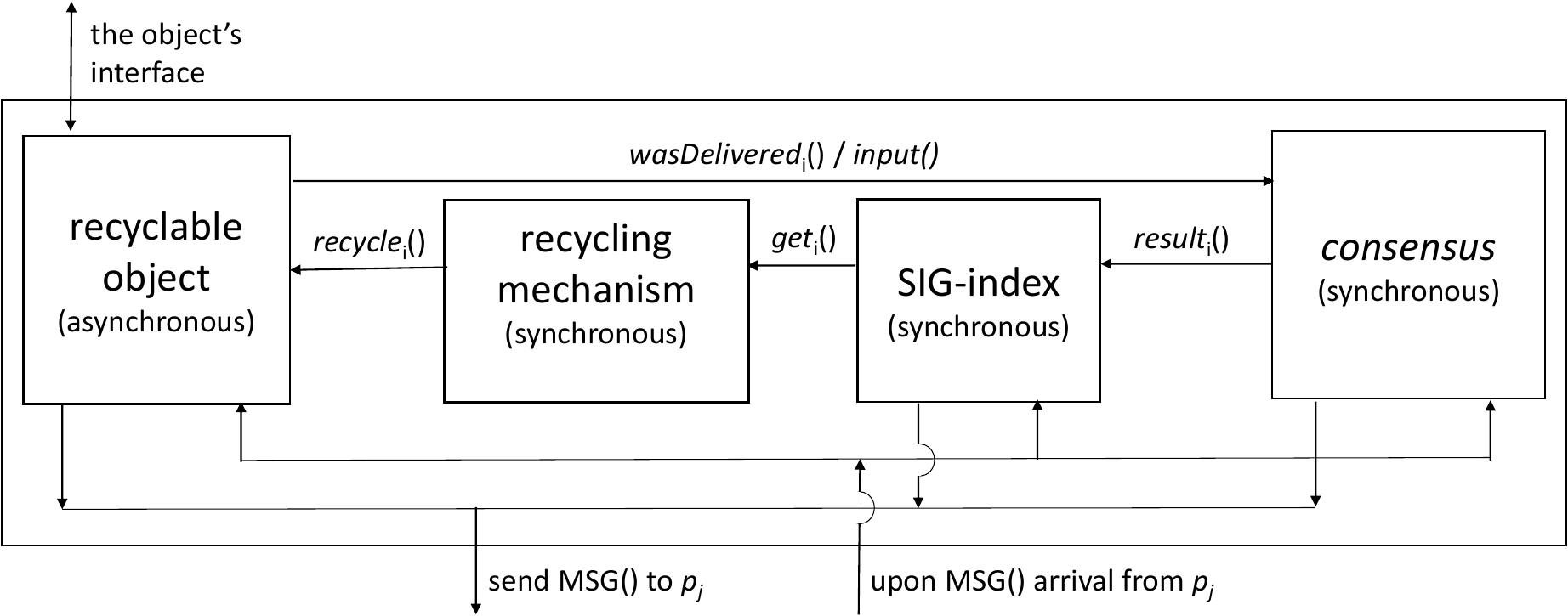}
	\end{center}
	\BBB
	\caption{\label{fig:arc}\small{The proposed solution using a recyclable object (Algorithm~\ref{alg:consensus}), the recycling mechanism (Algorithm~\ref{alg:logOneRecycling}),  a consensus protocol (Algorithm~\ref{alg:ssbftMVC}), and SIG-index  (Algorithm~\ref{alg:ssbftIndexViaSimRelInc}).}}
	\BBB
	%\end{wrapfigure}
\end{figure}

\begin{algorithm}[t!]
	\begin{\algSize}	
		
		\medskip
		\textbf{constants:} 
		$indexNum$ number of indices of recyclable objects\label{ln:indexNum}\;
		$logSize \in \{0,\ldots,indexNum-2\}$ user-defined bound on the object log size\label{ln:logSize}\;
		
		\medskip
		
		\textbf{variables:} 
		
		$\mathit{ssbftIndex}:$ an SSBFT index that marks the current object in use (Algorithm~\ref{alg:ssbftIndexViaSimRelInc})\;
		
		\smallskip
		
		$obj[indexNum]:$ array of recyclable objects, \eg Algorithm~\ref{alg:consensus}. Note that during legal execution only at most $(logSize+1)$ objects are stores at any given point of time\label{ln:objIndexNum}\; 
		
		\medskip

		\textbf{upon pulse} \texttt{/* signal from global pulse system */} \label{ln:vbbBradcastAAA} \Begin{
			\indent \ForEach{$x \notin \{y ~\bmod$ $ indexNum:y\in \{z-logSize,\ldots,z\}\}$ \textbf{\emph{where}} $z=indexNum+\mathit{ssbftIndex}.getIndex()$}{$obj[x].\mathsf{recycle}()$}}
		
		\medskip		
		
		\caption{\label{alg:logOneRecycling}SSBFT object recycling with a predefined log size; code for $p_i$}		
	\end{\algSize}
	
\end{algorithm}

\Subsection{Solution overview}
\label{sec:over}
The SSBFT recycling solution is a composition of several algorithms, see Figure~\ref{fig:arc}. Our recycling mechanism is presented in Algorithm~\ref{alg:logOneRecycling}. It allows every correct node to retrieve at least once the result of any object that is stored in a bounded array and yet over time that array can store an unbounded number of object instances. Algorithm~\ref{alg:logOneRecycling} guarantees that for every instance of the recyclable object, every correct node calls $\done()$ (line~\ref{ln:wasDelivered}) at least once before every correct node simultaneously invokes $\mathsf{recycle}()$ (line~\ref{ln:recycle}). 

%The retrieval of the object results is done by an entity external to the proposed solution. Thus, we assume such once at least $t+1$ correct nodes have performed this retrieval then this retrieval is performed by all $n-t$ correct nodes within a known constant number of synchronous rounds, which we denote by $logSize$ (line~\ref{ln:logSize}). The evidence for result retrieval by at least $t+1$ correct nodes is collected by at least $t+1$ node for which $\mathsf{wasDelivered}()$ returns '1' (line~\ref{ln:wasDelivered}), as we explain next.     

We consider the case in which the entity that retrieves the result of object $obj$ might be external (and perhaps, asynchronous) to the proposed solution. The proposed solution does not decide to recycle $obj$ before there is sufficient evidence that, within $logSize$ synchronous cycles, the system is going to reach a state in which $obj$ can be legitimately recycled. Specifically, Assumption~\ref{def:logSize} considers an event that can be locally learned about when $\mathsf{wasDelivered}()$ returns '1' (line~\ref{ln:wasDelivered}).

\begin{assumption}[Result retrial within a bounded time]
\label{def:logSize}
Let us consider the system state, $c[r]$, in which the result of object $obj$ was retrieved by at least $t+1$ correct nodes. We assume that, within $logSize$ synchronous cycles from $c[r]$, the system reaches a state, $c[r+logSize]$, in which all $n-t$ correct nodes have retrieved the result of $obj$ at least once.
\end{assumption}

\begin{algorithm}[t!]
	\begin{\algSize}

		\medskip
		\textbf{variables:} $currentResult$ stores the most recent result of $co$\label{ln:varResults}\; 
		
		$co$ a (non-self-stabilizing) BFT (multivalued) consensus object\label{ln:varCo}\;
		
		\medskip
		
		\textbf{interface required:}

		$input():$  defines the input value to be provided to the given consensus protocol\;
		
		\medskip
		
		\textbf{interface provided:}

		$\done()$\label{ln:doneSSBFTcon} \textbf{do} $\Return{(currentResult)}$ \texttt{//} the decided value of the most recent $co$'s invocation\;
		
		\medskip
		
		\textbf{message structure:} 
		$\langle appMsg \rangle$, where $appMsg$ is the application message, \ie a message sent by the given consensus protocol\; 
		
		\medskip
		
		\textbf{upon pulse} \texttt{/* signal from global pulse system */} \label{ln:vbbBradcastAAAB} \Begin{

			\textbf{let} $M$ be message that holds at $M[j]$ the arriving $\langle appMsg_j \rangle$ messages from $p_j$ for the current synchronous round and $M'=[\bot, \ldots,\bot]$\label{ln:messageArrival}\;

			\If{$clock(\kappa)=0$\label{ln:case0co}}{
				
				$currentResult \gets co.\done()$\label{ln:getResults}\;  
				$co.\mathit{restart}()$\label{ln:coRestart}\;				
				$M' \gets co.\mathit{propuse}(input())$\label{ln:coPrupuse}\tcc{for recycling $input()$ is $\mathsf{wasDelivered}()$\label{ln:coWasDelivered}}
				
			}
			
			\lElseIf{$clock(cycleSize) \in \{1,\ldots,t\}$\label{ln:processIf}}{$M' \gets co.\mathit{process}(M)$\label{ln:processThen}}
			
			\lForEach{$p_j \in \sP$\label{ln:sendMjToAllIFor}}{\textbf{send} $\langle M'[j] \rangle$ \textbf{to} $p_j$\label{ln:sendMjToAllIDo}}
			
		}

		\medskip

		\caption{\label{alg:ssbftMVC}SSBFT multivalued consensus in $\mathsf{BSMP_{n,t}[t < n/3,(t+1)\mathit{-}SGC]}$; code for node $p_i$}		
	\end{\algSize}
	
\end{algorithm}

\begin{algorithm}[t!]
	\begin{\algSize}	
		
		\medskip
		
		\textbf{constants:} $I:$ bound on the number of states an index may have\; 
		
		\medskip
		
		\textbf{variables:} 
		$\mathit{index} \in \{0,\ldots, I-1\}:$ a local copy of the global logical object index\;
		
		\smallskip
		
		$\mathit{ssbftCO}:$ an SSBFT consensus object (Algorithm~\ref{alg:ssbftMVC}) that is used for agreeing on the garbage collector state, \ie 1 when there is a need to recycle (otherwise 0)\; 
		
		\medskip
		
		\textbf{interfaces provided:} 
		$getIndex()$ \textbf{do} \Return{$index$}\;

		\medskip
		
		\textbf{message structure:} 
		$\langle index \rangle$: the logical object index\; 
		
		\medskip
		
		\textbf{upon pulse} \texttt{/* signal from global pulse system */} \label{ln:vbbBradcastAAAC} \Begin{
			
			\textbf{let} $M$ be the arriving $\langle index_j\rangle$ messages from $p_j$\label{ln:MarrivingIndex}\;
			
			\Switch{$clock(\kappa)$\label{ln:switchA} \emph{\texttt{/* consider $clock()$ at the beginning of the pulse */}}}{
				
				\lCase{$\kappa-4$\label{ln:case0}}{\textbf{broadcast} $\langle index=getIndex() \rangle$\label{ln:clockAZero}}
				
				\Case{$\kappa-3$}{
					
					\textbf{let} $propose := \bot$\label{ln:proposeGetsBotprp}\;
					\lIf{$\exists v \neq \bot:|\{  \langle v  \rangle \in M\}|\geq n-f$}{$propose \gets v$\label{ln:existsVNeqBotVBull}}
					
					\textbf{broadcast} $\langle propose \rangle$\label{ln:broadcastPropose}\;
				}
				
				\Case{$\kappa-2$\label{ln:case2}}{
					
					\textbf{let} $bit := 0$\label{ln:saveBitGetsBot}; $save \gets \bot$\label{ln:saveBitGetsBotS}\;
					
					\lIf{$\exists s \neq \bot:|\{  \langle s  \rangle \in M\}|> n/2$}{$save\gets s$\label{ln:sNeqBot}}
					
					\lIf{$|\{\langle save \neq \bot\rangle \in M\}|\geq n-f$}{$bit \gets 1$\label{ln:saveNeqBot}}
					
					\lIf{$save = \bot$}{$save \gets 0$\label{ln:saveBot}}
					
					\textbf{broadcast} $\langle bit\rangle$\label{ln:saveBotX}\;	
				}
				
				\Case{$\kappa-1$\label{ln:case3}}{
					\textbf{let} $\mathit{inc} := 0$\label{ln:glbSvV}\;
					
					\lIf{$\mathit{ssbftCO}.\done()$}{$\mathit{inc}\gets 1$\label{ln:glbSvVONE}}
					
					%					\lIf{$TEST$}{$\mathit{inc} \gets 1$\label{ln:glbSv}}
					\lIf{$|\{\langle 1\rangle \in M\}|\geq n-f$}{$\mathit{index} \gets (save + \mathit{inc})\bmod I$\label{ln:logicSavePlusSv}}
					\lElseIf{$|\{\langle 0\rangle \in M\}|\geq n-f$}{$\mathit{index} \gets 0$\label{ln:logicZero}}
					\lElse{$\mathit{index} \gets rand(save + \mathit{inc}) \bmod I$\label{ln:logicRandSavePlusSv}}
				}				
			}
		}
		\medskip

		\caption{\label{alg:ssbftIndexViaSimRelInc}SSBFT SIG-index in $\mathsf{BSMP_{n,t}[t < n/3,4\mathit{-}SGC]}$; code for node $p_i$}		
	\end{\algSize}	
\end{algorithm}

Algorithm~\ref{alg:logOneRecycling}'s recycling guarantees are facilitated by Algorithm~\ref{alg:ssbftMVC}, which agrees on a single evidence from all collected ones, and Algorithm~\ref{alg:ssbftIndexViaSimRelInc}, which uses the agreed evidence for updating the value of the index that points to the current entry in the array. Algorithm~\ref{alg:logOneRecycling}'s detailed presentation and correctness proof appear in Section~\ref{sec:ssbftBinCon}.

\Subsubsection{Evidence collection using an SSBFT (multivalued) consensus (Algorithm~\ref{alg:ssbftMVC})}
Algorithm~\ref{alg:ssbftMVC} offers an SSBFT multivalued consensus protocol that returns within $t+1$ synchronous rounds an agreed non-$\bot$ value as long as at least $t+1$ nodes proposed that value, \ie at least one correct node proposed that value. As mentioned, we use $\mathsf{wasDelivered}()$ (line~\ref{ln:wasDelivered}) for providing input to Algorithm~\ref{alg:ssbftMVC}. Thus, whenever '1' is decided, at least one correct node got an indication from at least $n-t$ nodes that they have retrieved the results of the current object. This implies that by at least $t+1$ correct nodes have retrieved the results and, by Assumption~\ref{def:logSize}, all $n-t$ correct nodes will retrieve the object result within a known number of synchronous rounds. Then, the object could be recycled. Algorithm~\ref{alg:ssbftMVC}'s detailed presentation and correctness proof appear in Section~\ref{sec:ssbftMVC}.

\Subsubsection{SSBFT simultaneous increment-or-get index (Algorithm~\ref{alg:ssbftIndexViaSimRelInc})}
Algorithm~\ref{alg:ssbftIndexViaSimRelInc} allows the proposed solution to keep track of the current object index that is currently used as well as facilitate synchronous increments to the index value. We call this task \emph{simultaneous increment-or-get index} (SIG-index). During legal executions of Algorithm~\ref{alg:ssbftIndexViaSimRelInc}, the correct nodes assert their agreement on the index value and update the index according to the result of Algorithm~\ref{alg:ssbftMVC}, which is an agreement on the value of $\mathsf{wasDelivered}()$. Algorithm~\ref{alg:ssbftIndexViaSimRelInc}'s detailed presentation and correctness proof appear in Section~\ref{sec:ssbftIndexViaSimRelInc}.

%As explained above, Algorithm~\ref{alg:ssbftMVC} contributes to the proposed solution the ability to uniformly decide that a given object should be recycled within $logSize$ synchronous rounds. Yet, there is a need to keep track of the current object index that is currently used and allow synchronous updates. We call this task \emph{simultaneous increment-or-get index} (SIG-index) and use Algorithm~\ref{alg:ssbftIndexViaSimRelInc} for achieving it. During legal executions, the correct nodes asserts that they all agree on the same index and then update the index according to the result of Algorithm~\ref{alg:ssbftMVC}.     

\begin{figure}[t!]
	%\begin{wrapfigure}{r}{0.25\textwidth}
	\begin{center}
		%					\BB%\BBB
		%\hspace*{-0.5em}
		%		\includegraphics[scale=0.25, clip]{arcCrop.pdf}
		\FF
		\includegraphics[scale=0.5, clip]{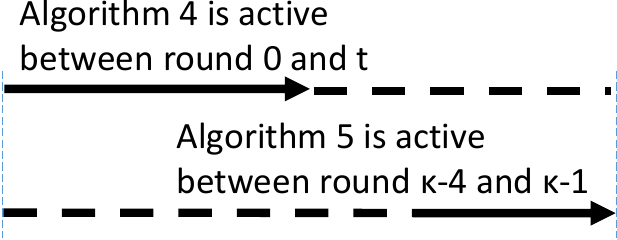}
	\end{center}
	\BBB
	\caption{\label{fig:schdule}\small{The schedule for algorithms~\ref{alg:ssbftMVC} to~\ref{alg:ssbftIndexViaSimRelInc} uses a cycle of $\kappa=\max\{t+1,logSize\}$ synchronous rounds.}}
	\BBB
	%\end{wrapfigure}
\end{figure} 

\Subsubsection{Scheduling strategy for algorithms~\ref{alg:ssbftMVC} to~\ref{alg:ssbftIndexViaSimRelInc}}
As mentioned, Algorithm~\ref{alg:ssbftMVC} requires $t+1$ synchronous rounds to complete and provide input to Algorithm~\ref{alg:ssbftIndexViaSimRelInc} and $\kappa-(t+1)$ synchronous rounds after that, any correct node can recycle the current object (according to Algorithm~\ref{alg:ssbftMVC}'s result), where $\kappa=\max\{t+1,logSize\}$. Thus, Algorithm~\ref{alg:ssbftIndexViaSimRelInc} has to defer its index updates until that time. Figure~\ref{fig:schdule} presents this scheduling strategy, which considers the schedule cycle s of $\kappa$. That is, algorithms~\ref{alg:ssbftMVC} and~\ref{alg:ssbftIndexViaSimRelInc} starting points are $0$ and $\kappa-4$, respectively. Note that Algorithm~\ref{alg:logOneRecycling} does not require scheduling since it accesses the index only via Algorithm~\ref{alg:ssbftIndexViaSimRelInc}'s interface of SIG-index, see Figure~\ref{fig:arc}.  

\Subsubsection{Communication piggybacking and multiplexing}
We use a piggybacking technique in order to facilitate the spread of the result (decision) values of the recyclable objects. As Figure~\ref{fig:arc} illustrates, all communications are piggybacked. Specifically, we consider a meta-message $MSG()$ that has a field for each message sent by algorithms~\ref{alg:consensus},~\ref{alg:ssbftMVC}, and~\ref{alg:ssbftIndexViaSimRelInc}. That is, when any of the algorithms~\ref{alg:ssbftMVC} and~\ref{alg:ssbftIndexViaSimRelInc} are active, its respective field in $MSG()$ includes a non-$\bot$ value. With respect to Algorithm~\ref{alg:consensus}'s field, $MSG()$ includes the most recent message that Algorithm~\ref{alg:consensus} has sent (or currently wishes to send). We note that this piggybacking technique allows the multiplexing of timed and reliable communications (assumed for $\mathsf{BSMP_{n,t}[\kappa\mathit{-}SGC,t < n/3,RCCs]}$) and fair communication (assumed for $\mathsf{BAMP_{n,t}[\mathit{-}FC, t < n/3]}$).

\Subsection{SSBFT recycling in $\mathsf{BSMP_{n,t}[t < n/3,(t+1)\mathit{-}SGC]}$ (Algorithm~\ref{alg:logOneRecycling})}
\label{sec:ssbftBinCon}
As mentioned, Algorithm~\ref{alg:logOneRecycling} considers an array, $obj[]$ (line~\ref{ln:objIndexNum}), of $indexNum$ recyclable objects (line~\ref{ln:indexNum}). We require the array size to be larger than $logSize$ (line~\ref{ln:logSize} and Assumption~\ref{def:logSize}). In addition to the array $obj[]$, Algorithm~\ref{alg:logOneRecycling}'s variable set includes $\mathit{ssbftIndex}$, which is an integer that holds the entry number of the latest object in use. Algorithm~\ref{alg:logOneRecycling} accesses the agreed current index by calling $\mathit{ssbftIndex}.getIndex()$. This lets the algorithm's code to nullify any entry in $obj[]$ that is not $\mathit{ssbftIndex}.getIndex()$ or at most $logSize$ older than $\mathit{ssbftIndex}.getIndex()$. Corollary~\ref{thm:ssbftBinCon} is directly implied by Assumption~\ref{def:logSize} and the properties of algorithms~\ref{alg:ssbftMVC} to~\ref{alg:ssbftIndexViaSimRelInc}, which we show in sections~\ref{sec:corrAlgorithmMVC} and~\ref{sec:corrIsSelf}, respectively.

\begin{corollary}
\label{thm:ssbftBinCon}
Algorithm~\ref{alg:logOneRecycling} is an SSBFT recycling mechanism that stabilizes within expected $\bigO(\kappa)$ synchronous rounds.
\end{corollary}

\Subsection{SSBFT multivalued consensus in $\mathsf{BSMP_{n,t}[t < n/3,(t+1)\mathit{-}SGC]}$}
\label{sec:ssbftMVC}
Algorithm~\ref{alg:ssbftMVC} assumes access to a deterministic (non-self-stabilizing) BFT (multivalued) consensus object, $co$, such as the ones proposed by Kowalski and Most{\'{e}}faoui~\cite{DBLP:conf/podc/KowalskiM13} or Abraham and Dolev~\cite{DBLP:conf/stoc/AbrahamD15}, for which completion is guaranteed to occur within $t+1$ synchronous rounds. We list our assumptions regarding the interface to the consensus object in Section~\ref{sec:assumptionsMVC}.

\Subsubsection{Required interface to the consensus object}
\label{sec:assumptionsMVC}
The proposed SSBFT solution uses the technique of recomputation of $co$'s floating output~\cite[Chapter 2.8]{DBLP:books/mit/Dolev2000}. In order to provide this, we assume that $co$ has the following interface:		

\begin{itemize}
\item $\mathit{restart}()$ sets $co$ to its initial state.		

\item $\mathit{propuse}(v)$ proposes the value $v$ when invoking (or re-invoking) $co$. This operation is effective only after $\mathit{restart}()$ was invoked. The returned value is a message vector, $M[]$, that includes all the messages, $M[j]$, that $co$ wishes to send to node $p_j$ for the current synchrony round.

\item $\mathit{process}(M)$ runs a single step of $co$. The input vector $M$ includes the arriving messages for the current synchronous round, where $M[j]$ is $p_j$'s message. The returned value is a message vector that includes all the messages that $co$ wishes to send for the current synchrony round. This operation is guaranteed to work correctly only after all correct nodes have simultaneously taken a consecutive sequence of steps that include invocations of either (i) $\mathit{process}()$, or (ii) $\mathit{restart}()$ immediately before proposing a non-$\bot$ value via the invocation of $\mathit{propuse}()$.

\item $\done()$ returns a non-$\bot$ results after the completion of $co$. The returned value is required to satisfy the consensus specifications only if all correct nodes have simultaneity taken a sequence of correct $\mathit{process}()$ invocations.
\end{itemize}

\Subsubsection{Detailed description}
Algorithm~\ref{alg:ssbftMVC}'s set of variables includes $co$ itself (line~\ref{ln:varCo}) and the current version of the result, \ie $currentResult$ (line~\ref{ln:varResults}). This way, the SSBFT version of $co$'s result can be retrieved via a call to $\done()$ (line~\ref{ln:doneSSBFTcon}). Algorithm~\ref{alg:ssbftMVC} proceeds in periodic rounds. At the start of any round, node $p_i$ stores all the arriving messages at the message vector $M$ (line~\ref{ln:messageArrival}).

When the clock value is zero (line~\ref{ln:case0co}), it is time to start the re-computation of $co$'s result. Thus, Algorithm~\ref{alg:ssbftMVC} first stores the current value of $co$'s result at $currentResult_i$ (line~\ref{ln:getResults}). Then, it restarts $co$'s local state and proposes a new value to $co$ (lines~\ref{ln:coRestart} and~\ref{ln:coPrupuse}). For the recycling solution presented in this paper, the proposed value is retrieved from $\mathsf{wasDelivered}()$ (line~\ref{ln:wasDelivered}).
For the case in which the clock value is not zero (line~\ref{ln:processIf}), Algorithm~\ref{alg:ssbftMVC} simply lets $co$ to process the arriving messages of the current round. 
Both for the case in which the clock value is zero and the case it is not, Algorithm~\ref{alg:ssbftMVC} broadcasts $co$'s messages for the current round (line~\ref{ln:sendMjToAllIFor}).

\Subsubsection{Correctness proof}
\label{sec:corrAlgorithmMVC}
Theorem~\ref{thm:algorithmMVC} shows that Algorithm~\ref{alg:ssbftMVC} stabilizes within $2(t+1)$ synchronous rounds.

\begin{theorem}
\label{thm:algorithmMVC}
Algorithm~\ref{alg:ssbftMVC} is an SSBFT deterministic (multivalued) consensus solution for $\mathsf{BSMP_{n,t}[t < n/3,(t+1)\mathit{-}SGC]}$ that recovers after the occurrence of the last transient-faults within $\max\{ \kappa,2(t+1)\}$ synchronous rounds.     
\end{theorem}
\renewcommand{\thmcnt}{\ref{thm:algorithmMVC}}
\begin{theoremProof}
Let $R$ be an execution of Algorithm~\ref{alg:ssbftMVC}. Within $\kappa$ synchronous rounds, the system reaches a state $c \in R$ in which $clock(\kappa)=0$ holds. Immediately after $c$, every correct node, $p_i$, simultaneously restarts $co_i$ and proposes the input (lines~\ref{ln:coRestart} and~\ref{ln:coPrupuse}) before sending the needed messages (line~\ref{ln:sendMjToAllIDo}). Then, for the $t$ synchronous rounds that follows, all correct nodes simultaneously process the arriving messages and send their replies (line~\ref{ln:processIf} and~\ref{ln:sendMjToAllIDo}). Thus, after $\max\{ \kappa,2(t+1)\}$ synchronous rounds from $c$, the system reaches a state $c'\in R$ in which $clock(\kappa)=0$ holds. Also, in the following synchronous round, all correct nodes store $co$'s results. That results in guaranteed to be correct due to Section~\ref{sec:assumptionsMVC}'s assumptions.  	   
\end{theoremProof}

\Subsection{SSBFT simultaneous increment-or-get index}
\label{sec:ssbftIndexViaSimRelInc}
The task of \emph{simultaneous increment-or-get index} (SGI-index) requires all correct nodes to maintain identical index values that all nodes can independently retrieve via $getIndex()$. We use the $\mathsf{BSMP_{n,t}[t < n/3,RCC,4\mathit{-}SGC]}$ model. The task assumes that all increments are performed according to the result of a consensus object, $\mathit{ssbftCO}$, such as Algorithm~\ref{alg:ssbftMVC}. Algorithm~\ref{alg:ssbftIndexViaSimRelInc} presents an SGI-index solution that recovers from disagreement on the index value using an RCC. That is, whenever a correct node receives $n-f$ reports from other nodes that they have each observed $n-f$ identical index values, an agreement on the index value is assumed and the index is incremented according to the most recent result of $\mathit{ssbftCO}$. Otherwise, a randomized strategy is taken for guaranteeing recovery from a disagreement on the index value. Our strategy is inspired by BDH~\cite{DBLP:conf/podc/Ben-OrDH08}'s SSBFT clock synchronization algorithm, which is in turn derived from non-self-stabilizing BFT solutions by Rabin~\cite{DBLP:conf/focs/Rabin83} as well as Turpin and Coan~\cite{DBLP:journals/ipl/TurpinC84}. 

%its structure resembles the SSBFT clock synchronization solution by Ben-Or, Dolev, and Hoch~\cite{DBLP:conf/podc/Ben-OrDH08}. Their solution lets every correct node to increment that clock on every pulse and in every fourth pulse         

\Subsubsection{Detailed description}
Algorithm~\ref{alg:ssbftIndexViaSimRelInc} is active during four clock phases of a common pulse, \ie $\kappa-4,\kappa-3,\kappa-2,$ and $\kappa-1$. Each phase starts with storing all arriving messages (from the previous synchronous round) in the array, $M$ (line~\ref{ln:MarrivingIndex}). The first phase broadcasts the local index value (line~\ref{ln:clockAZero}). The second phase lets each node vote on the majority arriving index value, or $\bot$ in case such value was not received (lines~\ref{ln:proposeGetsBotprp} to~\ref{ln:broadcastPropose}). The third phase resolves the case in which there is an arriving non-$\bot$ value, $save$,  that received sufficient support when voting during phase two (lines~\ref{ln:saveBitGetsBotS} to~\ref{ln:saveBot}). Specifically, if $save\neq \bot$ exists, then $\langle bit=1\rangle$ is broadcast. Otherwise, $\langle bit=0\rangle$ is broadcast. On the fourth phase (line~\ref{ln:glbSvV} to~\ref{ln:logicRandSavePlusSv}), the (possibly new) index is set either to be the majority-supported index value of phase two plus $inc$ (line~\ref{ln:glbSvV} to~\ref{ln:logicSavePlusSv}), where $inc$ is the output of $\mathit{ssbftCO}$, or (if there was insufficient support) to a randomly chosen output of the RCC (lines~\ref{ln:logicZero} and~\ref{ln:logicRandSavePlusSv}).

\Subsubsection{Correctness proof}
\label{sec:corrIsSelf}
%
%    presents the BDH solution for a self-stabilizing Byzantine-tolerant $\kappa$-bits clock synchronization by Ben{-}Or, Dolev, and Hoch~\cite{DBLP:conf/podc/Ben-OrDH08}. BDH has a constant overhead both in message complexity and in its expected stabilization time. BDH uses coin-flipping and $2$-bit solutions, which can be found at~\cite[figures 1 and 3]{DBLP:conf/podc/Ben-OrDH08}.
%
%BDH executes four phases. Each executed in a consecutive pulse. The first phase sends the clock value to everyone. In the second phase, each node votes on the majority clock value it received, or the $\bot$ value if no such value exists. The third phase determines whether enough nodes voted on a non-$\bot$ value, thus ensuring that those nodes that have voted, voted on the same value in phase 2. Lastly, in the fourth phase the new clock value is set either to be the majority clock value of phase 2, or (if there were not enough votes) is randomly selected using the output of BDH's coin-flipping solution. 
%
Theorem~\ref{thm:isSelf} shows that Algorithm~\ref{alg:ssbftIndexViaSimRelInc} stabilizes within expected $\bigO(\kappa)$ synchronous rounds.

\begin{theorem}
\label{thm:isSelf}
Let $R$ be an execution of algorithms~\ref{alg:ssbftMVC} and~\ref{alg:ssbftIndexViaSimRelInc} that is legal w.r.t. Algorithm~\ref{alg:ssbftMVC} (Theorem~\ref{thm:algorithmMVC}). Algorithm~\ref{alg:ssbftIndexViaSimRelInc} is an SSBFT SGI-index implementation that stabilizes within expected $\bigO(\kappa)$ synchronous rounds.
\end{theorem}
\renewcommand{\thmcnt}{\ref{thm:isSelf}}
\begin{theoremProof}
Corollaries~\ref{thm:vAvB} and~\ref{thm:atMostOneV} are needed for the proof of lemmas~\ref{thm:atLeast} and~\ref{thm:bmodKappa}. The pigeonhole principle implies Corollary~\ref{thm:vAvB} and Corollary~\ref{thm:atMostOneV} is implied by Corollary~\ref{thm:vAvB}.

\begin{corollary}%[Observation 3.1 in~\cite{DBLP:conf/podc/Ben-OrDH08}]
\label{thm:vAvB}
Let $V_x:x \in \{a,b\}$ be two $n$-length vectors that differ in at most $f<n/3$ entries. For any $x \in \{a,b\}$, suppose $V_x$ contains $n-f$ copies of $v_x$. Then $v_a = v_b$.
\end{corollary}

Corollary~\ref{thm:atMostOneV} is implied by Corollary~\ref{thm:vAvB}.

\begin{corollary}
\label{thm:atMostOneV}
Let $c[r] \in R$ be a system state in which $clock(\kappa)=\kappa-3$ and $X=\{x_i:i\in\Correct\}$ be the set of values encoded in the messages $\langle x_i\rangle$ that any correct node, $p_i \in \sP$, broadcasts in line~\ref{ln:broadcastPropose} at the end of $a_i[r]$. The set $X$ includes at most one non-$\bot$ value.
\end{corollary}

Lemma~\ref{thm:atLeast} implies that, within $O(\kappa)$ of expected rounds, all correct nodes have the identical $index$ values. Recall that $c[r] \in R$ is \emph{(progress) enabling} if $\exists x \in \{0,1\}: \forall i \in \Correct: rand_i=x$ holds at $c[r]$ (Section~\ref{sec:rcc}).

\begin{lemma}[Convergence]
\label{thm:atLeast}
Let $r> \kappa$. Suppose $c[r] \in R$ is (progress) enabling system state (Section~\ref{sec:rcc}) for which $clock(\kappa) = \kappa - 1$ holds. With probability at least $\min \{p_0, p_1\}$, all correct nodes have the same $index$ at $c[r+1]$.
\end{lemma}
\renewcommand{\lemcnt}{\ref{thm:atLeast}}
\begin{lemmaProof}
The proof is implied by claims~\ref{thm:atLeastNoVal} to~\ref{thm:atLeastOneVal}.

\begin{claim}
\label{thm:atLeastNoVal}
Suppose (i) there is no value $x \in \{0,1\}$ and (ii) there is no correct node $p_i \in \sP$ that receives at the start of step $a_i[r]$ the message $\langle x\rangle$ from at least $n-f$ different nodes. For any correct node, $p_j \in \sP$, it holds that step $a_j[r]$ assigns $0$ to $index_j$ with probability $p_0$.
\end{claim}
\renewcommand{\clmcnt}{\ref{thm:atLeastNoVal}}
\begin{claimProof}
The proof is implied directly from lines~\ref{ln:glbSvVONE} to~\ref{ln:logicRandSavePlusSv}.
\end{claimProof}

\begin{claim}
\label{thm:atLeastZeoVal}
Suppose there is a correct node $p_i \in \sP$ that receives at the start of step $a_i[r]$ the message $\langle 0\rangle$ from at least $n-f$ different nodes. Also, suppose there is $x \in \{0,1\}$ and a correct node $p_j \in \sP$ that receive at the start of step $a_j[r]$ the message $\langle x\rangle$ from at least $n-f$ different nodes, where $i=j$ may or may not hold. The step $a_j[r]$ assigns $0$ to $index_j$.
\end{claim}
\renewcommand{\clmcnt}{\ref{thm:atLeastZeoVal}}
\begin{claimProof}
Line~\ref{ln:logicZero} implies the proof since $x=0$ (Corollary~\ref{thm:vAvB}).
\end{claimProof}

\begin{claim}
\label{thm:atLeastOneValSame}
Suppose there is a correct node $p_i \in \sP$ that receives at the start of step $a_i[r]$ the message $\langle 1\rangle$ from at least $n-f$ different nodes. Let $p_j \in \sP$ be a correct node. At $c[r]$, $\mathit{ssbftCO}_i.\done() = \mathit{ssbftCO}_j.\done()$ and $save_i=save_j$ hold.
\end{claim}
\renewcommand{\clmcnt}{\ref{thm:atLeastOneValSame}}
\begin{claimProof}
At $c[r]$, $\mathit{ssbftCO}_i.\done() = \mathit{ssbftCO}_j.\done()$ holds (Algorithm~\ref{alg:ssbftMVC}'s agreement property). The rest of the proof shows that $save_i=save_j$ holds at $c[r]$.

Since $p_i$ has received $\langle 1\rangle$ from at least $n-f$ different nodes at the start of $a_i[r]$, we know that there is a correct node, $p_k \in \sP$, that has sent $\langle 1\rangle$ at the end of $a[r-1]$. By lines~\ref{ln:saveBitGetsBotS} to~\ref{ln:saveNeqBot}, node $p_j$ receives at the start of $a_j[r-1]$ the message $\langle x\rangle$ from at least $n-f$ nodes, where $x=save_j \neq \bot$. 
By Corollary~\ref{thm:atMostOneV}, any correct node broadcasts (line~\ref{ln:broadcastPropose}) either $\bot$ or $x$ at the end of step $a[r-2]$. This means that at the start of $a[r-1]$, correct nodes receive at most $f <n-2f$ messages with values that are neither $\bot$ nor $x\neq \bot$. Therefore, $save_i = save_j$ since, at the start of $a_i[r]$ and $a_j[r]$ both $p_i$, and resp., $p_j$ receive from at least $n-f$ different nodes the messages  $\langle x_i \rangle$, and resp., $\langle x_j \rangle$, where neither $x_i$ not $x_j$ are $\bot$.
\end{claimProof}

\begin{claim}
\label{thm:atLeastOneVal}
Suppose there is a correct node $p_i \in \sP$ that receives at the start of step $a_i[r]$ the message $\langle 1\rangle$ from at least $n-f$ different nodes. Suppose there is $x \in \{0,1\}$ and a correct node $p_j \in \sP$ that receives at the start of step $a_j[r]$ the message $\langle x\rangle$ from at least $n-f$ different nodes, where $i=j$ may or may not hold. With a probability of at least $\min \{p_0, p_1\}$, the steps $a_i[r]$ and $a_j[r]$ assign the same value to $index_j$, and resp., $index_j$. 
%		
%		Also, the updated values are either $0$ or $(save+inc) \bmod I$, where the values of $save$ and $inc$ are determined in lines~\ref{ln:glbSvV},~\ref{ln:glbSvVONE} and~\ref{ln:saveBitGetsBotS}, and respectively, lines~\ref{ln:sNeqBot} and~\ref{ln:saveBot}.
\end{claim}
\renewcommand{\clmcnt}{\ref{thm:atLeastZeoVal}}
\begin{claimProof}
By Corollary~\ref{thm:vAvB}, we know that $x=1$. Note that $x$'s value is determined during step $a[r-1]$ and $rand$ is chosen at the start of step $a[r]$. Due to $rand$'s unpredictability (Section~\ref{sec:rcc}), $rand$ and $x$ are two independent values.
Thus, with a probability of at least $\min \{p_0, p_1\}$, all correct nodes update $index$ in the same manner, \ie to either $0$ or $save+inc$ (Claim~\ref{thm:atLeastOneValSame}), where $save$ and $inc$ are values determined by lines~\ref{ln:glbSvV},~\ref{ln:glbSvVONE} and~\ref{ln:saveBitGetsBotS}, and resp., lines~\ref{ln:sNeqBot} and~\ref{ln:saveBot}.
\end{claimProof}
\end{lemmaProof}

\FF

Lemma~\ref{thm:bmodKappa} shows that all correct nodes forever agree on their index values and simultaneously increment the index by one (modulo $I$) only when $clock(\kappa)=\kappa-1$ and $\mathit{ssbftCO}_i.\done()=1$. Lemma~\ref{thm:bmodKappa} uses the following notation. Let $R=c[0],c[1],\ldots,c[r],\ldots$ an unbounded synchronous execution of Algorithm~\ref{alg:ssbftIndexViaSimRelInc}, where $c[r]$ is the system state that immediately precedes the arrival of the $r$-th common pulse. Denote by $indices^{start}_r$ and $indices^{end}_r$ the sets of all $index_i: i \in \Correct$ values of correct nodes at $c[r]$, and resp., $c[r+1]$, \ie the beginning, and resp., the end of step $a[r]$. Note that, for all $r$ and $x \in \{start, end\}$, we have $indices^{x}_r \subseteq \{0, 1,\ldots, I -1\}$.

\begin{lemma}[Closure] 
\label{thm:bmodKappa}
Let $c[r] \in R$, such that $clock(\kappa) = \kappa-1$ at $c[r]$. Suppose $indices^{end}_r = \{v\neq\bot\}$. For every $c[r'] \in R:r'  \in \{r+1,r+\kappa \}$ it holds that $indices^{start}_{r'} = \{ v + x \bmod  I \}$ where $x$ is $1$ when $r'=r+\kappa$ and $\mathit{ssbftCO}.\done()=1$ and $0$ when $r'  \in \{r+1,r+\kappa-1 \}$ or $\mathit{ssbftCO}.\done()\neq 1$.
\end{lemma}
\renewcommand{\lemcnt}{\ref{thm:bmodKappa}}
\begin{lemmaProof}
For $r' = r + 1$ the lemma holds since, by definition, $\forall r'': indices^{end}_{r''} = indices^{start}_{r''+1}$. Also, for any system state between $c[r']:r' \in \{r+1,r +\kappa-1\}$, no correct node, $p_i \in \sP$, updates $index_i$ during the step, $a_i[r']$, since $clock(\kappa)\neq\kappa-1$ at $c[r']:r' \in \{r+1,r +\kappa-1\}$ and thus lines~\ref{ln:logicSavePlusSv} to~\ref{ln:logicRandSavePlusSv} are not executed, which are the only lines that update $index_i$.

It remains to show that all correct nodes, $p_i \in \sP$, update $index_i$ in the same way during the steps $a_i[r']:r'=r+\kappa$ that immediately follow $c[r']$. This is due to the agreement property of Algorithm~\ref{alg:ssbftMVC}, the arguments above about $c[r']:r' \in \{r+1,r +\kappa-1\}$ as well as Claim~\ref{thm:theSameV}.

\begin{claim}
\label{thm:theSameV}
$indices^{start}_{r+\kappa} = \{v\}:v\neq\bot$.
\end{claim}
\renewcommand{\clmcnt}{\ref{thm:theSameV}}
\begin{claimProof}
By the schedule (Figure~\ref{fig:schdule}) and the length of the scheduling cycle, $\kappa$, we know that Algorithm~\ref{alg:ssbftIndexViaSimRelInc} is not active between $c[r+1]$ and $c[r+\kappa-3]$, but it is active during steps $a[r+\kappa-3]$, $a[r+\kappa-2]$, $a[r+\kappa-1]$, and $a[r+\kappa]$. During the steps $a[r+\kappa-3]$, all correct nodes broadcast $\langle v \rangle$ (line~\ref{ln:clockAZero}). Thus, at the start of steps $a[r+\kappa-3]$, all correct nodes receive $\langle v \rangle$ at least $n-f$ times. Thus, during $a[r+\kappa-2]$, all correct nodes assign $v$ to their $propose$ variables (line~\ref{ln:existsVNeqBotVBull}) and broadcast  $\langle v \rangle$ (line~\ref{ln:broadcastPropose}). By similar arguments, during $a[r+\kappa-1]$, all correct nodes assign $v$ and $1$ to their $save$, and resp., $bit$ variables (lines~\ref{ln:sNeqBot} to~\ref{ln:saveNeqBot}) and broadcast $\langle 1 \rangle$ (line~\ref{ln:saveBotX}). Therefore, all correct nodes receive $\langle 1 \rangle$ at least $n - f$ times. This implies that during $a[r+\kappa]$, the if-statement condition in line~\ref{ln:logicSavePlusSv} holds and thus $indices^{start}_{r+\kappa} = \{v\neq\bot\}$ holds.
\end{claimProof}
\end{lemmaProof}
\end{theoremProof}

\Section{Discussion}
\label{sec:conclusions}
We have presented a new loosely-self-stabilizing variation of the MMR algorithm~\cite{DBLP:conf/podc/MostefaouiMR14} for solving binary consensus \ems{for the $\mathsf{BAMP_{n,t}[\mathit{-}FC, t < n/3,RCCs]}$ model.} The proposed solution preserves the following properties of the studied algorithm: it does not require signatures, %synchrony assumptions are not used, 
it offers optimal fault-tolerance, and the expected time until completion is the same as the studied algorithm.
%
% It also to some extent preserves the expected communication costs. 
%
The proposed solution is able to achieve this using a new application of the design criteria of loosely-self-stabilizing systems, which requires the satisfaction of safety properties with a probability in $\bigO(1-2^{-M})$. For any practical purposes and in the absence of transient-faults, one can select $M$ to be sufficiently large so that the risk of violating safety is negligible. \ems{An SSBFT solution for recycling distributed objects and $\mathsf{BSMP_{n,t}[\kappa\mathit{-}SGC,t < n/3,RCCs]}$ is proposed in order to support an unbounded number of instances of our SSBFT MMR solution.} We believe that this work is preparing the groundwork needed to construct self-stabilizing (BFT) algorithms for distributed systems, such as Blockchains, that need to run in \ems{hostile environments.}

\subsection*{Acknowledgments}
The work of M. Raynal was partially supported by the French ANR project DESCARTES (16-CE40-0023-03). The work of I. Marcoullis was funded by the ONISILLOS postdoctoral funding scheme of the University of Cyprus.

\begin{table*}[t!]
	\begin{center}
		\begin{\algSizeVerySmall}
			\begin{tabular}{|l|l|}
				\hline
				\textbf{Notation} & \textbf{Meaning} \\ \hline \hline
				MMR & Mostéfaoui, Moumen, and Raynal ~\cite{DBLP:conf/podc/MostefaouiMR14}        \\ \hline
				BDH &     Ben{-}Or, Dolev, and Hoch~\cite{DBLP:conf/podc/Ben-OrDH08}    \\ \hline
				BFT &    non-self-stabilizing Byzantine fault-tolerant solutions     \\ \hline
				SSBFT &    self-stabilizing Byzantine fault-tolerant     \\ \hline
				$\mathsf{BAMP_{n,t}}$ &    Byzantine Asynchronous Message-Passing model    \\ \hline
				$\mathsf{BSMP_{n,t}}$ &     Byzantine synchronous message-passing model \\ \hline
				$RCCs$&  random common coins   \\ \hline
				$FC$& fair communication assumption        \\ \hline
				$\kappa\mathit{-}SGC$&    $\kappa$-state global clock     \\ \hline
			\end{tabular}
			\caption{\label{fig:Glossary}Glossary}
		\end{\algSizeVerySmall}
	\end{center}
\end{table*}

%\bibliographystyle{plain}
%\bibliography{referances}	

\begin{thebibliography}{100}
	
	\bibitem{DBLP:conf/stoc/AbrahamD15}
	Ittai Abraham and Danny Dolev.
	\newblock {Byzantine} agreement with optimal early stopping, optimal resilience
	and polynomial complexity.
	\newblock In {\em {STOC}}, pages 605--614. {ACM}, 2015.
	
	\bibitem{DBLP:journals/jcss/AlonADDPT15}
	Noga Alon, Hagit Attiya, Shlomi Dolev, Swan Dubois, Maria Potop{-}Butucaru, and
	S{\'{e}}bastien Tixeuil.
	\newblock Practically stabilizing {SWMR} atomic memory in message-passing
	systems.
	\newblock {\em J. Comput. Syst. Sci.}, 81(4):692--701, 2015.
	
	\bibitem{DBLP:series/synthesis/2019Altisen}
	Karine Altisen, St{\'{e}}phane Devismes, Swan Dubois, and Franck Petit.
	\newblock {\em Introduction to Distributed Self-Stabilizing Algorithms}.
	\newblock Synthesis Lectures on Distributed Computing Theory. Morgan {\&}
	Claypool Publishers, 2019.
	
	\bibitem{DBLP:conf/wdag/AnagnostouH93}
	Efthymios Anagnostou and Vassos Hadzilacos.
	\newblock Tolerating transient and permanent failures (extended abstract).
	\newblock In {\em Distributed Algorithms, 7th International Workshop, {WDAG}},
	volume 725 of {\em Lecture Notes in Computer Science}, pages 174--188.
	Springer, 1993.
	
	\bibitem{DBLP:conf/sss/AshkenaziDKKOW21}
	Yotam Ashkenazi, Shlomi Dolev, Sayaka Kamei, Yoshiaki Katayama, Fukuhito
	Ooshita, and Koichi Wada.
	\newblock Location functions for self-stabilizing byzantine tolerant swarms.
	\newblock In {\em {SSS}}, volume 13046 of {\em Lecture Notes in Computer
		Science}, pages 229--242. Springer, 2021.
	
	\bibitem{DBLP:conf/ic-nc/AshkenaziDKOW19}
	Yotam Ashkenazi, Shlomi Dolev, Sayaka Kamei, Fukuhito Ooshita, and Koichi Wada.
	\newblock Forgive {\&} forget: Self-stabilizing swarms in spite of byzantine
	robots.
	\newblock In {\em {CANDAR} Workshops}, pages 188--194. {IEEE}, 2019.
	
	\bibitem{DBLP:journals/jacm/Aspnes98}
	James Aspnes.
	\newblock Lower bounds for distributed coin-flipping and randomized consensus.
	\newblock {\em J. {ACM}}, 45(3):415--450, 1998.
	
	\bibitem{DBLP:journals/ijsysc/BeauquierK97}
	Joffroy Beauquier and Synn{\"{o}}ve Kekkonen{-}Moneta.
	\newblock Fault-tolerance and self-stabilization: impossibility results and
	solutions using self-stabilizing failure detectors.
	\newblock {\em Int. J. Systems Science}, 28(11):1177--1187, 1997.
	
	\bibitem{DBLP:conf/podc/Ben-Or83}
	Michael Ben{-}Or.
	\newblock Another advantage of free choice: Completely asynchronous agreement
	protocols (extended abstract).
	\newblock In {\em Proceedings of the Second Annual {ACM} {SIGACT-SIGOPS}
		Symposium on Principles of Distributed Computing}, pages 27--30, 1983.
	
	\bibitem{DBLP:conf/podc/Ben-OrDH08}
	Michael Ben{-}Or, Danny Dolev, and Ezra~N. Hoch.
	\newblock Fast self-stabilizing {Byzantine} tolerant digital clock
	synchronization.
	\newblock In {\em Proceedings of the Twenty-Seventh Annual {ACM} Symposium on
		Principles of Distributed Computing, {PODC} 2008, Toronto, Canada, August
		18-21, 2008}, pages 385--394. {ACM}, 2008.
	
	\bibitem{DBLP:conf/dsn/BessaniSA14}
	Alysson~Neves Bessani, Jo{\~{a}}o Sousa, and Eduardo Ad{\'{\i}}lio~Pelinson
	Alchieri.
	\newblock State machine replication for the masses with {BFT-SMART}.
	\newblock In {\em 44th Annual {IEEE/IFIP} International Conference on
		Dependable Systems and Networks, {DSN}}, pages 355--362. {IEEE} Computer
	Society, 2014.
	
	\bibitem{DBLP:conf/sss/BinunCDKLPYY16}
	Alexander Binun, Thierry Coupaye, Shlomi Dolev, Mohammed Kassi{-}Lahlou, Marc
	Lacoste, Alex Palesandro, Reuven Yagel, and Leonid Yankulin.
	\newblock Self-stabilizing {Byzantine}-tolerant distributed replicated state
	machine.
	\newblock In {\em Stabilization, Safety, and Security of Distributed Systems -
		18th International Symposium, {SSS}}, pages 36--53, 2016.
	
	\bibitem{DBLP:conf/cscml/BinunDH19}
	Alexander Binun, Shlomi Dolev, and Tal Hadad.
	\newblock Self-stabilizing {Byzantine} consensus for blockchain - (brief
	announcement).
	\newblock In {\em Cyber Security Cryptography and Machine Learning - Third
		International Symposium, {CSCML}}, pages 106--110, 2019.
	
	\bibitem{DBLP:conf/podc/BonomiDPR15}
	Silvia Bonomi, Shlomi Dolev, Maria Potop{-}Butucaru, and Michel Raynal.
	\newblock Stabilizing server-based storage in {Byzantine} asynchronous
	message-passing systems: Extended abstract.
	\newblock In {\em Proceedings of the 2015 {ACM} Symposium on Principles of
		Distributed Computing, {PODC}}, pages 471--479, 2015.
	
	\bibitem{DBLP:conf/ipps/BonomiPT15}
	Silvia Bonomi, Maria Potop{-}Butucaru, and S{\'{e}}bastien Tixeuil.
	\newblock Stabilizing {Byzantine}-fault tolerant storage.
	\newblock In {\em 2015 {IEEE} International Parallel and Distributed Processing
		Symposium, {IPDPS}}, pages 894--903, 2015.
	
	\bibitem{DBLP:conf/icdcn/BonomiPP16}
	Silvia Bonomi, Antonella~Del Pozzo, and Maria Potop{-}Butucaru.
	\newblock Tight self-stabilizing mobile byzantine-tolerant atomic register.
	\newblock In {\em {ICDCN}}, pages 6:1--6:10. {ACM}, 2016.
	
	\bibitem{DBLP:journals/tcs/BonomiPP18}
	Silvia Bonomi, Antonella~Del Pozzo, and Maria Potop{-}Butucaru.
	\newblock Optimal self-stabilizing synchronous mobile byzantine-tolerant atomic
	register.
	\newblock {\em Theor. Comput. Sci.}, 709:64--79, 2018.
	
	\bibitem{DBLP:conf/podc/BonomiPPT16}
	Silvia Bonomi, Antonella~Del Pozzo, Maria Potop{-}Butucaru, and S{\'{e}}bastien
	Tixeuil.
	\newblock Optimal mobile {Byzantine} fault tolerant distributed storage:
	Extended abstract.
	\newblock In {\em Proceedings of the 2016 {ACM} Symposium on Principles of
		Distributed Computing, {PODC}}, pages 269--278, 2016.
	
	\bibitem{DBLP:conf/srds/BonomiPPT17}
	Silvia Bonomi, Antonella~Del Pozzo, Maria Potop{-}Butucaru, and S{\'{e}}bastien
	Tixeuil.
	\newblock Optimal storage under unsynchronized mobile {Byzantine} faults.
	\newblock In {\em 36th {IEEE} Symposium on Reliable Distributed Systems,
		{SRDS}}, pages 154--163, 2017.
	
	\bibitem{DBLP:conf/sss/BonomiPPT18}
	Silvia Bonomi, Antonella~Del Pozzo, Maria Potop{-}Butucaru, and S{\'{e}}bastien
	Tixeuil.
	\newblock Brief announcement: Optimal self-stabilizing mobile
	{Byzantine}-tolerant regular register with bounded timestamps.
	\newblock In {\em Stabilization, Safety, and Security of Distributed Systems -
		20th International Symposium, {SSS}}, pages 398--403, 2018.
	
	\bibitem{DBLP:journals/tcs/BonomiPPT19}
	Silvia Bonomi, Antonella~Del Pozzo, Maria Potop{-}Butucaru, and S{\'{e}}bastien
	Tixeuil.
	\newblock Approximate agreement under mobile {Byzantine} faults.
	\newblock {\em Theor. Comput. Sci.}, 758:17--29, 2019.
	
	\bibitem{Bracha1987}
	Gabriel Bracha.
	\newblock Asynchronous byzantine agreement protocols.
	\newblock {\em Inf. Comput.}, 75(2):130--143, 1987.
	
	\bibitem{DBLP:books/daglib/0025983}
	Christian Cachin, Rachid Guerraoui, and Lu{\'{\i}}s E.~T. Rodrigues.
	\newblock {\em Introduction to Reliable and Secure Distributed Programming
		{(2.} ed.)}.
	\newblock Springer, 2011.
	
	\bibitem{Cachin2001}
	Christian Cachin, Klaus Kursawe, Frank Petzold, and Victor Shoup.
	\newblock Secure and efficient asynchronous broadcast protocols.
	\newblock {\em {IACR} Cryptol. ePrint Arch.}, 2001:6, 2001.
	
	\bibitem{DBLP:journals/joc/CachinKS05}
	Christian Cachin, Klaus Kursawe, and Victor Shoup.
	\newblock Random oracles in constantinople: Practical asynchronous {Byzantine}
	agreement using cryptography.
	\newblock {\em J. Cryptol.}, 18(3):219--246, 2005.
	
	\bibitem{DBLP:conf/wdag/CachinV17}
	Christian Cachin and Marko Vukolic.
	\newblock Blockchain consensus protocols in the wild (keynote talk).
	\newblock In {\em 31st International Symposium on Distributed Computing,
		{DISC}}, volume~91 of {\em LIPIcs}, pages 1:1--1:16. Schloss Dagstuhl -
	Leibniz-Zentrum f{\"{u}}r Informatik, 2017.
	
	\bibitem{DBLP:conf/wdag/CachinZ21}
	Christian Cachin and Luca Zanolini.
	\newblock Brief announcement: Revisiting signature-free asynchronous byzantine
	consensus.
	\newblock In {\em {DISC}}, volume 209 of {\em LIPIcs}, pages 51:1--51:4.
	Schloss Dagstuhl - Leibniz-Zentrum f{\"{u}}r Informatik, 2021.
	
	\bibitem{DBLP:conf/stoc/CanettiR93}
	Ran Canetti and Tal Rabin.
	\newblock Fast asynchronous byzantine agreement with optimal resilience.
	\newblock In {\em Proceedings of the Twenty-Fifth Annual {ACM} Symposium on
		Theory of Computing}, pages 42--51. {ACM}, 1993.
	
	\bibitem{DBLP:journals/tocs/CastroL02}
	Miguel Castro and Barbara Liskov.
	\newblock Practical {Byzantine} fault tolerance and proactive recovery.
	\newblock {\em {ACM} Trans. Comput. Syst.}, 20(4):398--461, 2002.
	
	\bibitem{DBLP:journals/jacm/ChandraT96}
	Tushar~Deepak Chandra and Sam Toueg.
	\newblock Unreliable failure detectors for reliable distributed systems.
	\newblock {\em J. {ACM}}, 43(2):225--267, 1996.
	
	\bibitem{DBLP:journals/cj/CorreiaNV06}
	Miguel Correia, Nuno~Ferreira Neves, and Paulo Ver{\'{\i}}ssimo.
	\newblock From consensus to atomic broadcast: Time-free {Byzantine}-resistant
	protocols without signatures.
	\newblock {\em Comput. J.}, 49(1):82--96, 2006.
	
	\bibitem{DBLP:journals/ijccbs/CorreiaVNV11}
	Miguel Correia, Giuliana~Santos Veronese, Nuno~Ferreira Neves, and Paulo
	Ver{\'{\i}}ssimo.
	\newblock {Byzantine} consensus in asynchronous message-passing systems: a
	survey.
	\newblock {\em Int. J. Crit. Comput. Based Syst.}, 2(2):141--161, 2011.
	
	\bibitem{DBLP:conf/podc/DaliotD06}
	Ariel Daliot and Danny Dolev.
	\newblock Self-stabilizing byzantine agreement.
	\newblock In {\em {PODC}}, pages 143--152. {ACM}, 2006.
	
	\bibitem{DBLP:conf/podc/DaliotDP04}
	Ariel Daliot, Danny Dolev, and Hanna Parnas.
	\newblock Brief announcement: linear time byzantine self-stabilizing clock
	synchronization.
	\newblock In {\em {PODC}}, page 379. {ACM}, 2004.
	
	\bibitem{DBLP:journals/dc/DefagoPP20}
	Xavier D{\'{e}}fago, Maria Potop{-}Butucaru, and Philippe~Raipin Parv{\'{e}}dy.
	\newblock Self-stabilizing gathering of mobile robots under crash or byzantine
	faults.
	\newblock {\em Distributed Comput.}, 33(5):393--421, 2020.
	
	\bibitem{DBLP:journals/corr/DefagoP0MPP16}
	Xavier D{\'{e}}fago, Maria~Gradinariu Potop{-}Butucaru, Julien Cl{\'{e}}ment,
	St{\'{e}}phane Messika, and Philippe~Raipin Parv{\'{e}}dy.
	\newblock Fault and byzantine tolerant self-stabilizing mobile robots gathering
	- feasibility study -.
	\newblock {\em CoRR}, abs/1602.05546, 2016.
	
	\bibitem{DBLP:journals/cacm/Dijkstra74}
	Edsger~W. Dijkstra.
	\newblock Self-stabilizing systems in spite of distributed control.
	\newblock {\em Commun. {ACM}}, 17(11):643--644, 1974.
	
	\bibitem{DBLP:conf/sss/DolevH07}
	Danny Dolev and Ezra~N. Hoch.
	\newblock Byzantine self-stabilizing pulse in a bounded-delay model.
	\newblock In {\em {SSS}}, volume 4838 of {\em Lecture Notes in Computer
		Science}, pages 234--252. Springer, 2007.
	
	\bibitem{DBLP:conf/wdag/DolevH07}
	Danny Dolev and Ezra~N. Hoch.
	\newblock On self-stabilizing synchronous actions despite byzantine attacks.
	\newblock In {\em {DISC}}, volume 4731 of {\em Lecture Notes in Computer
		Science}, pages 193--207. Springer, 2007.
	
	\bibitem{DBLP:conf/opodis/DolevHR07}
	Danny Dolev, Ezra~N. Hoch, and Robbert van Renesse.
	\newblock Self-stabilizing and byzantine-tolerant overlay network.
	\newblock In {\em {OPODIS}}, volume 4878 of {\em Lecture Notes in Computer
		Science}, pages 343--357. Springer, 2007.
	
	\bibitem{DBLP:conf/stoc/DolevS82}
	Danny Dolev and H.~Raymond Strong.
	\newblock Polynomial algorithms for multiple processor agreement.
	\newblock In {\em {STOC}}, pages 401--407. {ACM}, 1982.
	
	\bibitem{DBLP:books/mit/Dolev2000}
	Shlomi Dolev.
	\newblock {\em Self-Stabilization}.
	\newblock {MIT} Press, 2000.
	
	\bibitem{DBLP:journals/ipl/DolevDPT11}
	Shlomi Dolev, Swan Dubois, Maria Potop{-}Butucaru, and S{\'{e}}bastien Tixeuil.
	\newblock Stabilizing data-link over non-fifo channels with optimal
	fault-resilience.
	\newblock {\em Inf. Process. Lett.}, 111(18):912--920, 2011.
	
	\bibitem{DBLP:journals/jcss/DolevGMS18}
	Shlomi Dolev, Chryssis Georgiou, Ioannis Marcoullis, and Elad~Michael Schiller.
	\newblock Practically-self-stabilizing virtual synchrony.
	\newblock {\em J. Comput. Syst. Sci.}, 96:50--73, 2018.
	
	\bibitem{DBLP:conf/cscml/DolevGMS18}
	Shlomi Dolev, Chryssis Georgiou, Ioannis Marcoullis, and Elad~Michael Schiller.
	\newblock Self-stabilizing {Byzantine} tolerant replicated state machine based
	on failure detectors.
	\newblock In {\em Cyber Security Cryptography and Machine Learning - Second
		International Symposium, {CSCML}}, pages 84--100, 2018.
	
	\bibitem{DBLP:conf/sss/DolevHSS12}
	Shlomi Dolev, Ariel Hanemann, Elad~Michael Schiller, and Shantanu Sharma.
	\newblock Self-stabilizing end-to-end communication in (bounded capacity,
	omitting, duplicating and non-fifo) dynamic networks - (extended abstract).
	\newblock In {\em {SSS}}, volume 7596 of {\em LNCS}, pages 133--147. Springer,
	2012.
	
	\bibitem{DBLP:conf/netys/DolevLS13}
	Shlomi Dolev, Omri Liba, and Elad~Michael Schiller.
	\newblock Self-stabilizing {Byzantine} resilient topology discovery and message
	delivery - (extended abstract).
	\newblock In {\em Networked Systems - First International Conference, {NETYS}},
	pages 42--57, 2013.
	
	\bibitem{DBLP:conf/podc/DolevPS15}
	Shlomi Dolev, Thomas Petig, and Elad~Michael Schiller.
	\newblock Brief announcement: Robust and private distributed shared atomic
	memory in message passing networks.
	\newblock In {\em Proceedings of the 2015 {ACM} Symposium on Principles of
		Distributed Computing, {PODC}}, pages 311--313, 2015.
	
	\bibitem{DBLP:journals/corr/abs-1806-03498}
	Shlomi Dolev, Thomas Petig, and Elad~Michael Schiller.
	\newblock Self-stabilizing and private distributed shared atomic memory in
	seldomly fair message passing networks.
	\newblock {\em CoRR}, abs/1806.03498, 2018.
	\newblock Also to appear in Springer's Algorithmica.
	
	\bibitem{DBLP:journals/tpds/DolevS03}
	Shlomi Dolev and Elad Schiller.
	\newblock Communication adaptive self-stabilizing group membership service.
	\newblock {\em {IEEE} Trans. Parallel Distributed Syst.}, 14(7):709--720, 2003.
	
	\bibitem{DBLP:conf/podc/DolevW95}
	Shlomi Dolev and Jennifer~L. Welch.
	\newblock Self-stabilizing clock synchronization in the presence of {Byzantine}
	faults (abstract).
	\newblock In {\em Proceedings of the Fourteenth Annual {ACM} Symposium on
		Principles of Distributed Computing}, page 256, 1995.
	
	\bibitem{DBLP:journals/jacm/DolevW04}
	Shlomi Dolev and Jennifer~L. Welch.
	\newblock Self-stabilizing clock synchronization in the presence of byzantine
	faults.
	\newblock {\em J. {ACM}}, 51(5):780--799, 2004.
	
	\bibitem{DBLP:conf/sss/DongSIM21}
	Rongcheng Dong, Yuichi Sudo, Taisuke Izumi, and Toshimitsu Masuzawa.
	\newblock Loosely-stabilizing maximal independent set algorithms with
	unreliable communications.
	\newblock In {\em {SSS}}, volume 13046 of {\em Lecture Notes in Computer
		Science}, pages 335--349. Springer, 2021.
	
	\bibitem{BEAT2018}
	Sisi Duan, Michael~K. Reiter, and Haibin Zhang.
	\newblock {BEAT:} asynchronous {BFT} made practical.
	\newblock In {\em Proceedings of the 2018 {ACM} {SIGSAC} Conference on Computer
		and Communications Security, {CCS}}, pages 2028--2041. {ACM}, 2018.
	
	\bibitem{DBLP:journals/jpdc/DuboisPNT12}
	Swan Dubois, Maria Potop{-}Butucaru, Mikhail Nesterenko, and S{\'{e}}bastien
	Tixeuil.
	\newblock Self-stabilizing {Byzantine} asynchronous unison.
	\newblock {\em J. Parallel Distributed Comput.}, 72(7):917--923, 2012.
	
	\bibitem{DBLP:journals/tcs/DuboisPT11}
	Swan Dubois, Maria Potop{-}Butucaru, and S{\'{e}}bastien Tixeuil.
	\newblock Dynamic {FTSS} in asynchronous systems: The case of unison.
	\newblock {\em Theor. Comput. Sci.}, 412(29):3418--3439, 2011.
	
	\bibitem{DBLP:journals/corr/abs-2201-12880}
	Romaric Duvignau, Michel Raynal, and Elad~Michael Schiller.
	\newblock Self-stabilizing byzantine-tolerant broadcast.
	\newblock {\em CoRR}, abs/2201.12880, 2022.
	
	\bibitem{DBLP:conf/stoc/FeldmanM88}
	Paul Feldman and Silvio Micali.
	\newblock Optimal algorithms for byzantine agreement.
	\newblock In Janos Simon, editor, {\em Proceedings of the 20th Annual {ACM}
		Symposium on Theory of Computing, May 2-4, 1988, Chicago, Illinois, {USA}},
	pages 148--161. {ACM}, 1988.
	
	\bibitem{DBLP:conf/icalp/FeldmanM89}
	Paul Feldman and Silvio Micali.
	\newblock An optimal probabilistic algorithm for synchronous byzantine
	agreement.
	\newblock In {\em Automata, Languages and Programming, 16th International
		Colloquium, ICALP}, volume 372 of {\em Lecture Notes in Computer Science},
	pages 341--378. Springer, 1989.
	
	\bibitem{DBLP:conf/sss/0001GS19}
	Michael Feldmann, Thorsten G{\"{o}}tte, and Christian Scheideler.
	\newblock A loosely self-stabilizing protocol for randomized congestion control
	with logarithmic memory.
	\newblock In {\em Stabilization, Safety, and Security of Distributed Systems -
		21st International Symposium, {SSS}}, volume 11914 of {\em Lecture Notes in
		Computer Science}, pages 149--164. Springer, 2019.
	
	\bibitem{DBLP:journals/ipl/FischerL82}
	Michael~J. Fischer and Nancy~A. Lynch.
	\newblock A lower bound for the time to assure interactive consistency.
	\newblock {\em Inf. Process. Lett.}, 14(4):183--186, 1982.
	
	\bibitem{DBLP:journals/jacm/FischerLP85}
	Michael~J. Fischer, Nancy~A. Lynch, and Mike Paterson.
	\newblock Impossibility of distributed consensus with one faulty process.
	\newblock {\em J. {ACM}}, 32(2):374--382, 1985.
	
	\bibitem{DBLP:journals/siamcomp/GarayM98}
	Juan~A. Garay and Yoram Moses.
	\newblock Fully polynomial byzantine agreement for \emph{n} {\textgreater} 3t
	processors in \emph{t} + 1 rounds.
	\newblock {\em {SIAM} J. Comput.}, 27(1):247--290, 1998.
	
	\bibitem{DBLP:journals/corr/abs-1807-07901}
	Chryssis Georgiou, Robert Gustafsson, Andreas Lindhe, and Elad~Michael
	Schiller.
	\newblock Self-stabilization overhead: an experimental case study on coded
	atomic storage.
	\newblock {\em CoRR}, abs/1807.07901, 2018.
	
	\bibitem{DBLP:conf/netys/GeorgiouGLS19}
	Chryssis Georgiou, Robert Gustafsson, Andreas Lindh{\'{e}}, and Elad~Michael
	Schiller.
	\newblock Self-stabilization overhead: {A} case study on coded atomic storage.
	\newblock In {\em Networked Systems - 7th International Conference, {NETYS}},
	pages 131--147, 2019.
	
	\bibitem{DBLP:conf/netys/GeorgiouLS19}
	Chryssis Georgiou, Oskar Lundstr{\"{o}}m, and Elad~Michael Schiller.
	\newblock Self-stabilizing snapshot objects for asynchronous failure-prone
	networked systems.
	\newblock In {\em Networked Systems - 7th International Conference, {NETYS}
		2019, Marrakech, Morocco, June 19-21, 2019, Revised Selected Papers}, pages
	113--130, 2019.
	
	\bibitem{DBLP:conf/sss/HochDD06}
	Ezra~N. Hoch, Danny Dolev, and Ariel Daliot.
	\newblock Self-stabilizing byzantine digital clock synchronization.
	\newblock In {\em {SSS}}, volume 4280 of {\em Lecture Notes in Computer
		Science}, pages 350--362. Springer, 2006.
	
	\bibitem{DBLP:conf/sirocco/Izumi15}
	Taisuke Izumi.
	\newblock On space and time complexity of loosely-stabilizing leader election.
	\newblock In Christian Scheideler, editor, {\em Structural Information and
		Communication Complexity - 22nd International Colloquium, {SIROCCO}}, volume
	9439 of {\em Lecture Notes in Computer Science}, pages 299--312. Springer,
	2015.
	
	\bibitem{DBLP:journals/ipl/KeidarR03}
	Idit Keidar and Sergio Rajsbaum.
	\newblock A simple proof of the uniform consensus synchronous lower bound.
	\newblock {\em Inf. Process. Lett.}, 85(1):47--52, 2003.
	
	\bibitem{DBLP:journals/mst/KhanchandaniL19}
	Pankaj Khanchandani and Christoph Lenzen.
	\newblock Self-stabilizing {Byzantine} clock synchronization with optimal
	precision.
	\newblock {\em Theory Comput. Syst.}, 63(2):261--305, 2019.
	
	\bibitem{DBLP:conf/podc/KowalskiM13}
	Dariusz~R. Kowalski and Achour Most{\'{e}}faoui.
	\newblock Synchronous {Byzantine} agreement with nearly a cubic number of
	communication bits: synchronous byzantine agreement with nearly a cubic
	number of communication bits.
	\newblock In {\em {PODC}}, pages 84--91. {ACM}, 2013.
	
	\bibitem{DBLP:journals/tocs/Lamport98}
	Leslie Lamport.
	\newblock The part-time parliament.
	\newblock {\em {ACM} Trans. Comput. Syst.}, 16(2):133--169, 1998.
	
	\bibitem{DBLP:conf/wdag/Lamport11a}
	Leslie Lamport.
	\newblock Byzantizing paxos by refinement.
	\newblock In David Peleg, editor, {\em Distributed Computing - 25th
		International Symposium, {DISC}}, volume 6950 of {\em Lecture Notes in
		Computer Science}, pages 211--224. Springer, 2011.
	
	\bibitem{lamport2001paxos}
	Leslie Lamport et~al.
	\newblock Paxos made simple.
	\newblock {\em ACM Sigact News}, 32(4):18--25, 2001.
	
	\bibitem{DBLP:journals/toplas/LamportSP82}
	Leslie Lamport, Robert~E. Shostak, and Marshall~C. Pease.
	\newblock The {Byzantine} generals problem.
	\newblock {\em {ACM} Trans. Program. Lang. Syst.}, 4(3):382--401, 1982.
	
	\bibitem{DBLP:journals/jacm/LenzenR19}
	Christoph Lenzen and Joel Rybicki.
	\newblock Self-stabilising {Byzantine} clock synchronisation is almost as easy
	as consensus.
	\newblock {\em J. {ACM}}, 66(5):32:1--32:56, 2019.
	
	\bibitem{DBLP:conf/icdcs/LundstromRS20}
	Oskar Lundstr{\"{o}}m, Michel Raynal, and Elad~Michael Schiller.
	\newblock Self-stabilizing set-constrained delivery broadcast (extended
	abstract).
	\newblock In {\em 40th {IEEE} International Conference on Distributed Computing
		Systems, {ICDCS}}, pages 617--627, 2020.
	
	\bibitem{DBLP:conf/netys/LundstromRS20}
	Oskar Lundstr{\"{o}}m, Michel Raynal, and Elad~Michael Schiller.
	\newblock Self-stabilizing uniform reliable broadcast.
	\newblock In {\em Networked Systems - 8th International Conference, {NETYS}},
	pages 296--313, 2020.
	
	\bibitem{DBLP:conf/icdcn/LundstromRS21}
	Oskar Lundstr{\"{o}}m, Michel Raynal, and Elad~Michael Schiller.
	\newblock Self-stabilizing indulgent zero-degrading binary consensus.
	\newblock In {\em {ICDCN} '21: International Conference on Distributed
		Computing and Networking}, pages 106--115, 2021.
	
	\bibitem{DBLP:journals/corr/abs-2104-03129}
	Oskar Lundstr{\"{o}}m, Michel Raynal, and Elad~Michael Schiller.
	\newblock Self-stabilizing multivalued consensus in asynchronous crash-prone
	systems.
	\newblock {\em CoRR}, abs/2104.03129, 2021.
	
	\bibitem{DBLP:conf/sss/Malekpour06}
	Mahyar~R. Malekpour.
	\newblock A byzantine-fault tolerant self-stabilizing protocol for distributed
	clock synchronization systems.
	\newblock In {\em {SSS}}, volume 4280 of {\em Lecture Notes in Computer
		Science}, pages 411--427. Springer, 2006.
	
	\bibitem{DBLP:conf/opodis/MasuzawaT05}
	Toshimitsu Masuzawa and S{\'{e}}bastien Tixeuil.
	\newblock A self-stabilizing link-coloring protocol resilient to unbounded
	byzantine faults in arbitrary networks.
	\newblock In {\em {OPODIS}}, volume 3974 of {\em Lecture Notes in Computer
		Science}, pages 118--129. Springer, 2005.
	
	\bibitem{DBLP:conf/opodis/Maurer20}
	Alexandre Maurer.
	\newblock Self-stabilizing {Byzantine}-resilient communication in dynamic
	networks.
	\newblock In {\em {OPODIS}}, volume 184 of {\em LIPIcs}, pages 27:1--27:11.
	Schloss Dagstuhl - Leibniz-Zentrum f{\"{u}}r Informatik, 2020.
	
	\bibitem{DBLP:conf/srds/MaurerT14}
	Alexandre Maurer and S{\'{e}}bastien Tixeuil.
	\newblock Self-stabilizing {Byzantine} broadcast.
	\newblock In {\em 33rd {IEEE} International Symposium on Reliable Distributed
		Systems, {SRDS} 2014, Nara, Japan, October 6-9, 2014}, pages 152--160. {IEEE}
	Computer Society, 2014.
	
	\bibitem{DBLP:conf/ccs/MillerXCSS16}
	Andrew Miller, Yu~Xia, Kyle Croman, Elaine Shi, and Dawn Song.
	\newblock The honey badger of {BFT} protocols.
	\newblock In {\em {CCS}}, pages 31--42. {ACM}, 2016.
	
	\bibitem{DBLP:journals/tdsc/MonizNCV11}
	Henrique Moniz, Nuno~Ferreira Neves, Miguel Correia, and Paulo
	Ver{\'{\i}}ssimo.
	\newblock {RITAS:} services for randomized intrusion tolerance.
	\newblock {\em {IEEE} Trans. Dependable Secur. Comput.}, 8(1):122--136, 2011.
	
	\bibitem{DBLP:conf/podc/MostefaouiMR14}
	Achour Most{\'{e}}faoui, Moumen Hamouma, and Michel Raynal.
	\newblock Signature-free asynchronous {Byzantine} consensus with t{\textless}
	n/3, {O}(n\({}^{\mbox{2}}\)) messages.
	\newblock In {\em {ACM} Symposium on Principles of Distributed Computing,
		{PODC}}, pages 2--9, 2014.
	
	\bibitem{DBLP:journals/jacm/MostefaouiMR15}
	Achour Most{\'{e}}faoui, Hamouma Moumen, and Michel Raynal.
	\newblock Signature-free asynchronous binary {Byzantine} consensus with
	t{\textless} n/3, {O}(n\({}^{\mbox{2}}\)) expected time.
	\newblock {\em J. {ACM}}, 62(4):31:1--31:21, 2015.
	
	\bibitem{DBLP:journals/acta/MostefaouiR17}
	Achour Most{\'{e}}faoui and Michel Raynal.
	\newblock Signature-free asynchronous {Byzantine} systems: from multivalued to
	binary consensus with t{\textless} n/3, {O}(n\({}^{\mbox{2}}\)) messages, and
	constant time.
	\newblock {\em Acta Informatica}, 54(5):501--520, 2017.
	
	\bibitem{DBLP:conf/eurocrypt/NaorPR99}
	Moni Naor, Benny Pinkas, and Omer Reingold.
	\newblock Distributed pseudo-random functions and kdcs.
	\newblock In Jacques Stern, editor, {\em Advances in Cryptology - {EUROCRYPT}
		'99, International Conference on the Theory and Application of Cryptographic
		Techniques}, volume 1592 of {\em Lecture Notes in Computer Science}, pages
	327--346. Springer, 1999.
	
	\bibitem{DBLP:journals/tpds/NesterenkoT09}
	Mikhail Nesterenko and S{\'{e}}bastien Tixeuil.
	\newblock Discovering network topology in the presence of {Byzantine} faults.
	\newblock {\em {IEEE} Trans. Parallel Distributed Syst.}, 20(12):1777--1789,
	2009.
	
	\bibitem{DBLP:journals/jacm/PeaseSL80}
	Marshall~C. Pease, Robert~E. Shostak, and Leslie Lamport.
	\newblock Reaching agreement in the presence of faults.
	\newblock {\em J. {ACM}}, 27(2):228--234, 1980.
	
	\bibitem{perner2013byzantine}
	Martin Perner, Martin Sigl, Ulrich Schmid, and Christoph Lenzen.
	\newblock Byzantine self-stabilizing clock distribution with hex:
	Implementation, simulation, clock multiplication.
	\newblock In {\em 6th Conference on Dependability (DEPEND)}. Citeseer, 2013.
	
	\bibitem{DBLP:conf/ftcs/Powell92}
	David Powell.
	\newblock Failure mode assumptions and assumption coverage.
	\newblock In {\em Digest of Papers: FTCS-22, The Twenty-Second Annual
		International Symposium on Fault-Tolerant Computing}, pages 386--395, 1992.
	
	\bibitem{DBLP:conf/focs/Rabin83}
	Michael~O. Rabin.
	\newblock Randomized {Byzantine} generals.
	\newblock In {\em 24th Annual Symposium on Foundations of Computer Science},
	pages 403--409, 1983.
	
	\bibitem{DBLP:books/sp/Raynal18}
	Michel Raynal.
	\newblock {\em Fault-Tolerant Message-Passing Distributed Systems - An
		Algorithmic Approach}.
	\newblock Springer, 2018.
	
	\bibitem{DBLP:journals/tkde/RodriguesR03}
	Lu{\'{\i}}s E.~T. Rodrigues and Michel Raynal.
	\newblock Atomic broadcast in asynchronous crash-recovery distributed systems
	and its use in quorum-based replication.
	\newblock {\em {IEEE} Trans. Knowl. Data Eng.}, 15(5):1206--1217, 2003.
	
	\bibitem{DBLP:conf/opodis/SakuraiOM04}
	Yusuke Sakurai, Fukuhito Ooshita, and Toshimitsu Masuzawa.
	\newblock A self-stabilizing link-coloring protocol resilient to byzantine
	faults in tree networks.
	\newblock In {\em {OPODIS}}, volume 3544 of {\em Lecture Notes in Computer
		Science}, pages 283--298. Springer, 2004.
	
	\bibitem{DBLP:conf/netys/SalemS18}
	Iosif Salem and Elad~Michael Schiller.
	\newblock Practically-self-stabilizing vector clocks in the absence of
	execution fairness.
	\newblock In {\em Networked Systems - 6th International Conference, {NETYS}},
	pages 318--333, 2018.
	
	\bibitem{DBLP:journals/cacm/Shamir79}
	Adi Shamir.
	\newblock How to share a secret.
	\newblock {\em Commun. {ACM}}, 22(11):612--613, 1979.
	
	\bibitem{DBLP:journals/tcs/SudoNYOKM12}
	Yuichi Sudo, Junya Nakamura, Yukiko Yamauchi, Fukuhito Ooshita, Hirotsugu
	Kakugawa, and Toshimitsu Masuzawa.
	\newblock Loosely-stabilizing leader election in a population protocol model.
	\newblock {\em Theor. Comput. Sci.}, 444:100--112, 2012.
	
	\bibitem{DBLP:journals/ieicet/SudoOKM20}
	Yuichi Sudo, Fukuhito Ooshita, Hirotsugu Kakugawa, and Toshimitsu Masuzawa.
	\newblock Loosely stabilizing leader election on arbitrary graphs in population
	protocols without identifiers or random numbers.
	\newblock {\em {IEICE} Trans. Inf. Syst.}, 103-D(3):489--499, 2020.
	
	\bibitem{DBLP:journals/tpds/SudoOKMDL19}
	Yuichi Sudo, Fukuhito Ooshita, Hirotsugu Kakugawa, Toshimitsu Masuzawa, Ajoy~K.
	Datta, and Lawrence~L. Larmore.
	\newblock Loosely-stabilizing leader election for arbitrary graphs in
	population protocol model.
	\newblock {\em {IEEE} Trans. Parallel Distributed Syst.}, 30(6):1359--1373,
	2019.
	
	\bibitem{DBLP:journals/tcs/SudoOKMDL20}
	Yuichi Sudo, Fukuhito Ooshita, Hirotsugu Kakugawa, Toshimitsu Masuzawa, Ajoy~K.
	Datta, and Lawrence~L. Larmore.
	\newblock Loosely-stabilizing leader election with polylogarithmic convergence
	time.
	\newblock {\em Theor. Comput. Sci.}, 806:617--631, 2020.
	
	\bibitem{DBLP:journals/corr/abs-1909-07453}
	Pierre Tholoniat and Vincent Gramoli.
	\newblock Certifying blockchain byzantine fault tolerance.
	\newblock {\em CoRR}, abs/1909.07453, 2019.
	
	\bibitem{DBLP:conf/podc/Toueg84}
	Sam Toueg.
	\newblock Randomized {Byzantine} agreements.
	\newblock In Tiko Kameda, Jayadev Misra, Joseph~G. Peters, and Nicola Santoro,
	editors, {\em Proceedings of the Third Annual {ACM} Symposium on Principles
		of Distributed Computing}, pages 163--178. {ACM}, 1984.
	
	\bibitem{DBLP:journals/ipl/TurpinC84}
	Russell Turpin and Brian~A. Coan.
	\newblock Extending binary {Byzantine} agreement to multivalued {Byzantine}
	agreement.
	\newblock {\em Inf. Process. Lett.}, 18(2):73--76, 1984.
	
	\bibitem{DBLP:journals/csur/RenesseA15}
	Robbert van Renesse and Deniz Altinbuken.
	\newblock Paxos made moderately complex.
	\newblock {\em {ACM} Comput. Surv.}, 47(3):42:1--42:36, 2015.
	
	\bibitem{DBLP:journals/comsur/XiaoZLH20}
	Yang Xiao, Ning Zhang, Wenjing Lou, and Y.~Thomas Hou.
	\newblock A survey of distributed consensus protocols for blockchain networks.
	\newblock {\em {IEEE} Commun. Surv. Tutorials}, 22(2):1432--1465, 2020.
	
	\bibitem{DBLP:conf/icpads/YuZY21}
	Shaolin Yu, Jihong Zhu, and Jiali Yang.
	\newblock Efficient two-dimensional self-stabilizing byzantine clock
	synchronization in {WALDEN}.
	\newblock In {\em {ICPADS}}, pages 723--730. {IEEE}, 2021.
	
	\bibitem{DBLP:journals/corr/abs-2203-14016}
	Shaolin Yu, Jihong Zhu, Jiali Yang, and Wei Lu.
	\newblock Expected constant time self-stabilizing byzantine pulse
	resynchronization.
	\newblock {\em CoRR}, abs/2203.14016, 2022.
	
\end{thebibliography}

\end{document}